\documentclass[reprint, amsmath,amssymb, aps, prd, nofootinbib, superscriptaddress ]{revtex4-2}

\usepackage{graphicx}
\usepackage{bm}
\usepackage{hyperref}
\usepackage{physics}
\usepackage{mathtools}
\usepackage{booktabs}
\usepackage{placeins}
\newcommand{\be}{\begin{equation}}
\newcommand{\ee}{\end{equation}}

\usepackage{color}
\usepackage{ulem}
	
\hypersetup{colorlinks=true, linkcolor=blue, citecolor=blue, urlcolor=blue}

\begin{document}
	
	\title{Gravitational-Wave Echoes from Layered Compact Objects: A Double-Shell Model}
	
	\author{Qi Su}
	\email{SQphysics@outlook.com}
	\affiliation{
		School of Physics and Optoelectronic Engineering,
		Beijing University of Technology,
		Beijing 100124, China
	}

	\author{Ding-Fang Zeng}
	\email{dfzeng@bjut.edu.cn}
	\affiliation{
		School of Physics and Optoelectronic Engineering,
		Beijing University of Technology,
		Beijing 100124, China
	}

	\begin{abstract}
		
		Layering is a ubiquitous feature of astrophysical objects. Motivated by the fact that physical black holes retain the layered structure of their progenitor stars when viewed in the time concepts synchronizable with the clock of an outside fixed-position probe, we investigate linear perturbations and gravitational-wave (GW) echoes from a compact object composed of two concentric thin shells. Compared with the single-shell case, the double-shell structure introduces an extra barrier in the effective potential and partitions the wave propagation space into four coupled effective cavities. As the mass ratio of the inner shell increases, new spectral peaks enter from the high-frequency side of the echo spectrum; the second and later peaks shift toward higher frequencies; and the lowest-frequency peak first shifts toward lower frequencies and then returns to its $q=0$ position. We call this variation pattern spectral-peak queueing (SQ). Its existence suggests that GW echoes can be used as probes for the internal structure of compact objects under consideration.
		
	\end{abstract}
	
	\maketitle
	
	\section{\label{sec:intro}INTRODUCTION}
	
	From the iron cores of massive stars to the crustal layers of neutron stars, nature repeatedly organizes matter into distinct concentric strata.
	This structural principle raises a pressing question for the most extreme objects in the universe: do compact objects formed by gravitational collapse retain analogous internal layering, and if so, can that layering be read off from GWs?
	The compact objects we have in mind are what we call physical black holes (PhBHs): objects whose gravitational collapse, as detected by exterior probes, never completes in finite time.
	In the time definition synchronizable with the clock of these probes through light signals, collapsing material asymptotically approaches but never crosses the event horizon (EH) radius, so the matter radius freezes just outside the would-be horizon.
	The Schwarzschild and Kerr solutions are idealized mathematical limits reached only after infinite duration by this time definition, and cannot be straightforwardly identified with the collapsed object that any external probe actually detects. So the frozen collapsar is the physically relevant object on observable timescales~\cite{Vachaspati2007,Barcelo2008}.
	
	Exact BH geometries such as Schwarzschild and Kerr are powerful idealizations, but treating them as the literal, direct outcome of finite-time collapse is precisely what leads to trouble: identifying a formed EH with the endpoint of collapse is what forces the conclusion that infalling information is destroyed as the EH evaporates~\cite{Hawking1976}.
	This information-loss puzzle has motivated a range of quantum-gravity (QG) scenarios and microscopic-state descriptions of black-hole entropy~\cite{Strominger:1996sh,Rovelli1996,Domagala2004}, in which the EH or interior is replaced, modified, or supplemented by microscopic degrees of freedom.
	The fuzzball proposal provides a prominent example of this viewpoint~\cite{Mathur2005,Mathur2009Fuzzballs}, while later developments such as the firewall argument and Page-curve calculations further sharpened the tension between semiclassical EHs and unitary evaporation~\cite{Almheiri2013,Almheiri2021Entropy,Raju2022Lessons}.
	Reframing the collapse endpoint as a PhBH, rather than a literal mathematical EH, is in this sense not merely a technical refinement: it is consistent with, and motivated by, the broader expectation from QG that near-horizon microstructure survives where the classical picture predicts none.
	
	Zeng's interpretation of BH entropy based on the complementarity idea \cite{complementarity1990thooft,complementarity1993,complementarity1993schoutens,dfzeng2025} provides the central motivation for the present work~\cite{Zeng2017,Zeng2018,Zeng2022,Zeng2023,Zeng2024}.
	By this idea \cite{dfzeng2025}, an ensemble of collapsars with different radial mass profiles in the Schwarzschild time definition and an over-cross oscillatory  ball ergodically passing through the central singularity in the Lemaître-time definition are two equally valid and complete descriptions of gravitational collapse. Exterior probes can only detect the first picture. The BH entropy simply equals the logarithm of the number of possible radial mass profiles measurable by the exterior probes, compatible with a unitary description of Hawking radiation.
	According to this interpretation, the layered structure of PhBHs is not an exotic addition but a thermodynamic necessity.
	
	GW echoes, as developed in some horizonless compact-object (ECO) framework, offer a concrete way to probe near-horizon structure observationally.
	Cardoso et al.~\cite{Cardoso2016PRL,Cardoso2016PRD} and subsequent studies \cite{Maselli2017,Maggio2019,Cardoso2019} showed that any ECO whose matter radius lies within the photon sphere acts as a resonant cavity, building on the earlier result that a thin-shell gravastar with no EH produces GW signatures qualitatively different from a Schwarzschild black hole~\cite{Pani2009}.
	GWs that would have been absorbed by an EH instead bounce between the photon-sphere potential barrier and the central reflective boundary, producing a train of GW echoes after the initial ringdown, probed through both time-domain echoes and frequency-domain resonances; a general template for such echo waveforms was developed in~\cite{Mark2017}.
	These echoes carry information about the near-horizon geometry that quasi-normal modes (QNMs) alone cannot resolve~\cite{Kokkotas1999,Berti2009}.
	Echo signals have been searched for in LIGO/Virgo data, with tentative claims and subsequent reanalyses or morphology-independent searches illustrating the current observational uncertainty \cite{Abedi2017,Westerweck2018,Tsang2020}, and their detectability with future detectors such as LISA has been assessed~\cite{Deppe2025}.
	The spectral structure of echo signals, in particular the comb of resonance peaks, has been studied analytically and numerically~\cite{Conklin2019,Maggio2019Echoes}. The broader question of how echo observations can constrain the nature of compact objects is an active area of research~\cite{Cardoso2019,DeLaurentis2025}, where gravastars and thin-shell mimickers remain useful theoretical laboratories~\cite{Mazur2004,Visser2004,Cattoen2005,Pani2009,Giri2025}.
	However, existing echo studies have largely focused on single-shell or single effective cavity models, which produce only a single comb of nearly equally spaced resonance peaks and are not designed to distinguish layered internal mass distributions.
	A directly preceding study by Cao, Liu and Zeng~\cite{Cao2026} investigated echo waveforms from PhBHs within this single-shell framework.
	
	Motivated by the layered picture, we extend the single thin-shell setup to a double-shell configuration, the minimal model capable of representing layered structure, and investigate how internal layering modifies GW echoes in both the time and frequency domains. 
	We introduce a mass-distribution parameter $q = m_1/M$ that continuously interpolates between the two single-shell limits ($q = 0$ and $q = 1$). 
	This construction has a built-in consistency check: at both endpoints the model must reproduce the single-shell waveform and spectrum, as verified in the analysis below.
	For intermediate $q$, the double-shell geometry introduces an additional potential peak, partitions the spacetime into multiple effective cavities with four characteristic length scales ($2L_1$, $2L_2$, $L_1+L_2$, $L_2-L_1$), and produces echo waveforms and spectra that are distinguishable for different
	values of the mass-distribution parameter $q$, even at fixed total mass and
	fixed number of shells.
	
	Our central result emerges when the spectral peaks are continuously tracked as q varies from $0$ to $1$. Although the $q=0$ and $q=1$ spectra are identical as a whole, the individual resonance peaks do not generally return to their $q=0$ positions as $q$ varies from 0 to 1. 
	We find that (i) inner effective cavity resonances produce high-frequency series entry; (ii) the second and later peaks undergo spectral-peak migration toward higher frequencies; and (iii) the first peak shifts toward lower frequencies before returning to its $q=0$ position as $q\to1$.
	These three behaviors collectively constitute spectral-peak queueing.
	The same phenomena persist for the scalar $\ell=2$ and Regge--Wheeler (RW) $\ell=2$ perturbations, indicating that they are generic consequences of the coupling among multiple effective cavities rather than artifacts of particular perturbation channels. 
	
	The paper is organized as follows.
	Section~\ref{sec:model} introduces the double-shell background metric and perturbation equations.
	Section~\ref{sec:potential} analyzes the effective potential.
	Section~\ref{sec:waveform} presents echo waveforms and their comparison with the single-shell baseline.
	Section~\ref{sec:spectrum} identifies the three constituent behaviors of spectral-peak queueing and relates them to their characteristic propagation lengths.
	Section~\ref{sec:robust} verifies robustness under higher angular momentum and RW perturbations.
	Section~\ref{sec:conclusion} summarizes conclusions and future directions.
	We use natural units $G = c = 1$ throughout.
	
	\section{\label{sec:model}Double-shell background and perturbation equations}
	
	We generalize the single thin-shell model of Cardoso et
	al.~\cite{Cardoso2016PRD}, itself building on the earlier thin-shell
	gravastar construction of~\cite{Pani2009}, to the double-shell case. The
	inner and outer shells are located at
	\begin{equation}
		r_1 = (2+\epsilon)\,m_1, \qquad r_2 = (2+\epsilon)\,M,
		\label{eq:shell_radii}
	\end{equation}
	where $M = m_1 + m_2$ is the total mass. We define the mass-distribution
	parameter
	\begin{equation}
		q = \frac{m_1}{M}.
		\label{eq:q_def}
	\end{equation}
	At fixed $M$, varying $q$ changes both the mass distribution and the inner-shell radius $r_1$, while $r_2$ remains fixed.
	At $q=0$ or $q=1$ the model reduces to the single-shell case; the waveform and spectrum at these two values are computed with the single-shell procedure and used throughout as the single-shell baseline.
	
	The background metric takes the piecewise Schwarzschild form
	\begin{equation}
		ds^2 = -f(r)\,dt^2 + \frac{dr^2}{f(r)} + r^2 d\Omega^2,
		\label{eq:metric}
	\end{equation}
	with
	\begin{equation}
		f(r) = 1 - \frac{2m(r)}{r}, \qquad
		m(r) =
		\begin{cases}
			0, & 0 < r < r_1, \\
			m_1, & r_1 \le r < r_2, \\
			M, & r \ge r_2.
		\end{cases}
		\label{eq:f_and_m}
	\end{equation}
	For the genuine double-shell configurations $0<q<1$, the tortoise coordinate is defined by $dr_*/dr=1/f(r)$. Integrating this relation in each smooth region and fixing the additive constants so that the exterior coordinate takes the standard Schwarzschild form gives
	\begin{equation}
		r_*(r) =
		\begin{cases}
			r - C_1, & 0 \le r < r_1, \\[2pt]
			r + 2m_1 \ln\!\left(\dfrac{r}{2m_1}-1\right) + C_2, & r_1 \le r < r_2, \\[6pt]
			r + 2M \ln\!\left(\dfrac{r}{2M}-1\right), & r \ge r_2,
		\end{cases}
		\label{eq:tortoise}
	\end{equation}
	where the constants
	\begin{align}
		C_1 &= 2m_1 \ln\!\left[\frac{r_2/(2m_1)-1}{r_1/(2m_1)-1}\right] - 2M \ln\!\left(\frac{r_2}{2M}-1\right), \label{eq:C1}\\
		C_2 &= 2M \ln\!\left(\frac{r_2}{2M}-1\right) - 2m_1 \ln\!\left(\frac{r_2}{2m_1}-1\right) \label{eq:C2}
	\end{align}
	enforce continuity of $r_*$ across the two shells.
	
	This is a fixed-background wave-propagation toy model, rather than a full thin-shell construction based on Israel junction conditions~\cite{Israel1966}. 
	The two shells serve only as boundaries between different Schwarzschild regions. We do not evolve the shell matter, include shell stress-energy perturbations, or model any interaction between shell matter and the gravitational perturbation. The regions are joined directly at the shells, with no smooth transition layer and no requirement that the metric coefficients themselves be continuous; we require only that $r_*$ be continuous, with matching conditions imposed at the level of the perturbation field. Since no additional $\delta$-function potential is placed at the shells, the effective potential is computed separately in each smooth region and joined through one-sided limits. Equivalently, integrating the master equation across an infinitesimal neighborhood of each shell gives the matching conditions
	\begin{equation}
		[\Psi]_{r_i} = 0, \qquad [\partial_{r_*}\Psi]_{r_i} = 0, \qquad i=1,2,
		\label{eq:matching}
	\end{equation}
	where $[A]_{r_i} \equiv A(r_i^+) - A(r_i^-)$. In the radial coordinate $r$, $[f(r)\,\partial_r\Psi]_{r_i} = 0$.
	
	We solve the following one-dimensional master equation for the perturbative probe field,
\begin{equation}
\Big[-\frac{\partial^2}{\partial t^2} +\frac{\partial^2}{\partial r_\star^2} -V_\ell(r)\Big]\Psi(t,r_\star)=0,
\label{masterEq}
\end{equation}
where the effective potential $V_\ell(r)$ depends on the probe field type. For the scalar probe, it is given by
\begin{equation}
V^{\rm scalar}\ell(r)
=f(r)\left[\frac{\ell(\ell+1)}{r^2}
+\frac{f'(r)}{r}\right],
\label{scalarPotential}
\end{equation}
where $f'(r)$ denotes the ordinary derivative within each smooth region and carries no distributional contribution at the shells. For the odd-parity gravitational perturbation described by the Regge–Wheeler (RW) equation, the corresponding effective potential is
\begin{equation}
V^{\rm RW}\ell(r)
=f(r)\left[\frac{\ell(\ell+1)}{r^2}
-\frac{6m(r)}{r^3}\right].
\label{RWPotential}
\end{equation}
	
	The same Gaussian pulse is used as initial data for all values of $q$. We evolve $\Psi$ numerically using the diamond scheme~\cite{diamond} in double-null coordinates $u = t-r_*$, $v = t+r_*$,
	\begin{equation}
		\Psi_N = \Psi_W + \Psi_E - \Psi_S - \frac{\Delta u \Delta v}{8}\,V_S(\Psi_W+\Psi_E).
		\label{eq:diamond}
	\end{equation}
	Regularity at $r=0$ requires, for the $\ell$-th harmonic, $\Psi_\ell(t,r) \sim r^{\ell+1}$ as $r\to 0$; we implement this numerically as a reflecting boundary, $\Psi(t,r=0)=0$. The outer boundary is placed at sufficiently large $r_*$ with an outgoing-wave condition, avoiding contamination of the echo signal by boundary reflections within the time window considered.
	
	Throughout this work we fix $\epsilon = 10^{-6}$ and observe the waveform at $r_*^{\rm obs} = 20M$, on a double-null grid with $\Delta u=\Delta v = 0.05$ for $40000$ evolution steps. To check numerical convergence, we repeated the evolution with $\Delta u=\Delta v=0.025$; the echo waveform and the positions of the main spectral peaks show no visible change within the time window relevant to this paper.

	\section{\label{sec:potential}Effective potential and effective cavity structure}
	
	\begin{figure*}[t]
		\centering
		
		\begin{minipage}{0.32\textwidth}
			\centering
			\includegraphics[width=\linewidth]{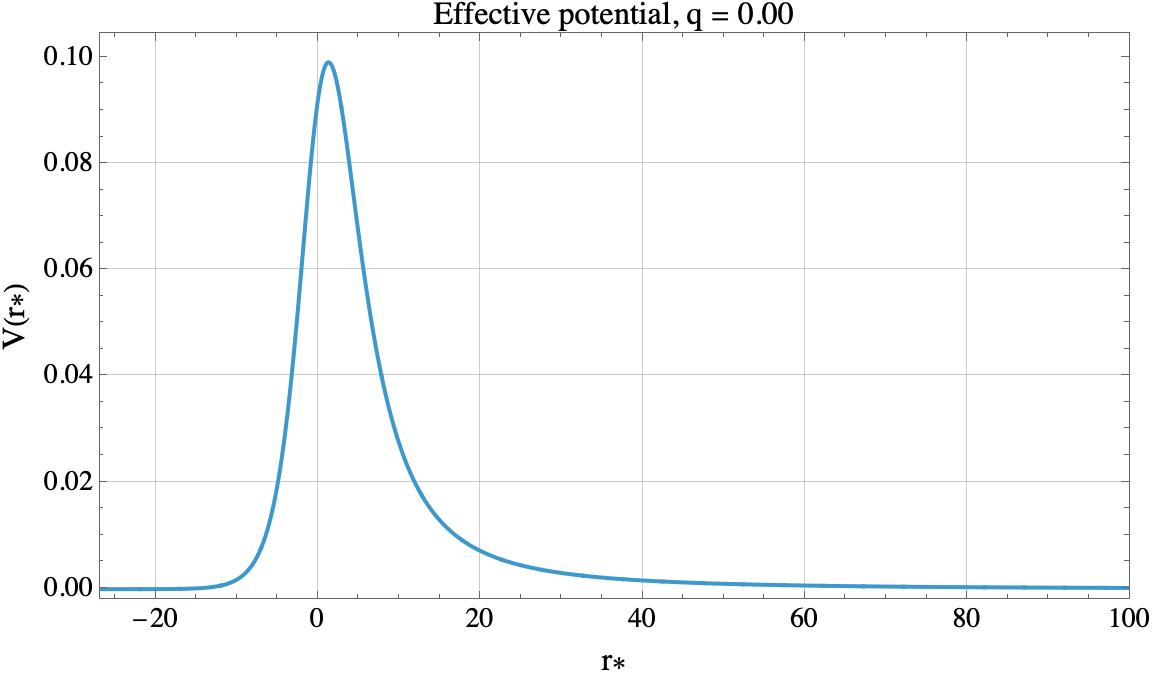}
		\end{minipage}
		\hfill
		\begin{minipage}{0.32\textwidth}
			\centering
			\includegraphics[width=\linewidth]{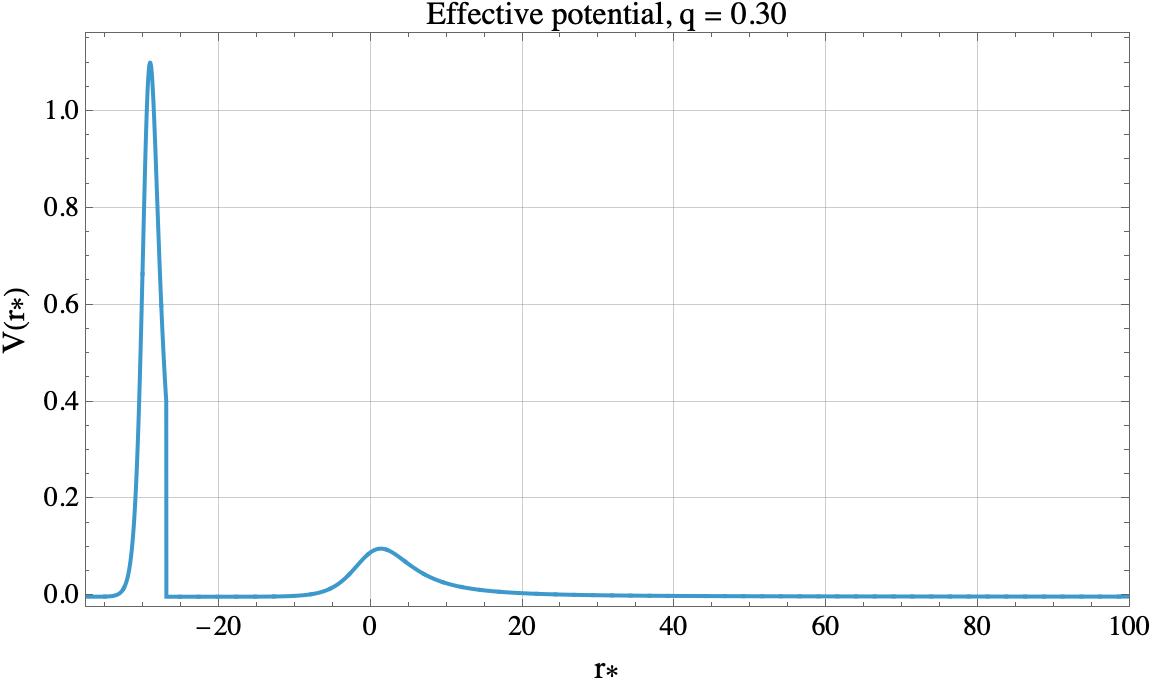}
		\end{minipage}
		\hfill
		\begin{minipage}{0.32\textwidth}
			\centering
			\includegraphics[width=\linewidth]{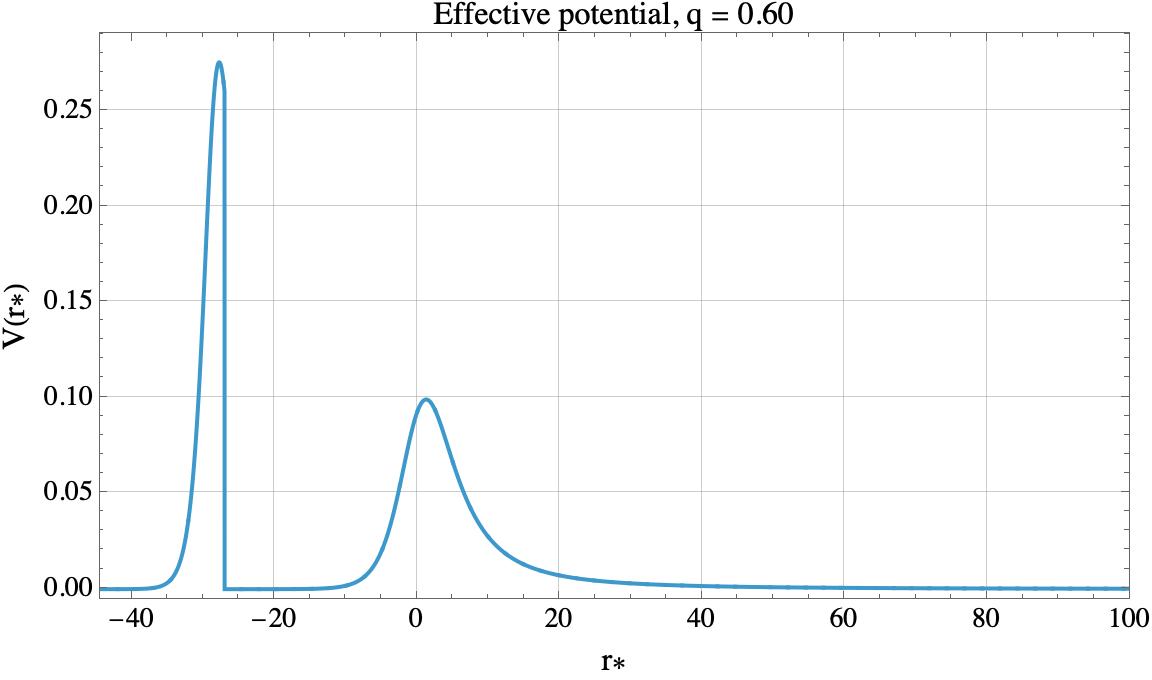}
		\end{minipage}
		
		\vspace{0.35cm}
		\begin{minipage}{0.32\textwidth}
			\centering
			\includegraphics[width=\linewidth]{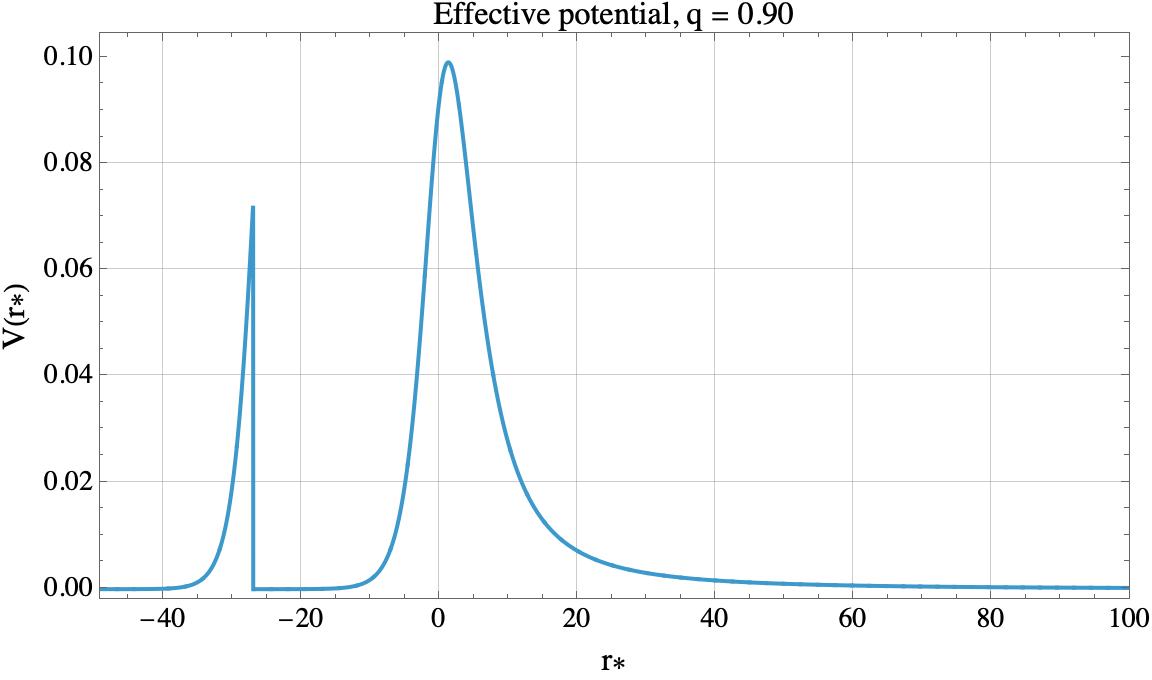}
		\end{minipage}
		\hfill
		\begin{minipage}{0.32\textwidth}
			\centering
			\includegraphics[width=\linewidth]{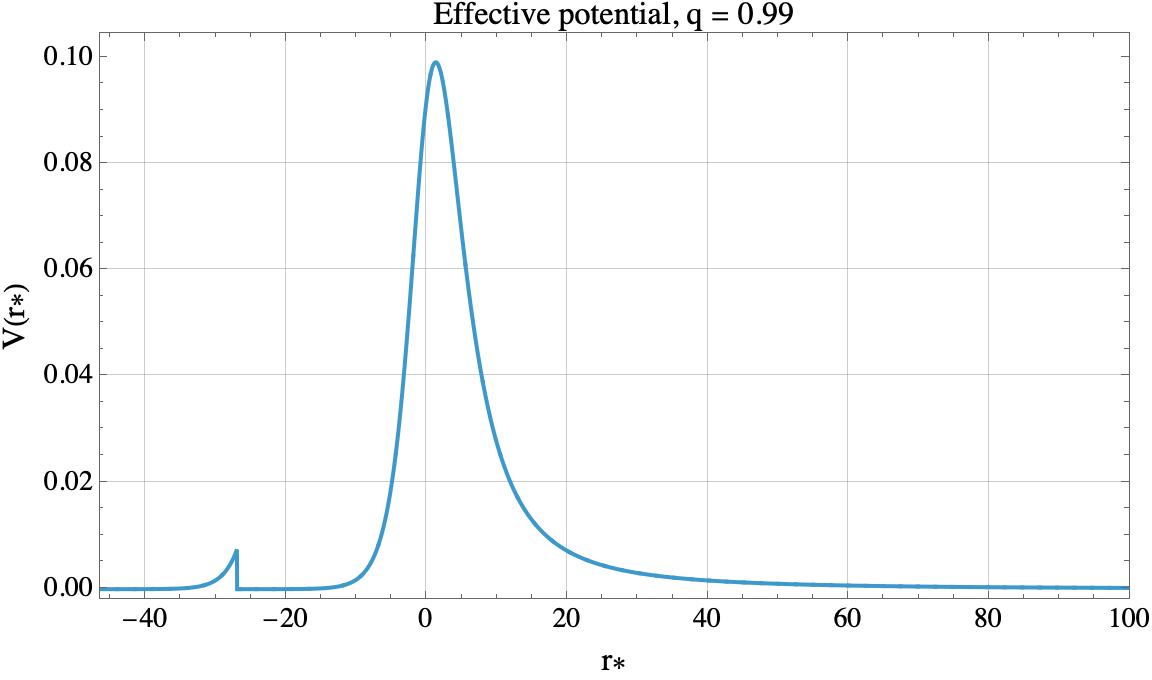}
		\end{minipage}
		\hfill
		\begin{minipage}{0.32\textwidth}
			\centering
			\includegraphics[width=\linewidth]{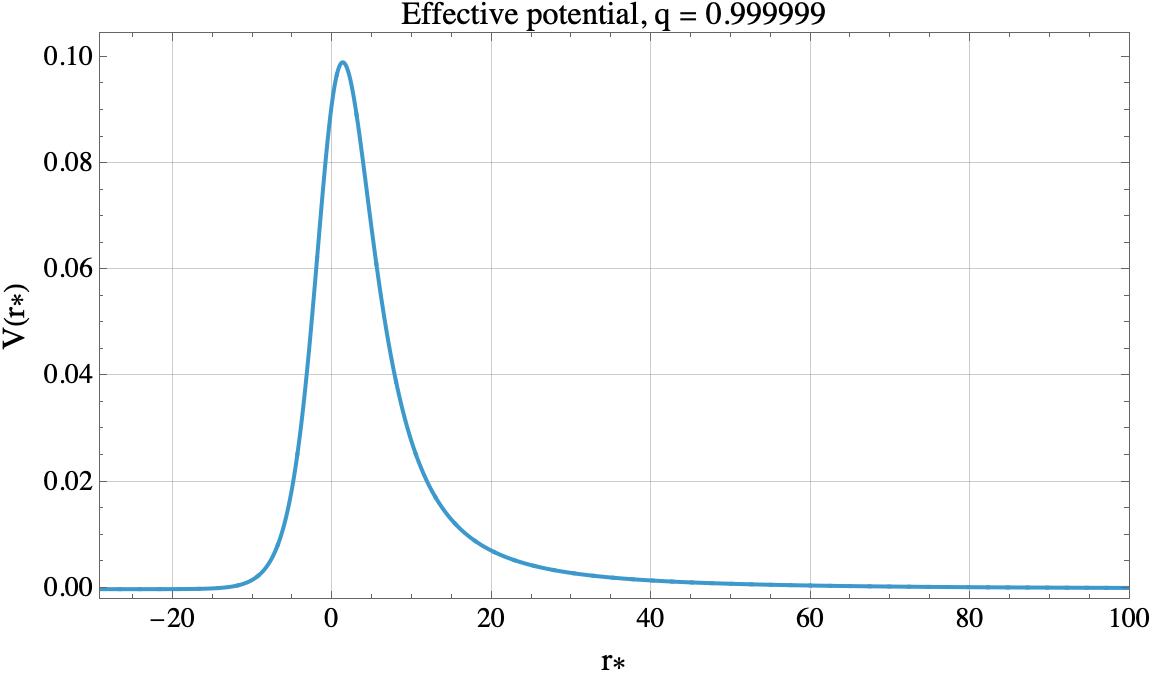}
		\end{minipage}
		
		\caption{
			Scalar effective potential $V_\ell(r_*)$ for several values of $q$ in the fiducial $\ell=1$ channel.
		}
		\label{fig:V_s}
	\end{figure*}
	
	Using the background and tortoise coordinate introduced in Sec.~\ref{sec:model}, we first analyze the scalar $\ell=1$ effective potential. The scalar $\ell=2$ and RW $\ell=2$ perturbations are discussed later in Sec.~\ref{sec:robust} as robustness checks. Figure~\ref{fig:V_s} shows $V_\ell(r_*)$ for several values of $q$.
	
	In the single-shell limit $q=0$, the potential has a single dominant barrier: the photon-sphere barrier of the exterior Schwarzschild region. The center $r=0$ is a regular reflecting boundary, and the apparent divergence of the scalar potential there is a centrifugal artifact of the multipole decomposition rather than a physical singularity.
	
	For $0<q<1$, the double-shell structure introduces an additional potential peak. It  changes the system from a single-barrier configuration into a coupled double-barrier system. As $q$ increases, this inner potential peak separates from the center and moves toward larger $r_*$ while its height decreases, consistent with the photon-sphere mass scaling
	\begin{equation}
		V_{\rm peak} \propto \frac{1}{m(r)^2}.
		\label{eq:Vpeak_scaling}
	\end{equation}
	For larger $q$, the inner potential peak is increasingly truncated by the outer shell and the exterior Schwarzschild region. This truncation becomes important when the nominal inner photon-sphere radius $r\simeq 3m_1$ approaches or exceeds the outer shell position $r_2=(2+\epsilon)M$, namely around $q\sim(2+\epsilon)/3\simeq2/3$ for $\epsilon\ll1$.
	In this regime the shape, height, and width of the inner potential peak are no longer set by a single, complete photon-sphere barrier, and the coupling between the associated effective cavities becomes particularly important.
	As shown in later sections, this is also the region where the late-echo differences and spectral changes are most pronounced. 
	
	In the limit $q\to 1$, the double-shell configuration reduces to a single shell and the inner potential peak disappears as an independent structure. Across the full range $q\in[0,1]$, the effective potential thus passes through the appearance, displacement, truncation, and disappearance of the additional potential peak --- the geometric mechanism underlying the waveform changes and spectral-peak queueing discussed below.
	
	\begin{figure}[t]
	\centering
	\begin{minipage}{0.48\textwidth}
		\centering
		\includegraphics[width=\linewidth]{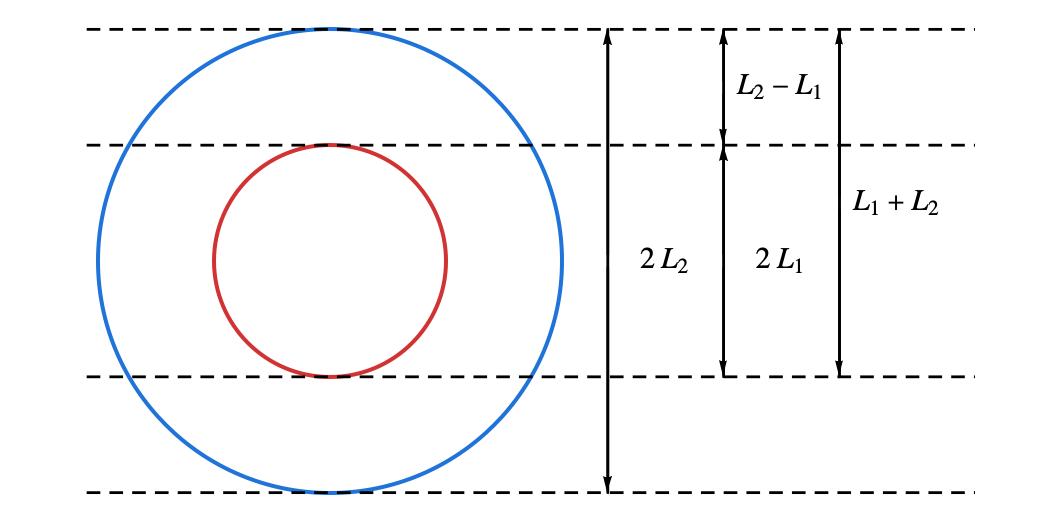}
	\end{minipage}
	
	\caption{
		Characteristic propagation lengths of the double-shell system. The red and blue circles denote the locations of the inner and outer effective-potential barriers, respectively.
	}
	\label{fig:cavity}
	\end{figure}
	
	To characterize the propagation structure set by the effective potential, let $r_{*,1}$ and $r_{*,2}$ denote the tortoise-coordinate positions of the inner and outer potential peaks. We define their distances from the central $r=0$ as
	\begin{equation}
		L_i=r_{*,i}-r_*(0),\qquad i=1,2,
	\end{equation}
	with $L_1<L_2$. Unfolding the central boundary gives four characteristic propagation lengths,
	\begin{equation}
		2L_1, \quad 2L_2, \quad L_1+L_2, \quad L_2-L_1.
		\label{eq:cavity_lengths}
	\end{equation}
	Here $2L_1$ is the round-trip propagation length between the central reflecting boundary and the inner barrier; $L_2-L_1$ is the inter-barrier separation. $2L_2$ and $L_1+L_2$ represent global or composite propagation paths.
	These are not eigenlengths of strictly isolated effective cavities, but characteristic propagation scales jointly set by the finite-height barriers and the central reflecting boundary. They provide the geometric scales used to interpret high-frequency series entry, spectral-peak migration, and the first-peak return analyzed in Sec.~\ref{sec:spectrum}.

	\section{\label{sec:waveform}Waveform and echo feature analysis}
	
\begin{figure*}[t]
		\centering
		
		\begin{minipage}{0.32\textwidth}
			\centering
			\includegraphics[width=\linewidth]{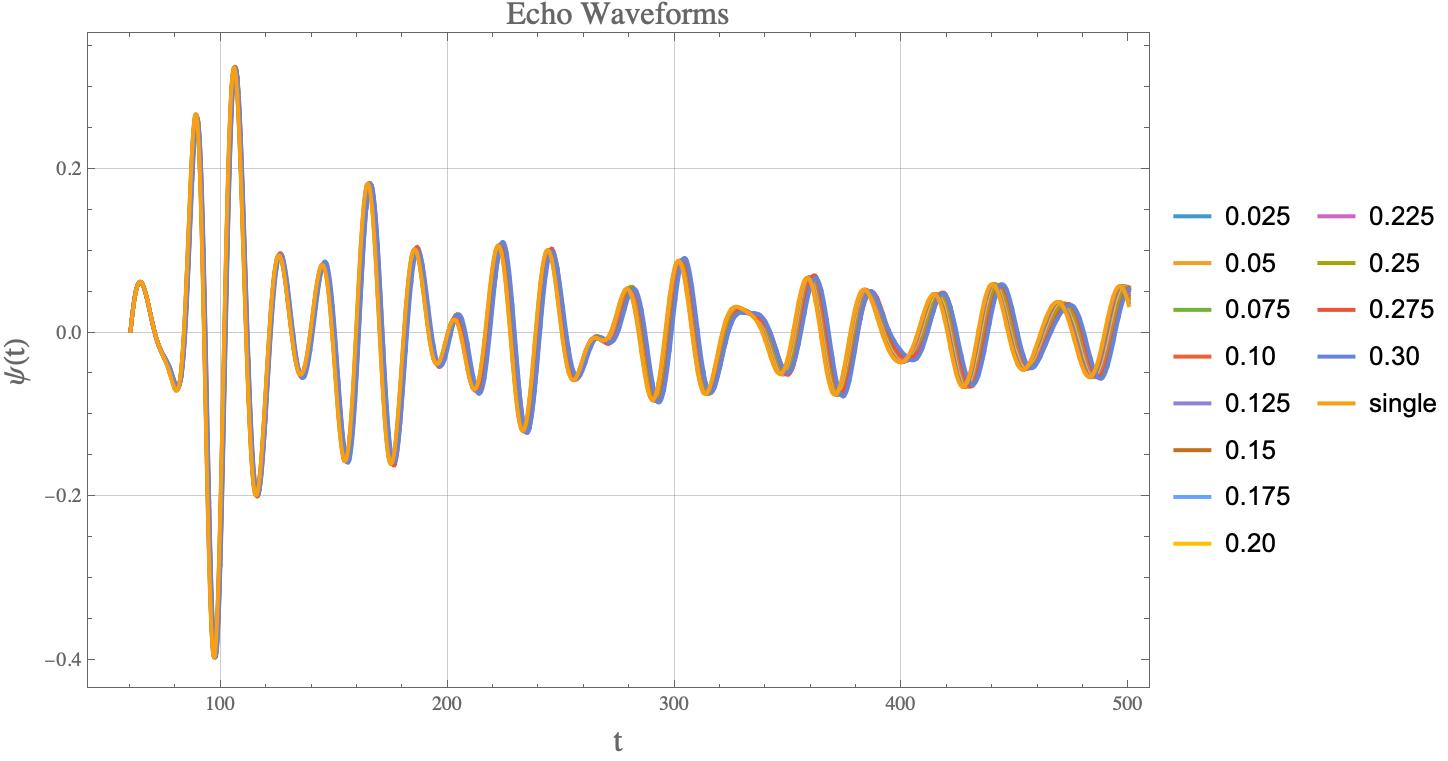}
		\end{minipage}
		\hfill
		\begin{minipage}{0.32\textwidth}
			\centering
			\includegraphics[width=\linewidth]{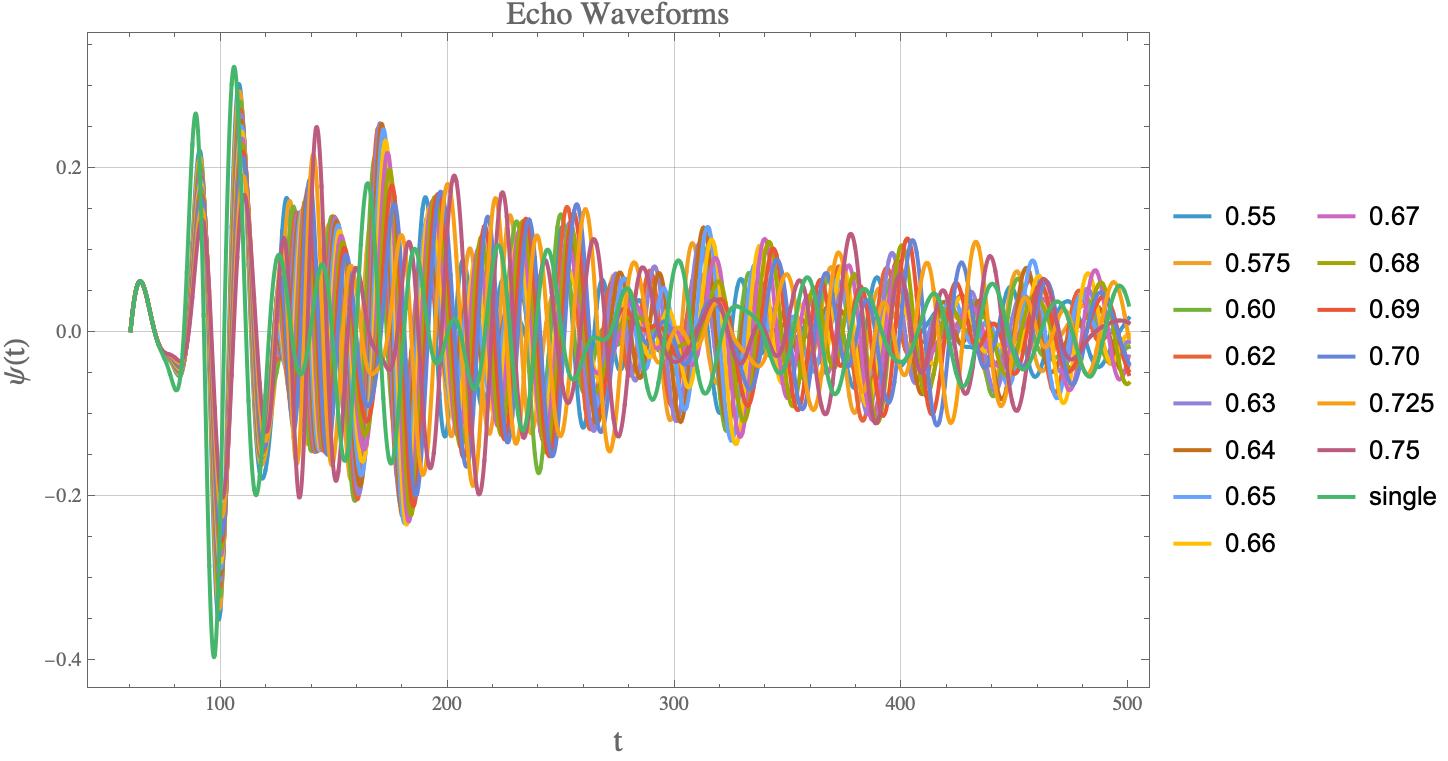}
		\end{minipage}
		\hfill
		\begin{minipage}{0.32\textwidth}
			\centering
			\includegraphics[width=\linewidth]{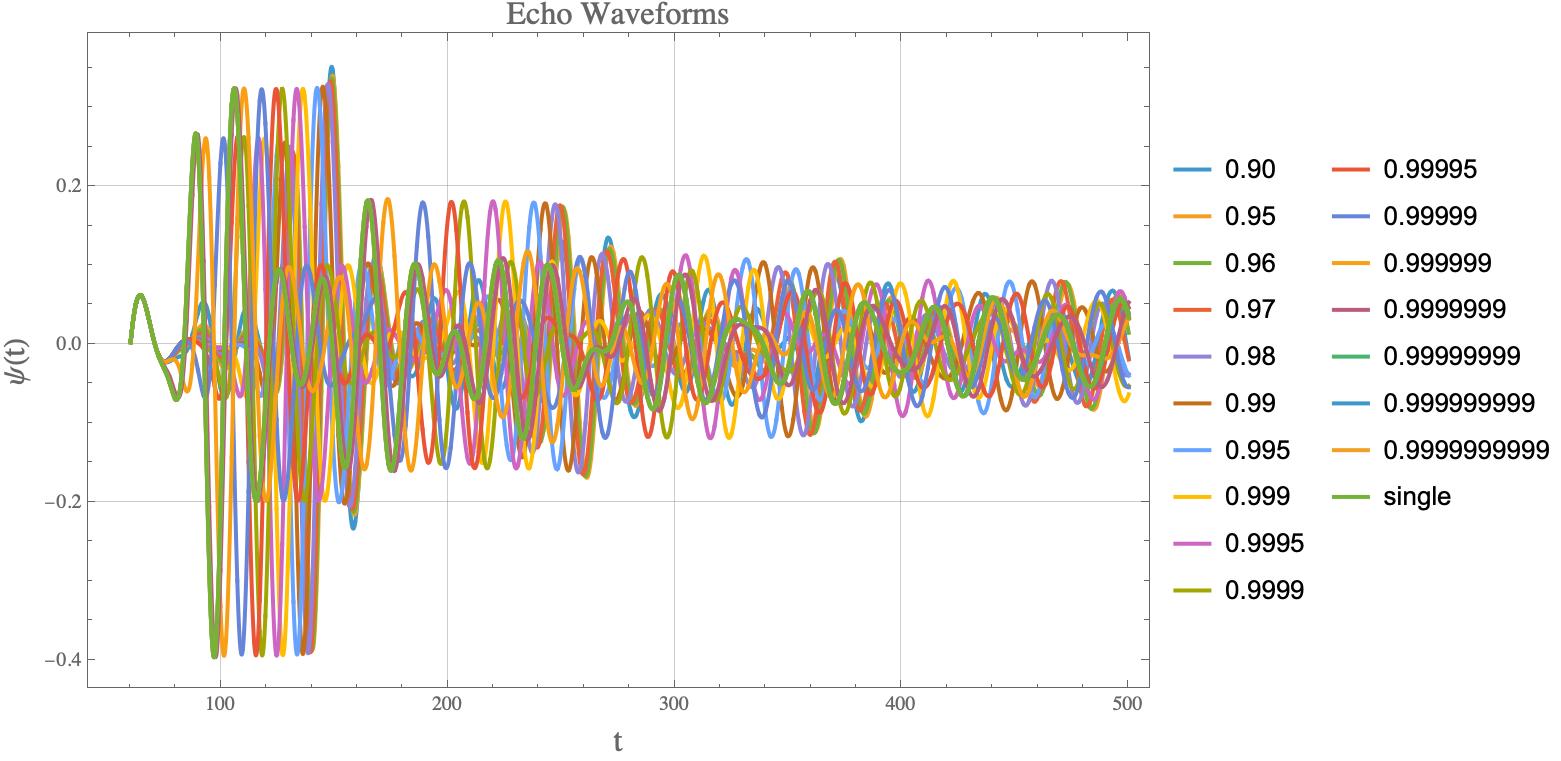}
		\end{minipage}
		
		\vspace{0.35cm}
		\begin{minipage}{0.32\textwidth}
			\centering
			\includegraphics[width=\linewidth]{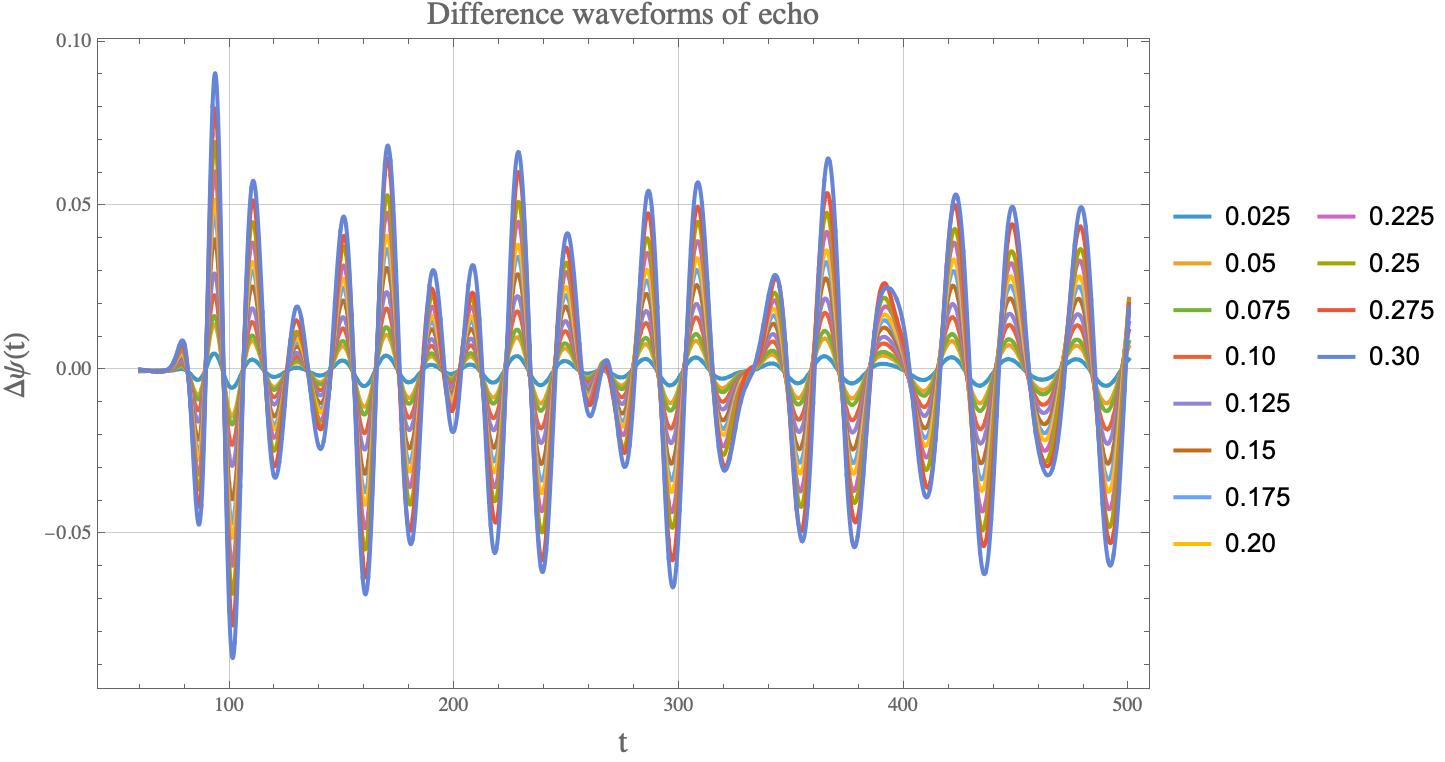}
		\end{minipage}
		\hfill
		\begin{minipage}{0.32\textwidth}
			\centering
			\includegraphics[width=\linewidth]{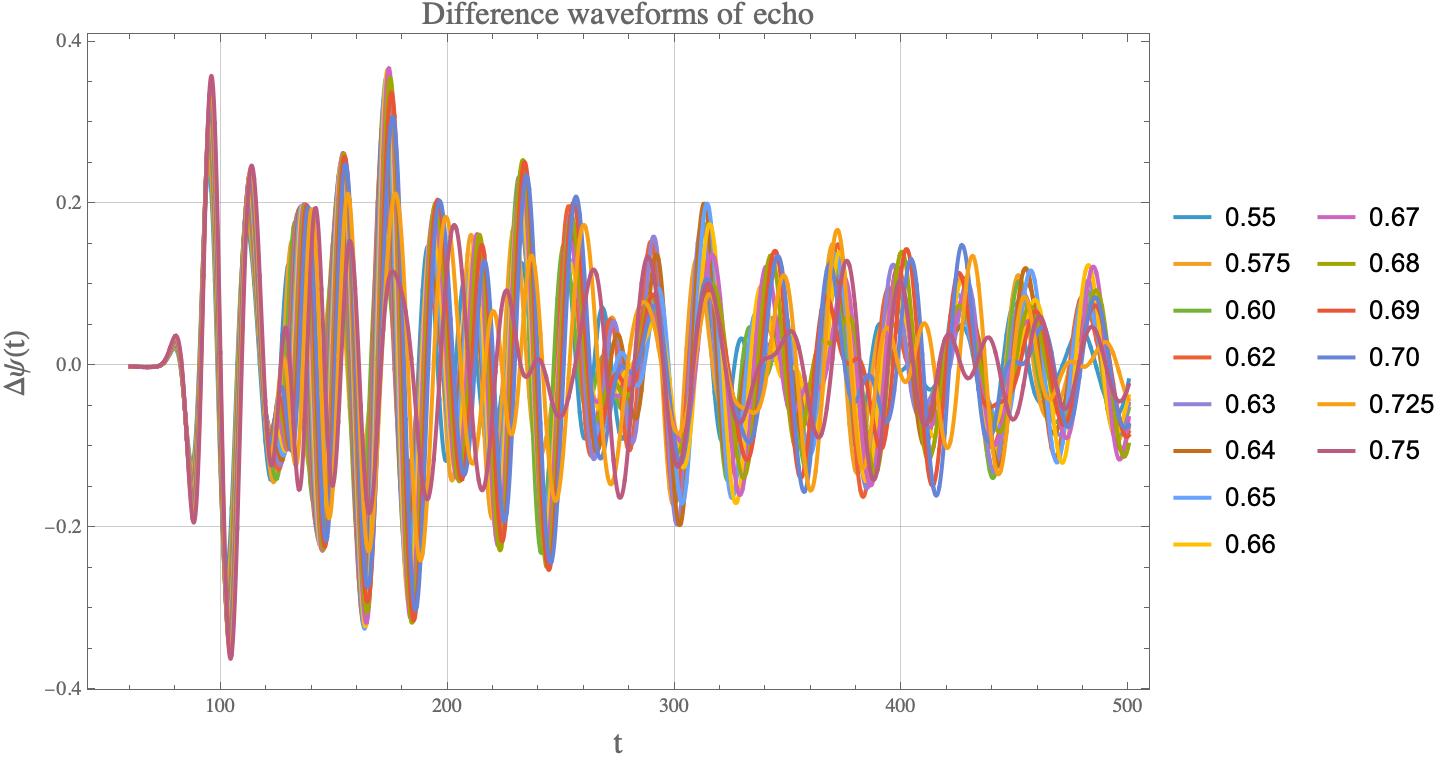}
		\end{minipage}
		\hfill
		\begin{minipage}{0.32\textwidth}
			\centering
			\includegraphics[width=\linewidth]{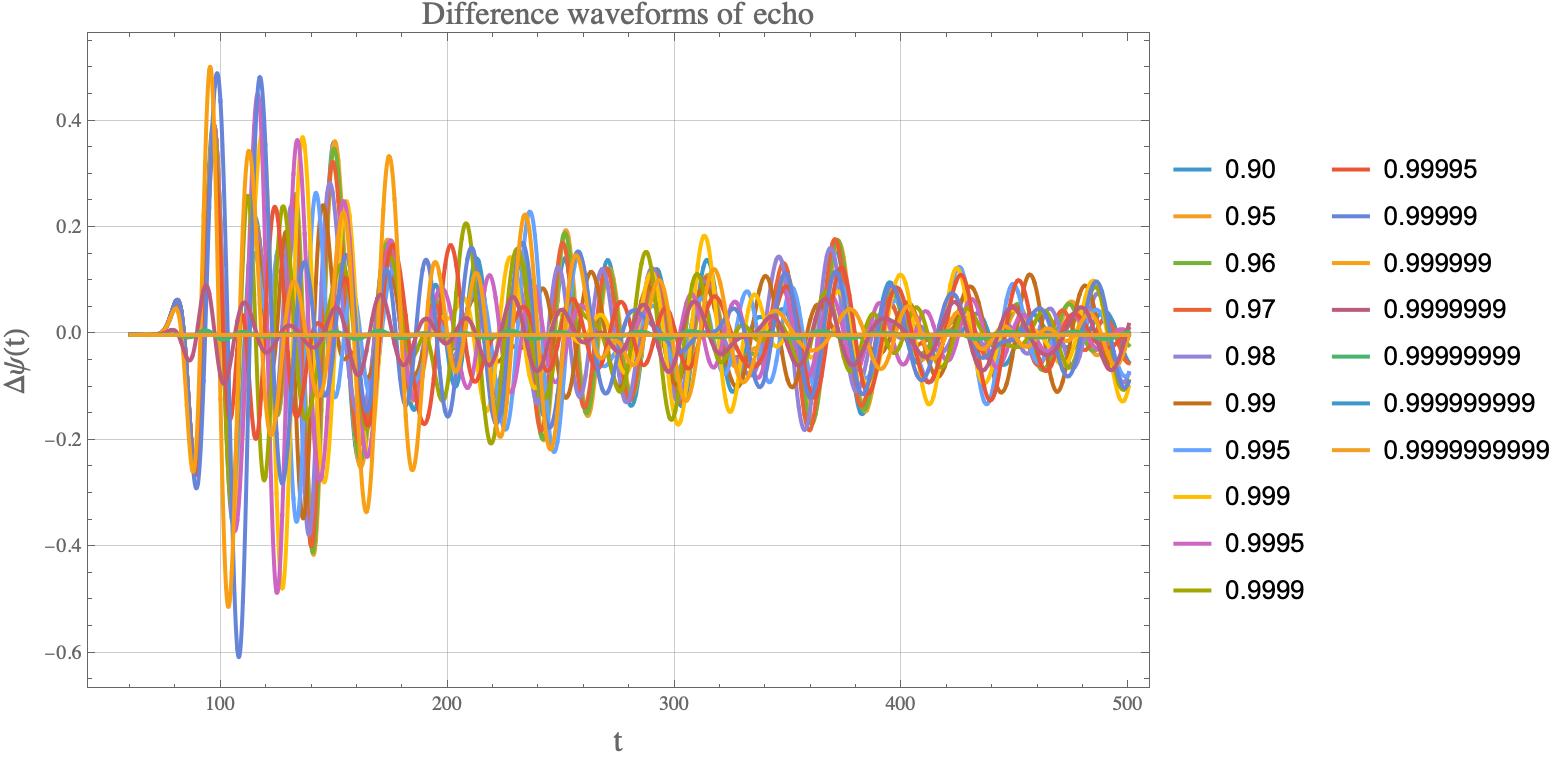}
		\end{minipage}
		
		\caption{
			Echo waveforms (top) and difference waveforms between the double-shell
			and single-shell cases (bottom).
		}
		\label{fig:WF_01}
	\end{figure*}

	We compare the double-shell waveforms with the common single-shell baseline obtained at $q=0$ and $q=1$. The interval $80M\leq t\leq120M$ defines the first echo, while $t > 120M$ defines the late-echo sector. Figure~\ref{fig:WF_01} provides a qualitative overview: for small $q$, the dominant differences are a time-scale change and an overall phase delay, whereas larger $q$ modifies both the amplitudes and shapes of the echoes. The single-shell waveform is recovered only extremely close to $q=1$.
	
	For a time window $I=[t_{\min},t_{\max}]$, we define the difference waveform
	\begin{equation}
		\Delta\Psi_q(t)
		=
		\Psi_q(t,r_*^{\rm obs})
		-
		\Psi_{\rm single}(t,r_*^{\rm obs}).
	\end{equation}
	We characterize this difference through the maximum residual $\max_{t\in I}|\Delta\Psi_q(t)|$ and the root-mean-square residual
	\begin{equation}
		{\rm RMS}_I(q)
		=
		\left[
		\frac{1}{t_{\max}-t_{\min}}
		\int_{t_{\min}}^{t_{\max}}
		|\Delta\Psi_q(t)|^2\,dt
		\right]^{1/2}.
	\end{equation}
	The former measures the largest local deviation from the single-shell waveform, whereas the latter measures the overall difference across the full time window. We also track $\max_{t\in I}|\Psi_q(t)|$ to determine whether the first and late echoes are suppressed or enhanced.
	
	None of these amplitude-based quantities distinguishes a genuine change of waveform shape from an overall delay or a rescaling of the time axis. We therefore introduce an affine time-aligned mismatch (TAM). With
	\begin{equation}
		t_c=\frac{t_{\min}+t_{\max}}{2},
		\qquad
		t_d(t)=t_c+a(t-t_c)+\tau,
	\end{equation}
	the normalized overlap is
	\begin{equation}
		{\cal O}_I(a,\tau;q)
		=
		\frac{
			\displaystyle
			\int_I
			\Psi_q\!\left(t_d(t),r_*^{\rm obs}\right)
			\Psi_{\rm single}\!\left(t,r_*^{\rm obs}\right)\,dt
		}{
			\displaystyle
			\left[
			\int_I
			\Psi_q^2\!\left(t_d(t),r_*^{\rm obs}\right)\,dt
			\right]^{1/2}
			\left[
			\int_I
			\Psi_{\rm single}^2\!\left(t,r_*^{\rm obs}\right)\,dt
			\right]^{1/2}
		},
	\end{equation}
	and
	\begin{equation}
		{\cal M}_I(q)
		=
		1-\max_{a,\tau}{\cal O}_I(a,\tau;q).
	\end{equation}
	A small ${\cal M}_I$ means that, after an overall time shift and time-scale rescaling, the normalized double-shell waveform retains nearly the same morphology as the single-shell waveform. Overall amplitude changes are measured separately through $\max_{t\in I}|\Psi_q(t)|$. A large ${\cal M}_I$ instead identifies a genuine deformation of the normalized waveform shape that cannot be removed by an affine transformation of time.
	
\begin{figure}[t]
		\centering
		
		\begin{minipage}{0.48\textwidth}
			\centering
			\includegraphics[width=\linewidth]{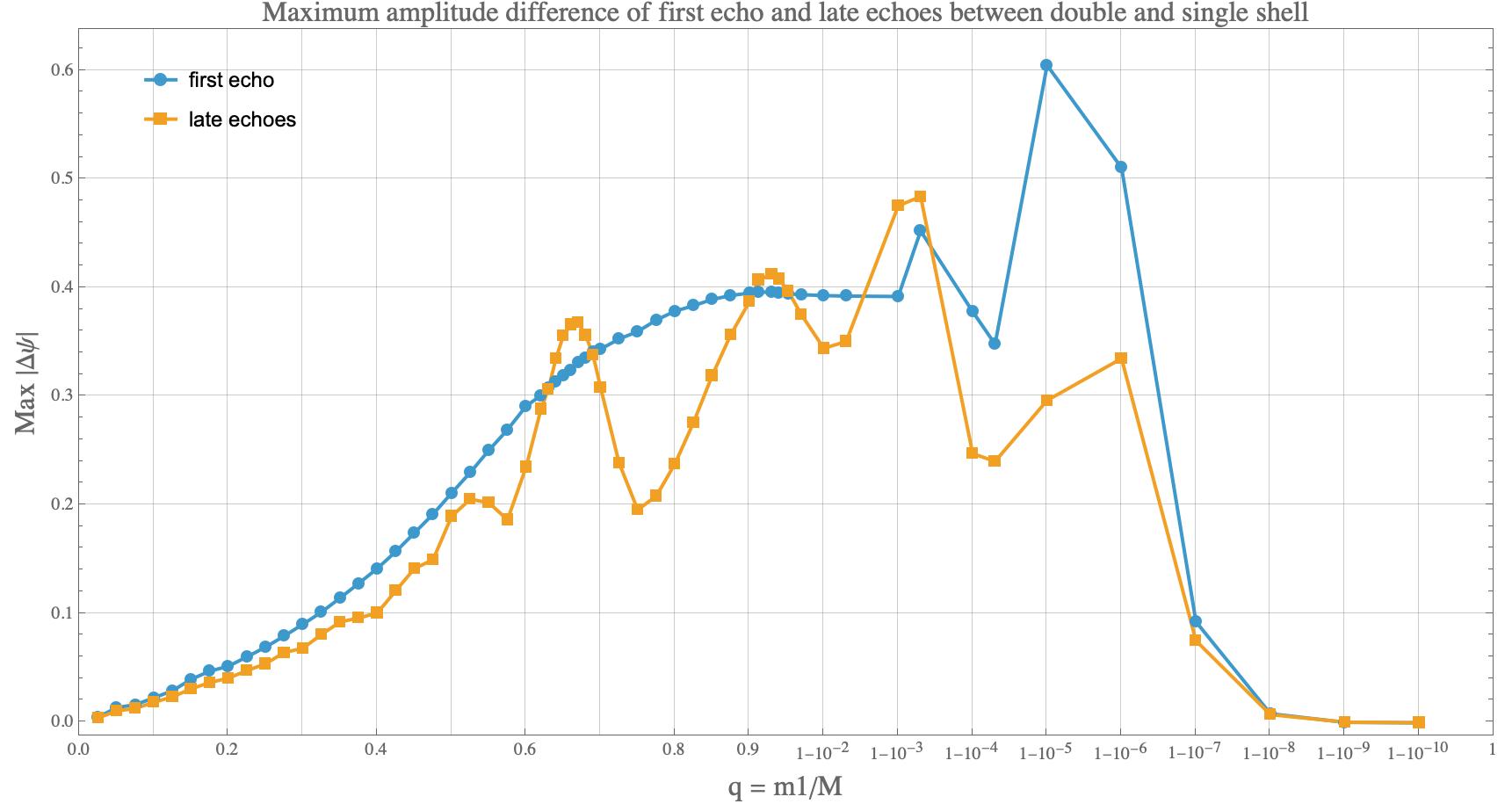}
		\end{minipage}
		
		\vspace{0.35cm}
		\begin{minipage}{0.48\textwidth}
			\centering
			\includegraphics[width=\linewidth]{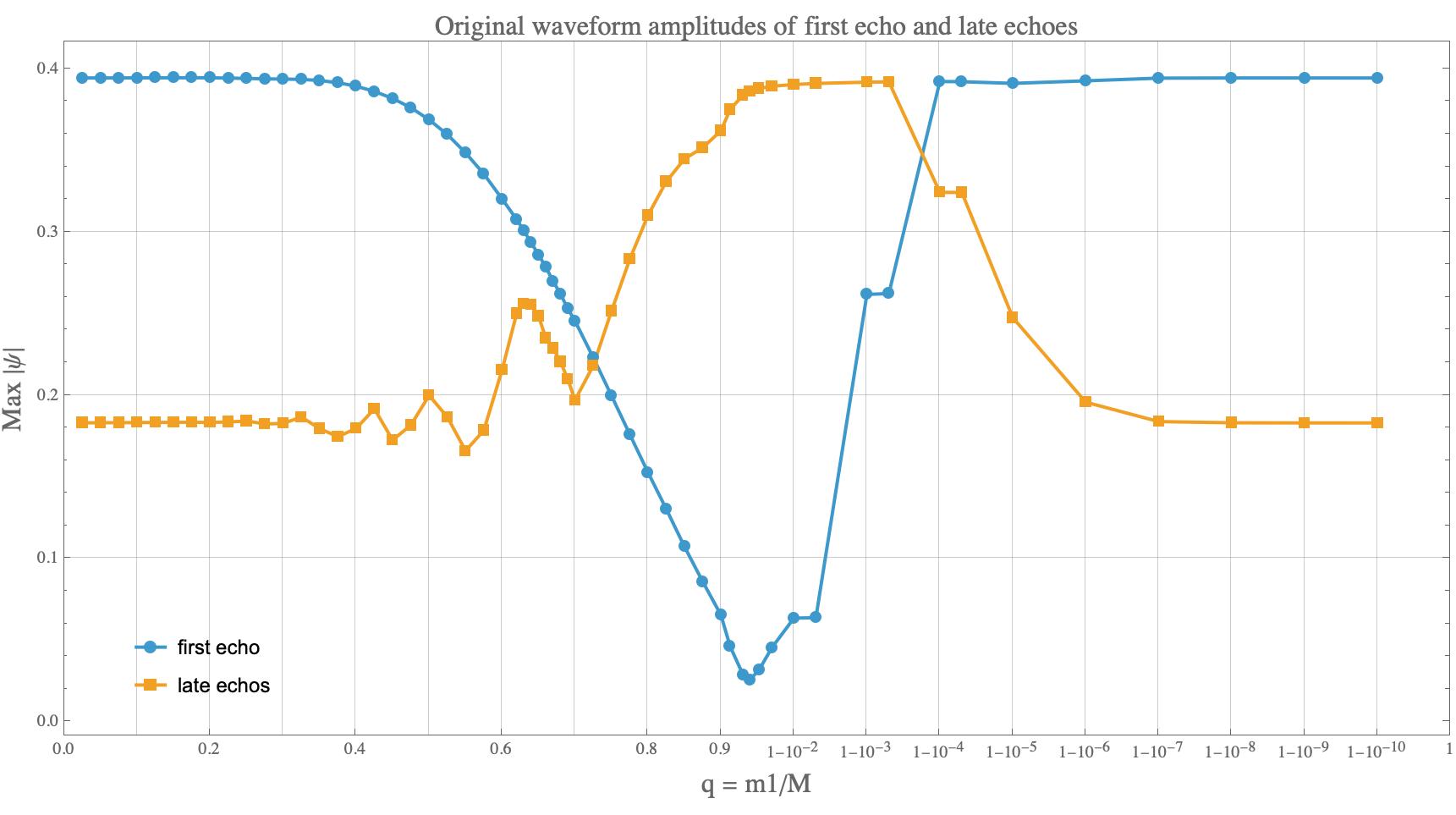}
		\end{minipage}
		
		\caption{
			Maximum waveform differences and echo amplitudes as functions
			of the mass-distribution parameter $q$. The upper panel shows
			$\max_{t\in I}|\Delta\Psi_q(t)|$ for the first-echo and late-echo
			windows, where $\Delta\Psi_q=\Psi_q-\Psi_{\rm single}$. The lower panel
			shows the corresponding maximum waveform amplitudes
			$\max_{t\in I}|\Psi_q(t)|$.
		}
		\label{fig:psi_01}
	\end{figure}

	The upper panel of Fig.~\ref{fig:psi_01} shows the maximum difference between the double-shell and single-shell waveforms in the first- and late-echo windows. As $q$ increases from zero, both differences exhibit an overall increasing trend, showing that the redistribution of a fixed total mass progressively separates the double-shell waveform from the single-shell baseline. The dependence on $q$ is non-monotonic and contains several local variations. The differences remain substantial over most of the parameter range and decrease rapidly toward zero only when $q$ becomes extremely close to unity.
	
	The lower panel of Fig.~\ref{fig:psi_01} compares the maximum amplitudes of the first and late echoes. For $q<0.4$, both remain close to their single-shell values, indicating that the relative strengths of the two echo sectors are nearly unchanged. As $q$ increases further, the first echo is progressively suppressed, whereas the late echoes are enhanced overall, with non-monotonic variations superposed on this trend. A local enhancement of the late-echo amplitude appears near $q\simeq2/3$. The contrast between the two sectors becomes strongest near $q\simeq0.95$, where the first-echo amplitude reaches a pronounced minimum. Beyond this region, the first echo begins to recover and the late echoes decrease toward their single-shell value. The double-shell structure therefore produces a redistribution of the observable waveform amplitude between the first-echo and late-echo sectors, rather than a uniform enhancement or suppression of the entire echo train.
	
\begin{figure}[t]
		\centering
		
		\begin{minipage}{0.48\textwidth}
			\centering
			\includegraphics[width=\linewidth]{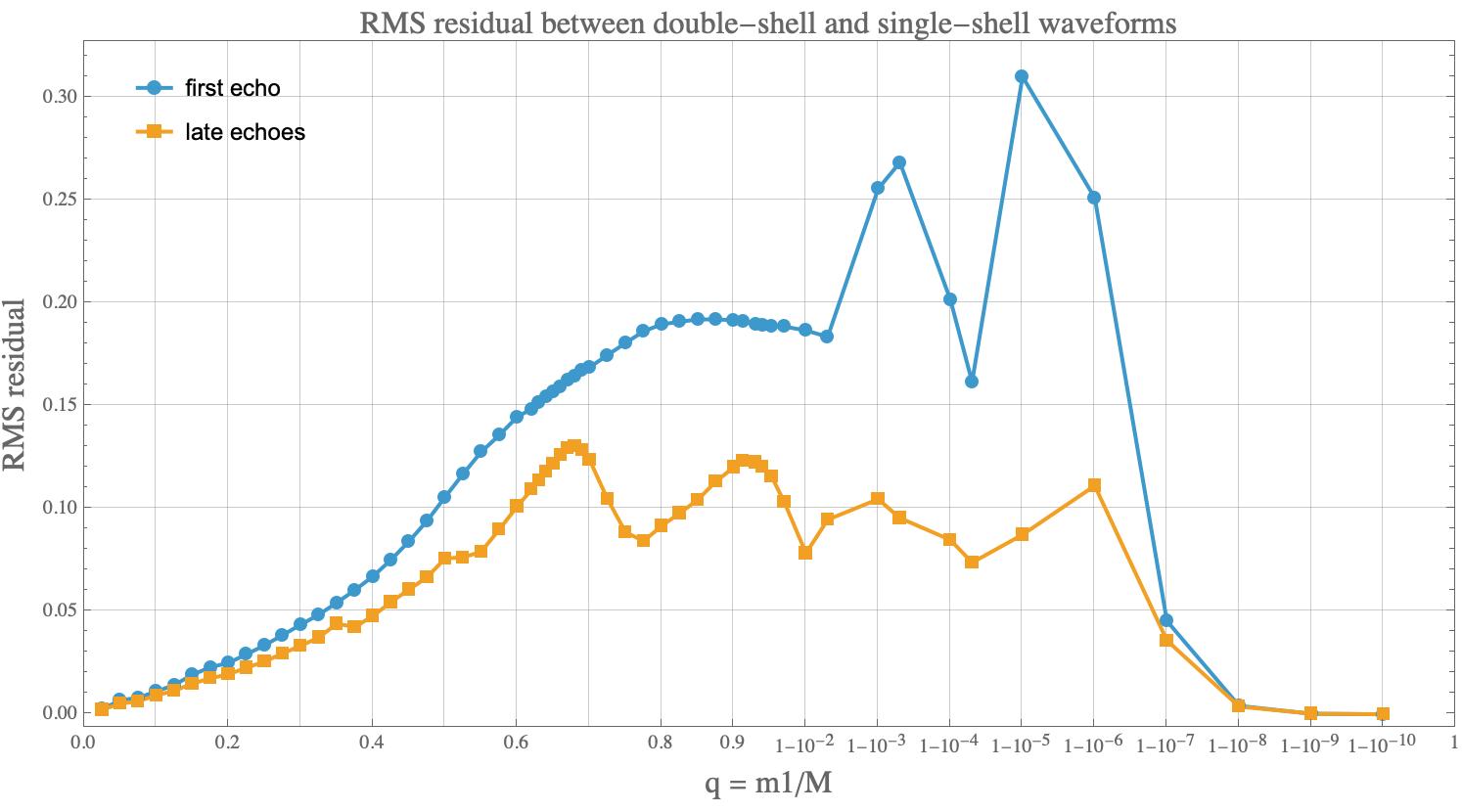}
		\end{minipage}
		
		\vspace{0.35cm}
		\begin{minipage}{0.48\textwidth}
			\centering
			\includegraphics[width=\linewidth]{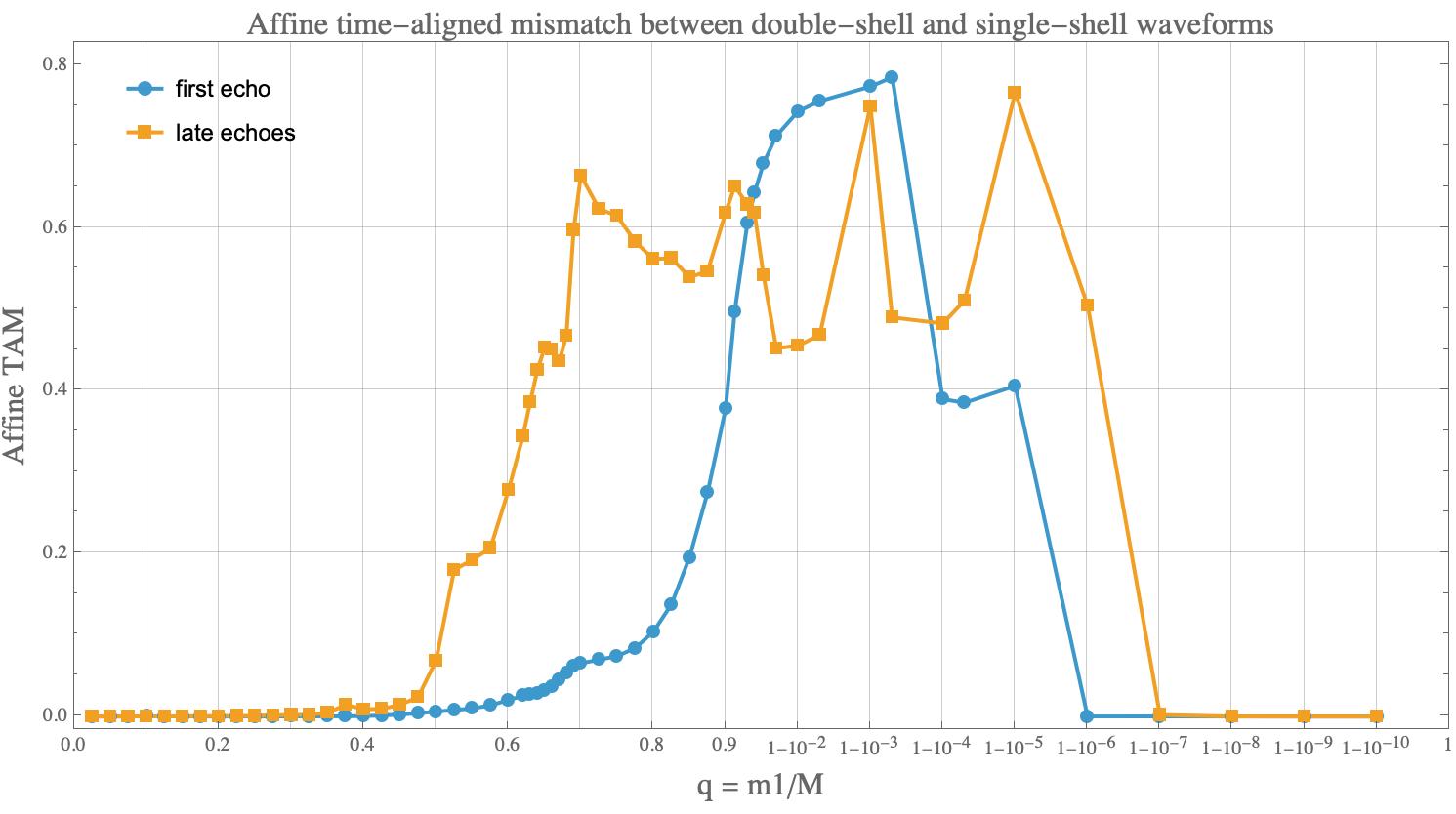}
		\end{minipage}
		
		\caption{
			RMS waveform residuals and TAM as
			functions of the mass-distribution parameter $q$. The upper panel shows
			$\mathrm{RMS}_{I}(q)$ for the difference
			$\Delta\Psi_q=\Psi_q-\Psi_{\rm single}$ in the first-echo and late-echo
			windows. The lower panel shows the corresponding TAM $\mathcal{M}_{I}(q)$ after maximizing the normalized overlap
			over an overall time shift and time rescaling.
		}
		\label{fig:RMS}
	\end{figure}

	The upper panel of Fig.~\ref{fig:RMS} shows ${\rm RMS}_I(q)$, which measures the overall difference across each echo window. Its dependence on $q$ closely follows that of the maximum residual, confirming that the observed deviations are distributed throughout the waveform rather than being produced only by isolated local extrema. A clear local maximum appears in the late-echo RMS near $q\simeq2/3$, in the same region where the late-echo maximum difference and amplitude are locally enhanced. This correspondence links the late-echo feature to the onset of truncation of the inner photon-sphere-like potential peak identified in Sec.~\ref{sec:potential}.
	
	The lower panel of Fig.~\ref{fig:RMS} shows the TAM, which determines whether the normalized waveform shape itself changes. For $q<0.4$, the TAM remains close to zero for both time windows, even though the maximum and RMS residuals are already increasing. The double-shell waveform in this regime therefore differs from the single-shell waveform mainly through a time-scale rescaling and an accumulated phase delay, while its intrinsic morphology remains nearly unchanged. For larger $q$, the late-echo TAM rises before the first-echo TAM, showing that the late echoes provide the first clear time-domain indication of a genuine waveform deformation. This earlier response is consistent with the accumulated effect of repeated propagation and scattering through the double-barrier geometry. As $q$ approaches $2/3$, truncation of the inner potential peak becomes important, and the waveform difference can no longer be removed by an affine transformation of time.
	
	The approach to the $q=1$ single-shell limit is highly non-monotonic: the residuals, amplitudes, and TAM exhibit pronounced variations over several decades in $1-q$, and the first- and late-echo sectors respond differently to the $q$-dependent double-shell configuration. Sizable double-shell effects persist until $q$ is extremely close to unity because the relevant separations are logarithmically stretched in the tortoise coordinate. The near-single-shell region is therefore displayed logarithmically in $1-q$ in Figs.~\ref{fig:psi_01} and~\ref{fig:RMS}; the waveform recovers the single-shell result only within the narrow range $1-q\sim10^{-7}$--$10^{-8}$.
	
	The time-domain diagnostics therefore show that internal mass redistribution first modifies the propagation time while leaving the normalized waveform shape nearly unchanged. At larger $q$, it produces a genuine deformation and a pronounced redistribution of amplitude between the first and late echoes. These results motivate the spectral analysis of the underlying propagation scales in the next section.

	\section{\label{sec:spectrum}$q$-Dependent Spectral-Peak Queueing}
	
	The spectra at $q=0$ and $q=1$ are identical because both endpoints reduce to the same single-shell configuration. Nevertheless, when individual spectral peaks are tracked continuously across the family of stationary configurations parameterized by $q$, the peaks do not generally return to their $q=0$ positions. 
	The resolved peaks exhibit three distinct behaviors. 
	First, a series of peaks associated with high-order resonances of the inner propagation region enters the analyzed range from the high-frequency side. These peaks eventually occupy lower frequencies in the final single-shell spectrum.
	Second, the second and later peaks undergo spectral-peak migration toward higher frequencies and ultimately occupy higher final peak indices.
	Third, the first peak initially shifts toward lower frequencies and subsequently returns to its $q=0$ position as $q\to1$. 
	Here $q$ labels distinct stationary backgrounds rather than physical time. We collectively refer to the combined pattern formed by these three spectral behaviors as spectral-peak queueing (SQ).
	
\begin{figure*}[t]
		\centering
		
		\begin{minipage}{0.24\textwidth}
			\centering
			\includegraphics[width=\linewidth]{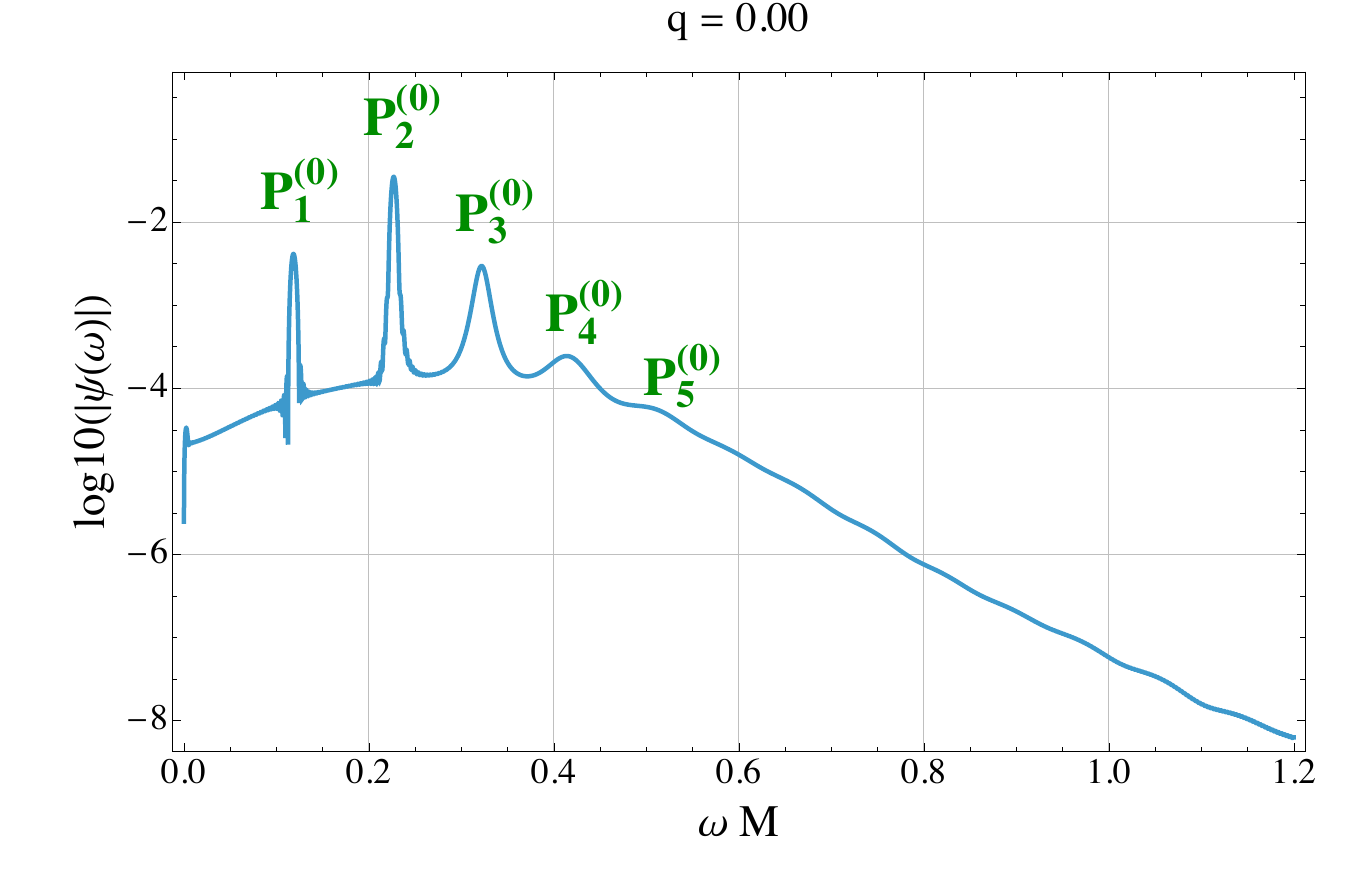}
		\end{minipage}
		\hfill
		\begin{minipage}{0.24\textwidth}
			\centering
			\includegraphics[width=\linewidth]{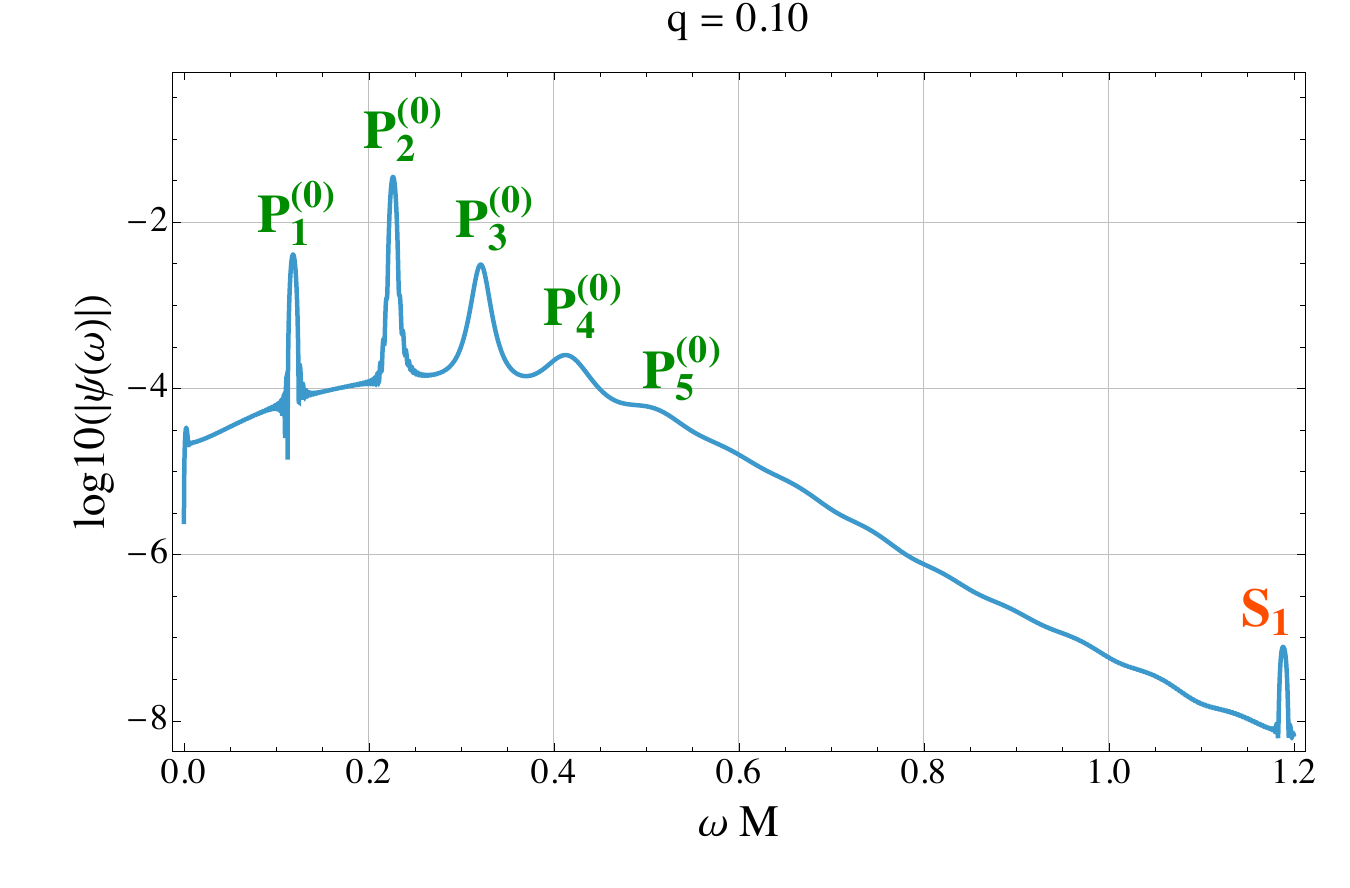}
		\end{minipage}
		\hfill
		\begin{minipage}{0.24\textwidth}
			\centering
			\includegraphics[width=\linewidth]{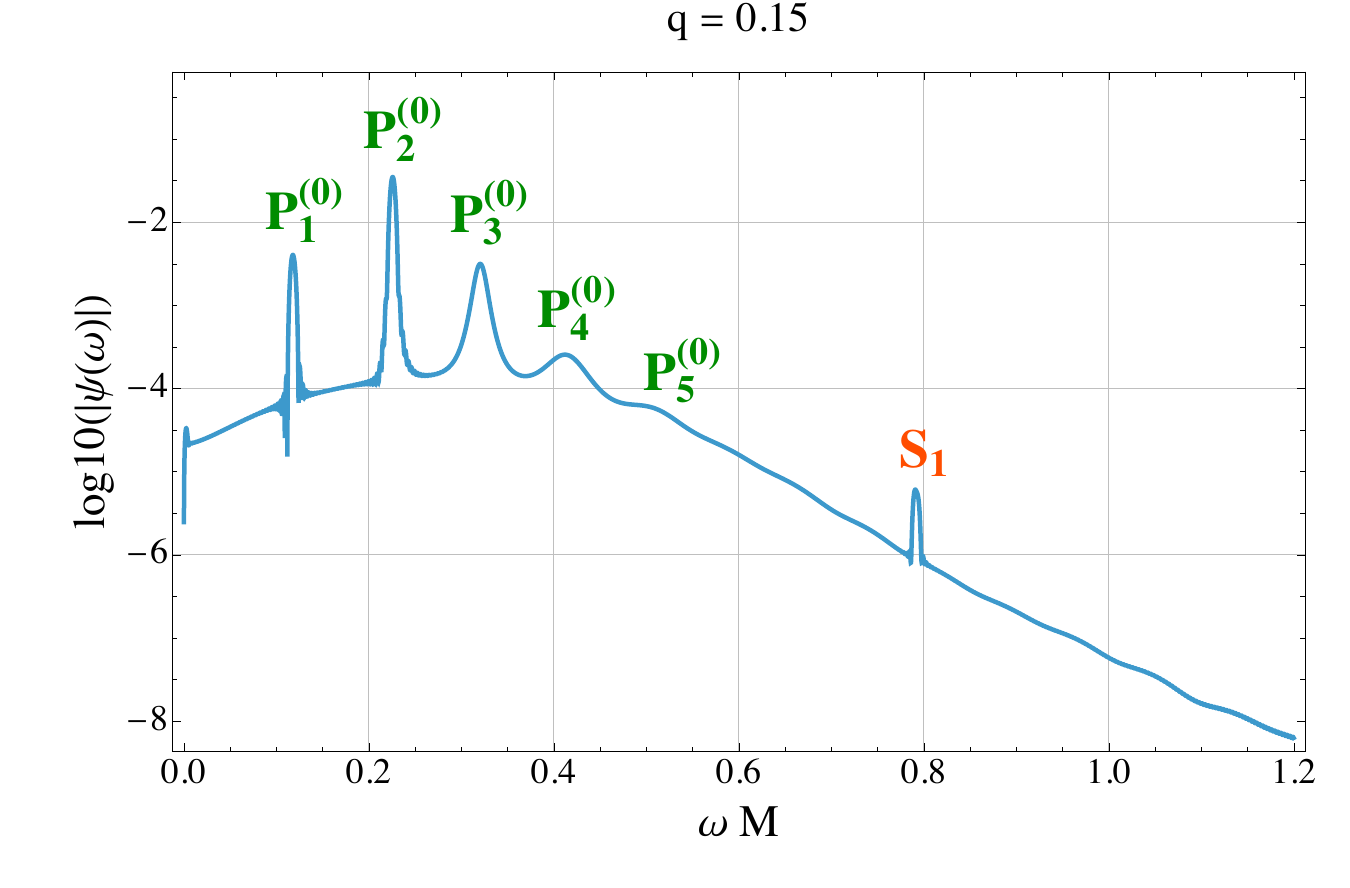}
		\end{minipage}
		\hfill
		\begin{minipage}{0.24\textwidth}
			\centering
			\includegraphics[width=\linewidth]{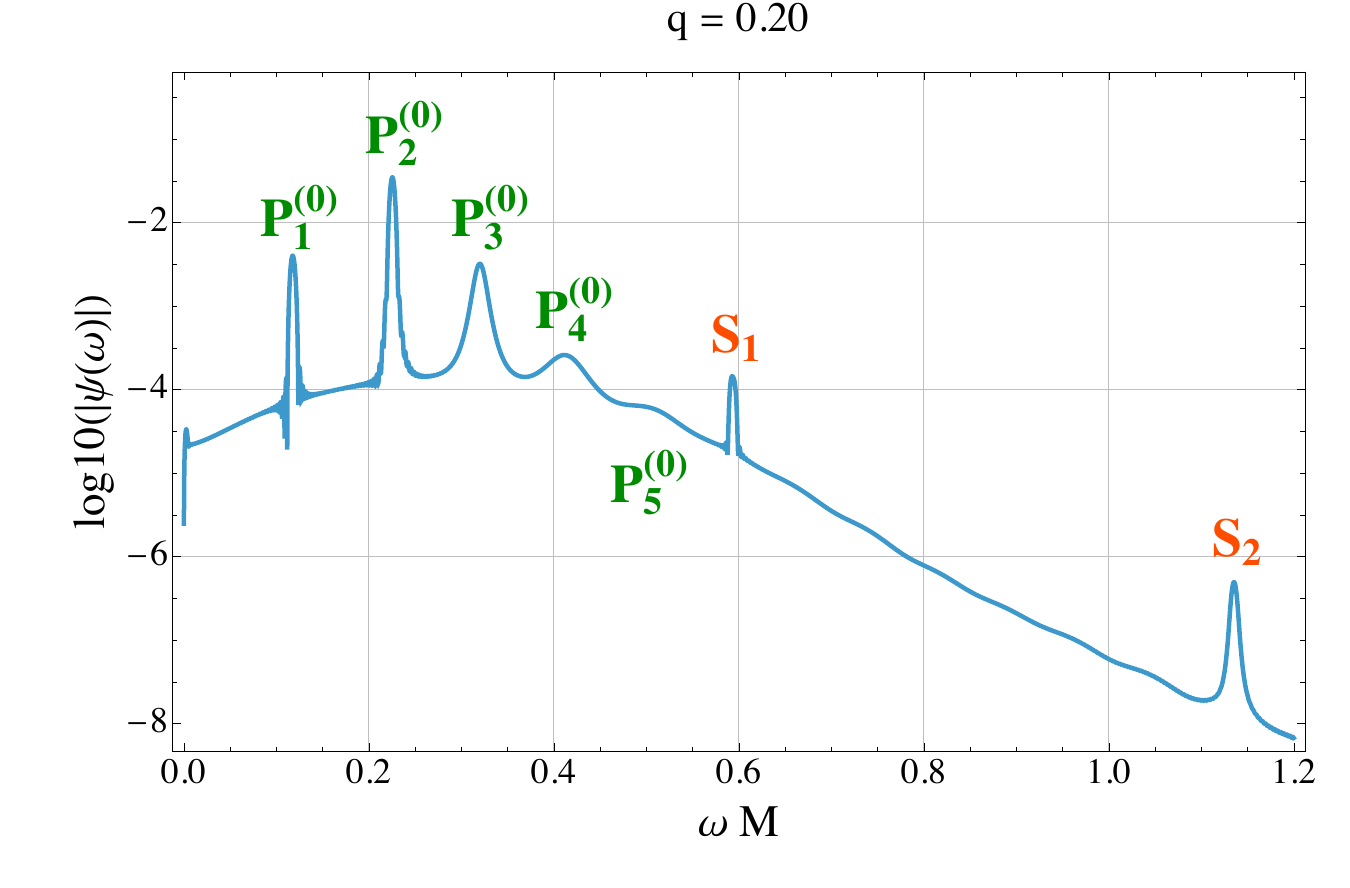}
		\end{minipage}
		
		\vspace{0.35cm}
		\begin{minipage}{0.24\textwidth}
			\centering
			\includegraphics[width=\linewidth]{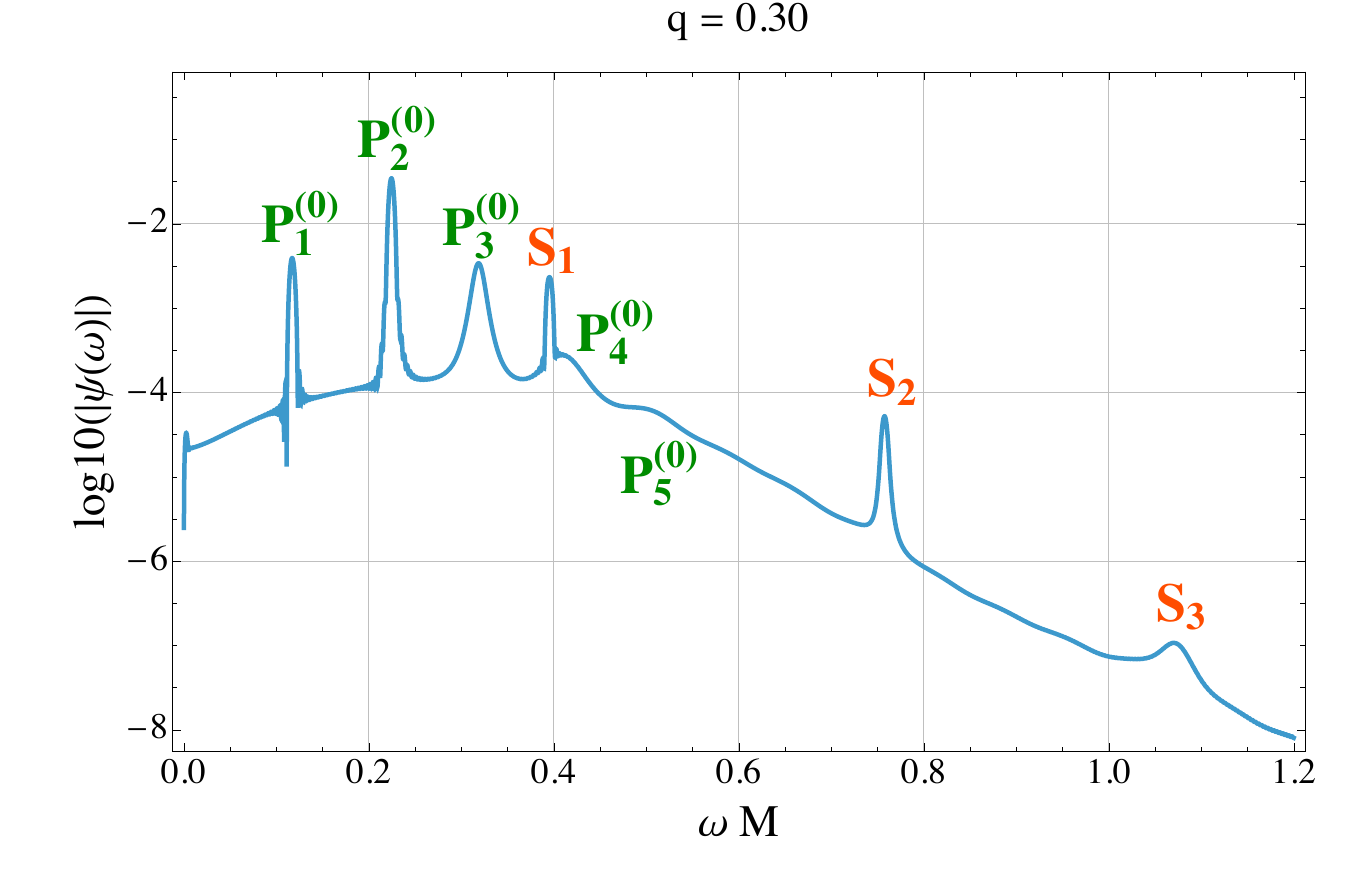}
		\end{minipage}
		\hfill
		\begin{minipage}{0.24\textwidth}
			\centering
			\includegraphics[width=\linewidth]{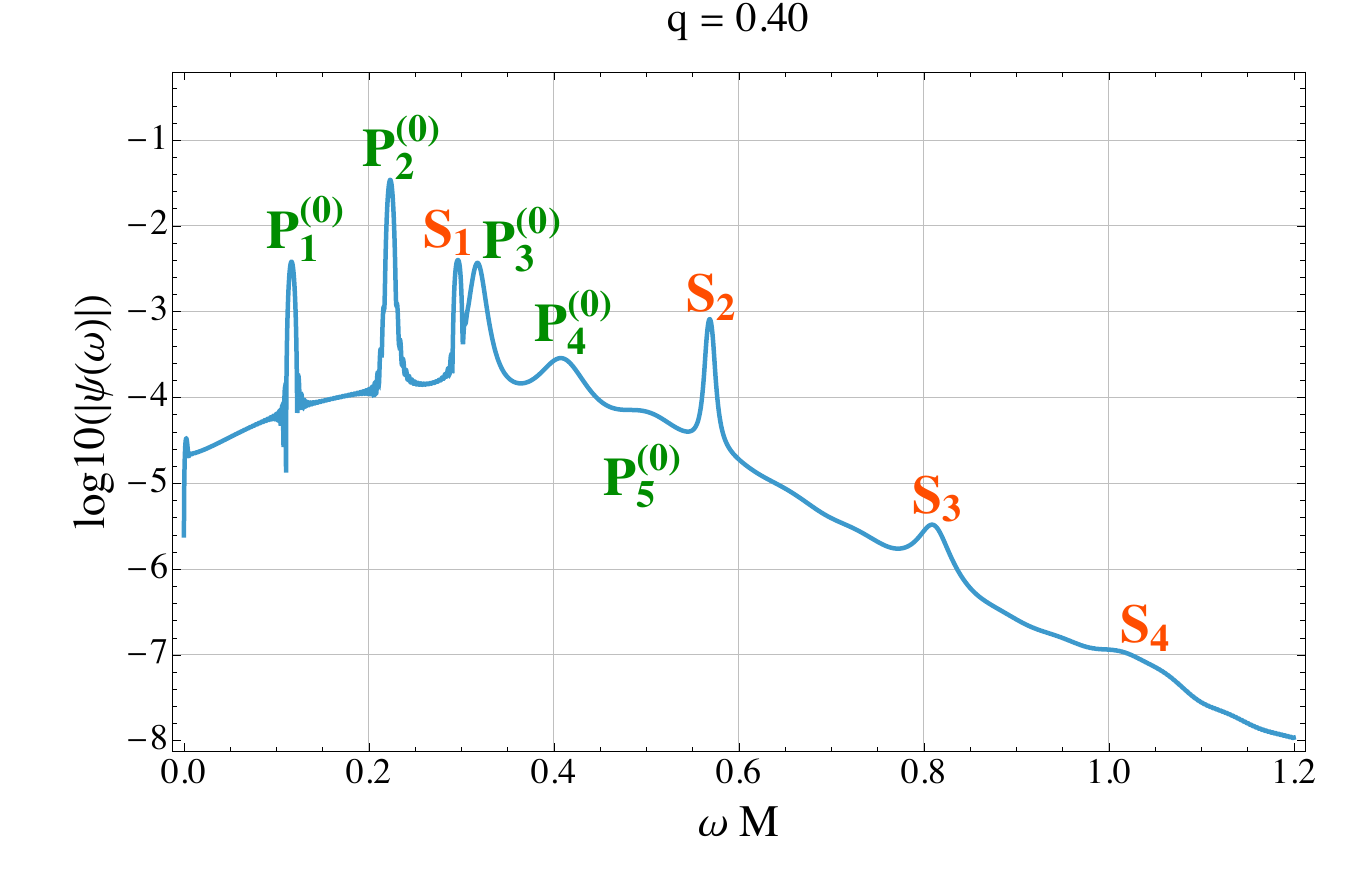}
		\end{minipage}
		\hfill
		\begin{minipage}{0.24\textwidth}
			\centering
			\includegraphics[width=\linewidth]{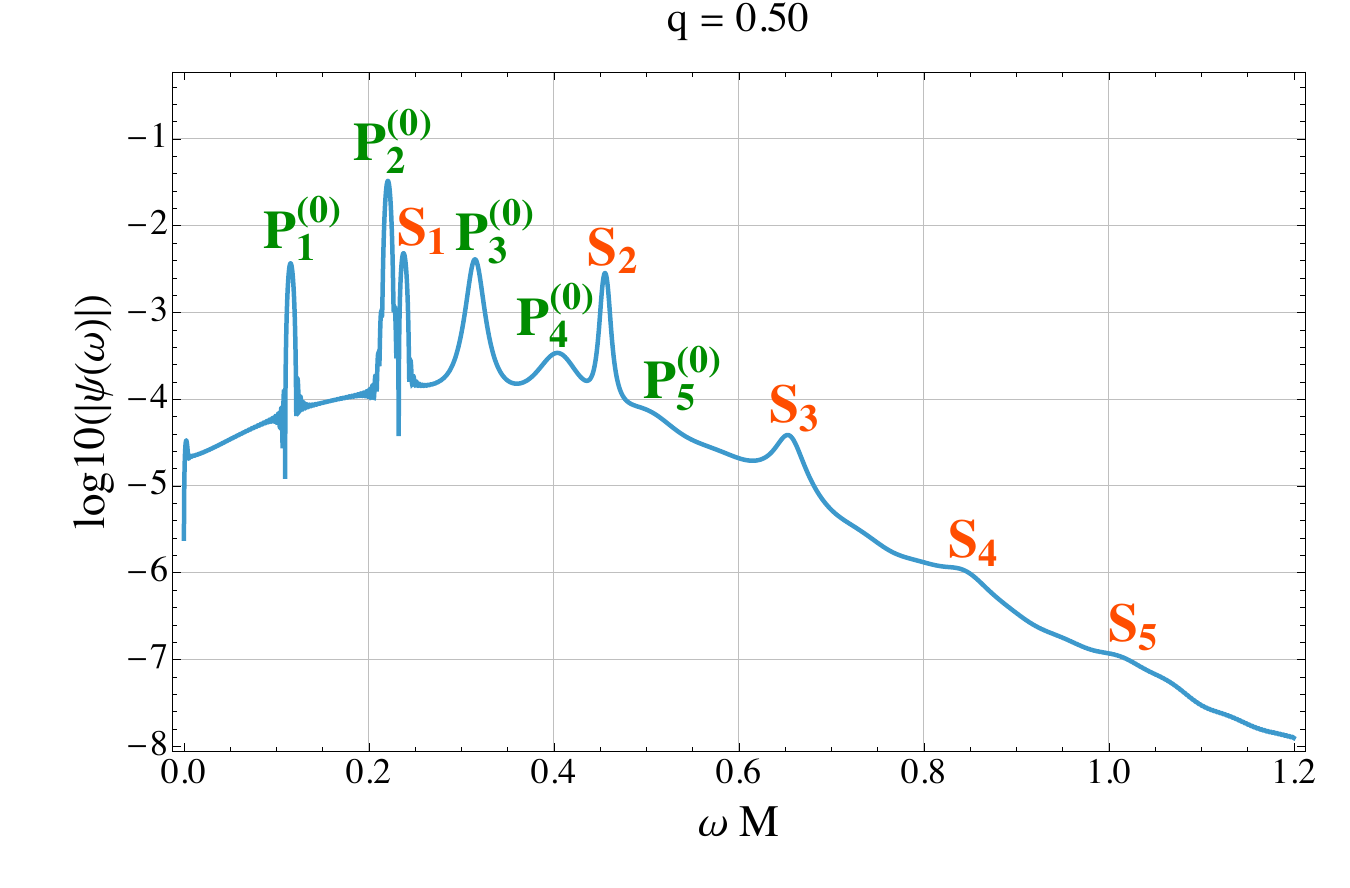}
		\end{minipage}
		\hfill
		\begin{minipage}{0.24\textwidth}
			\centering
			\includegraphics[width=\linewidth]{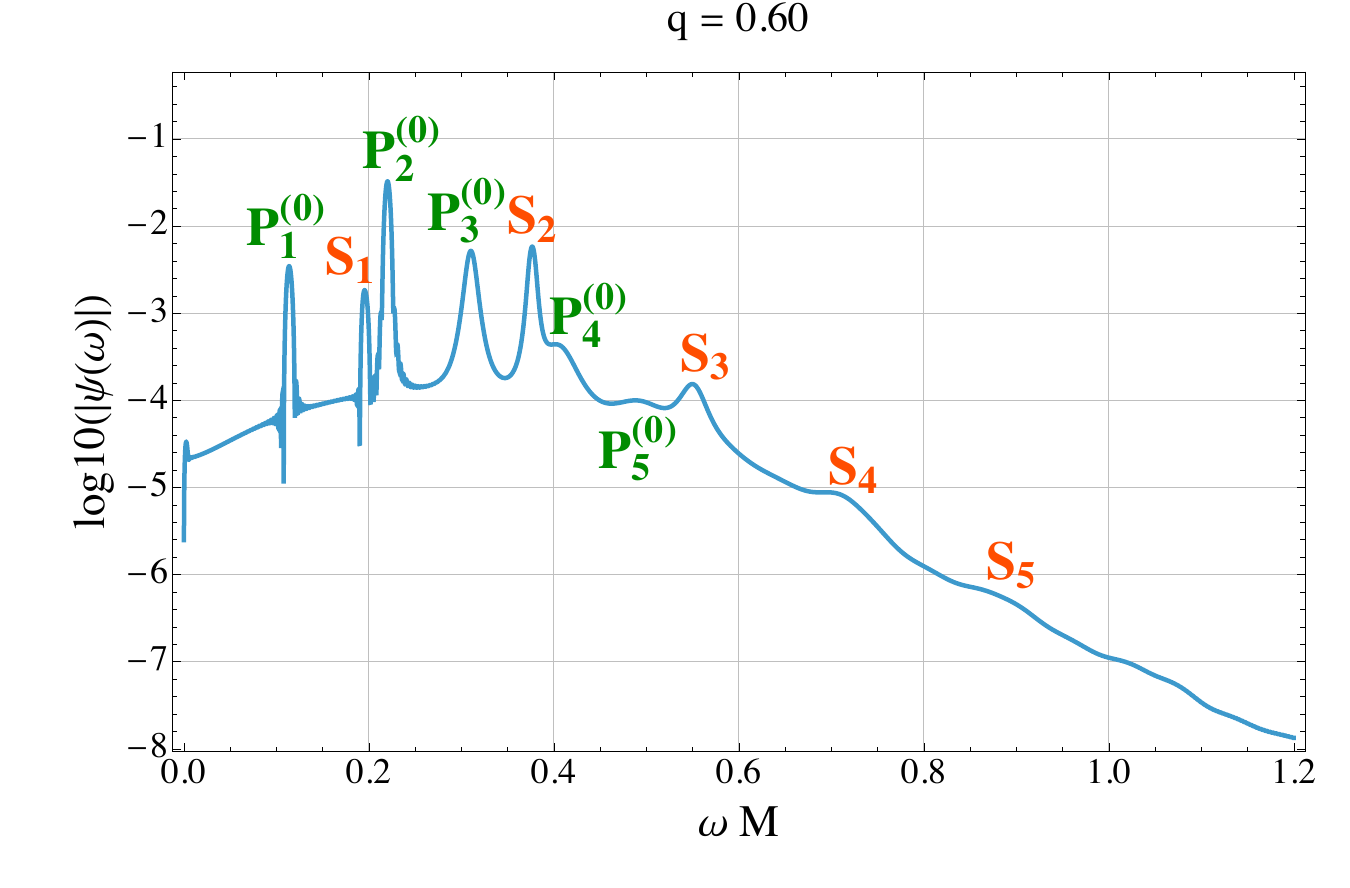}
		\end{minipage}
		
		\vspace{0.35cm}
		\begin{minipage}{0.24\textwidth}
			\centering
			\includegraphics[width=\linewidth]{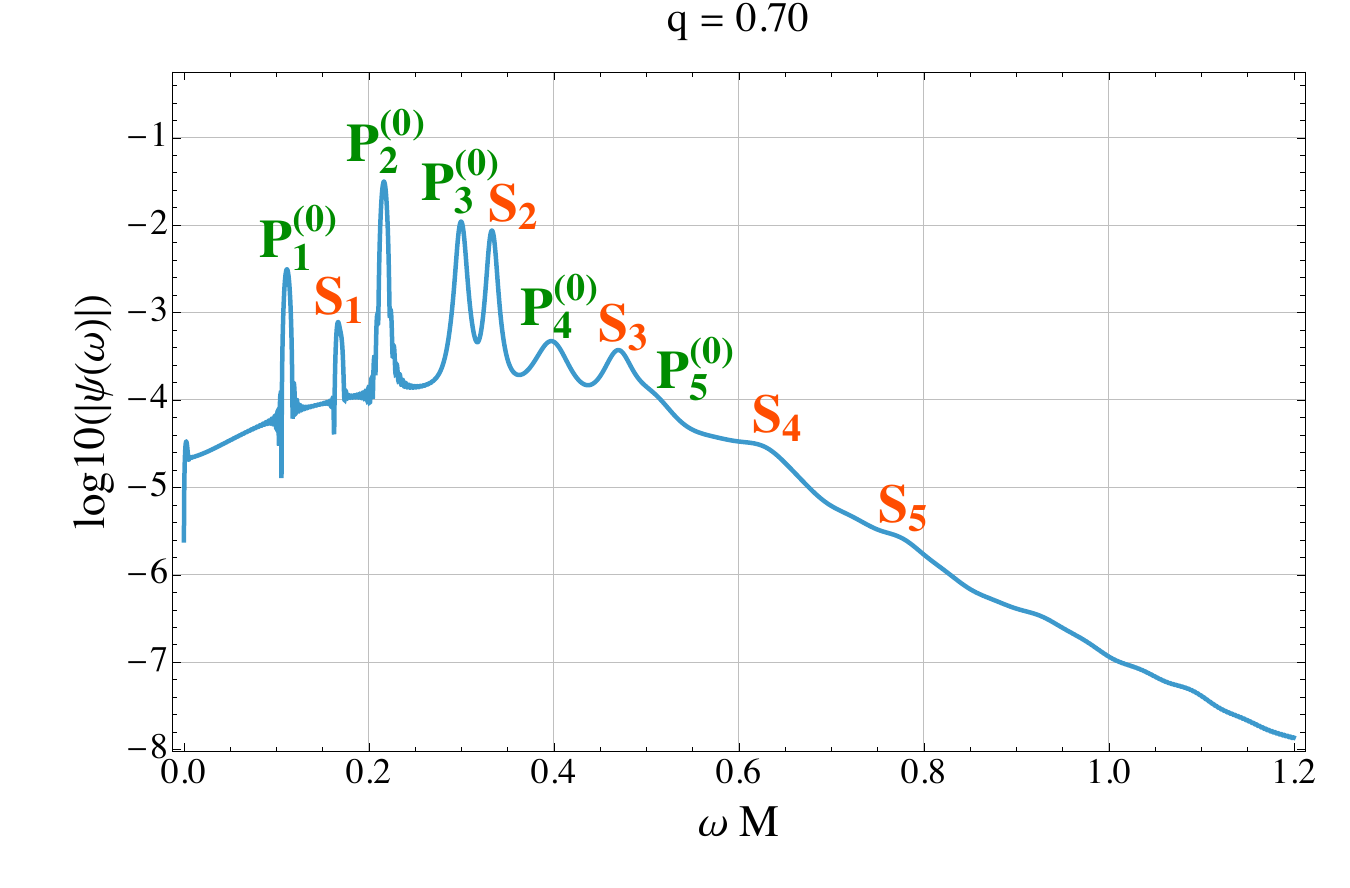}
		\end{minipage}
		\hfill
		\begin{minipage}{0.24\textwidth}
			\centering
			\includegraphics[width=\linewidth]{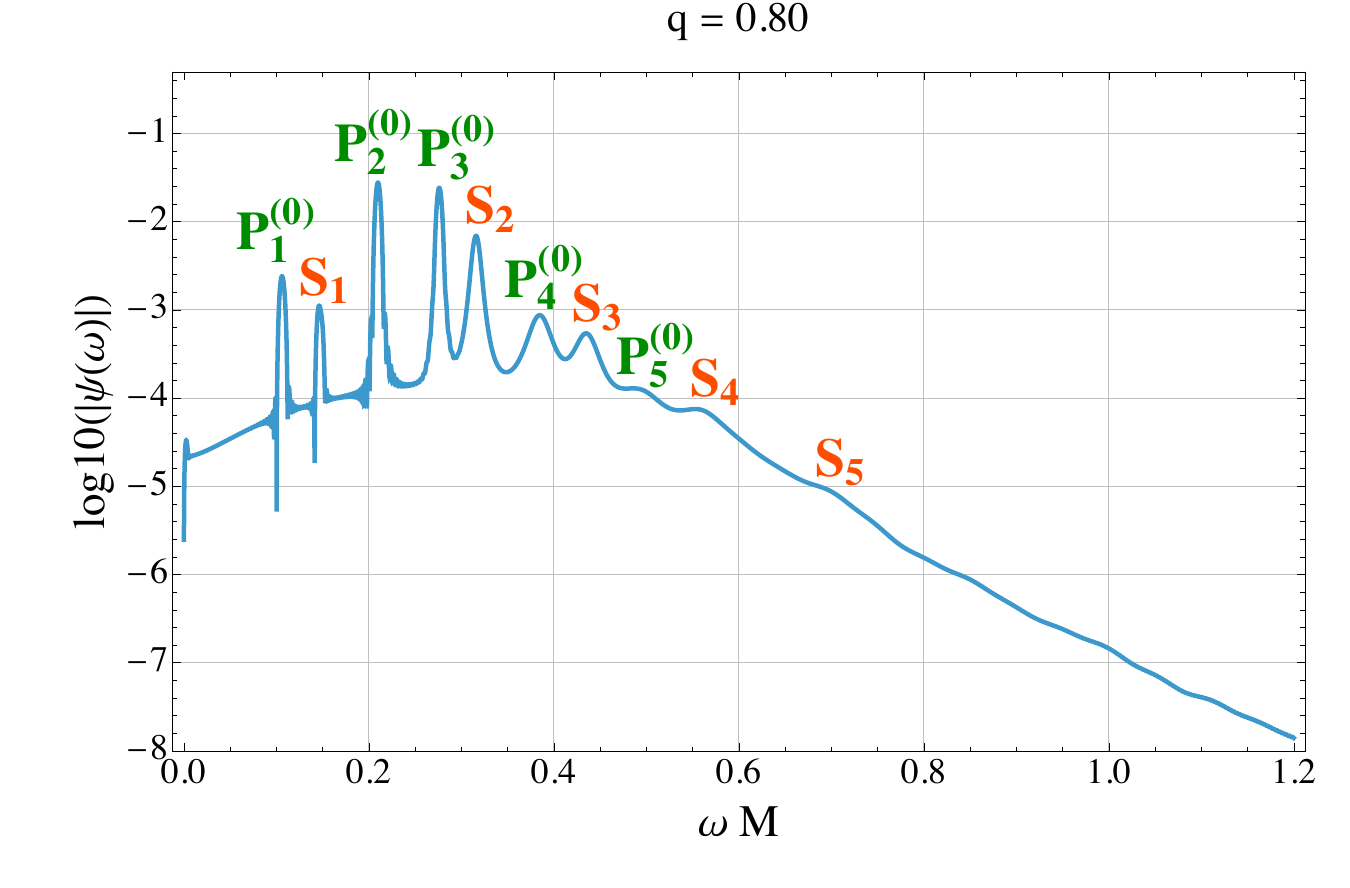}
		\end{minipage}
		\hfill
		\begin{minipage}{0.24\textwidth}
			\centering
			\includegraphics[width=\linewidth]{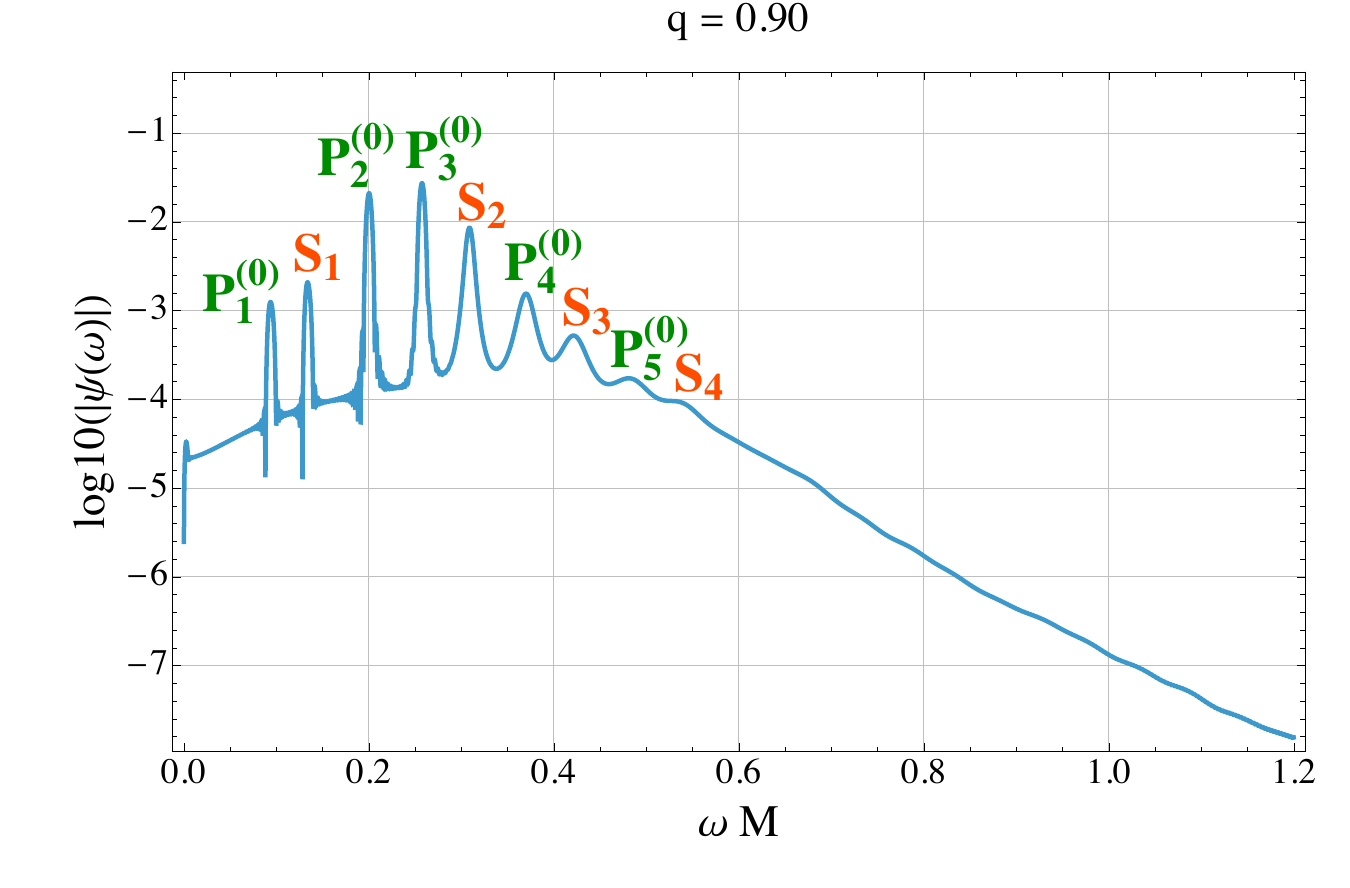}
		\end{minipage}
		\hfill
		\begin{minipage}{0.24\textwidth}
			\centering
			\includegraphics[width=\linewidth]{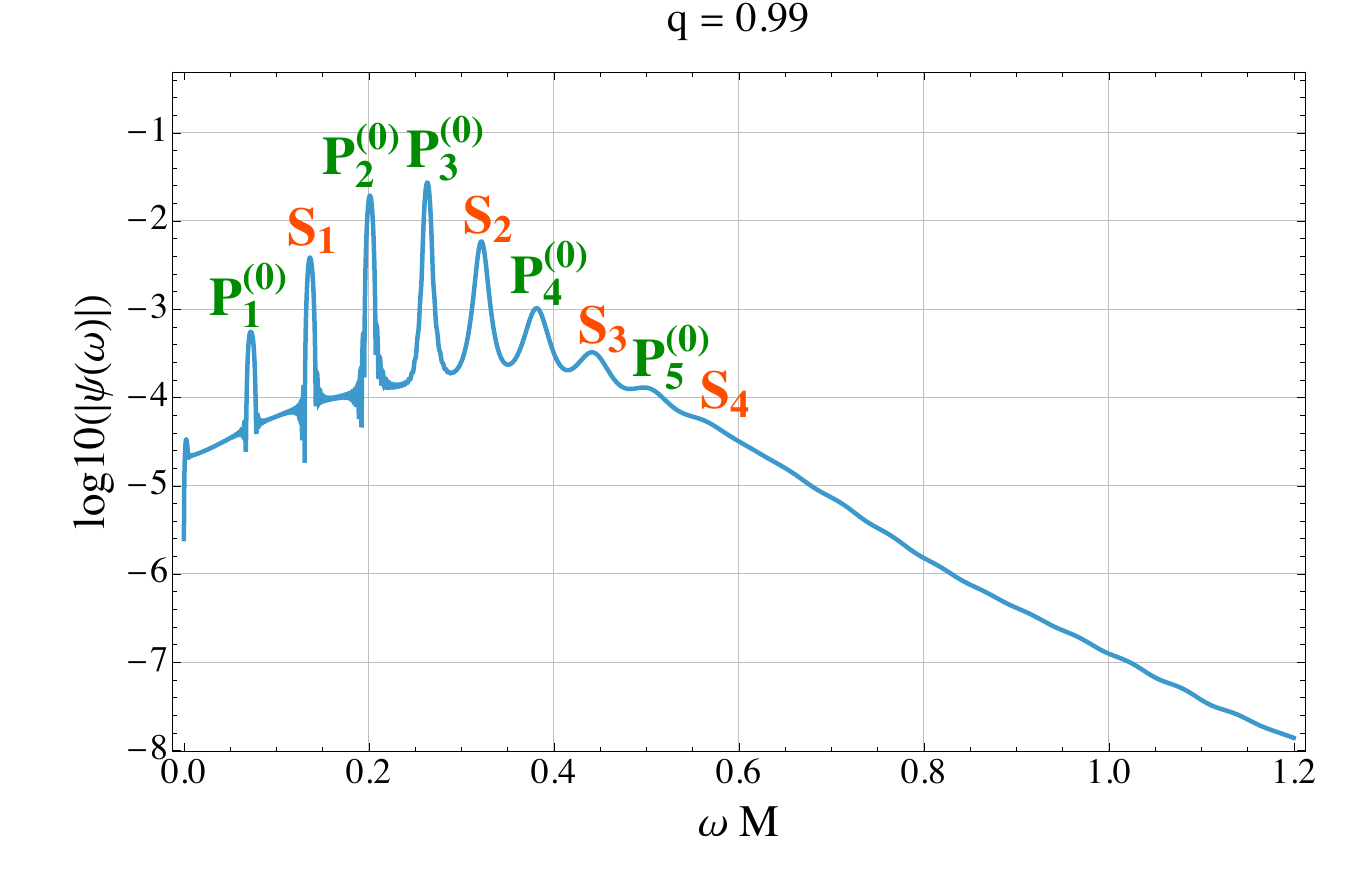}
		\end{minipage}
		
		\vspace{0.35cm}
		\begin{minipage}{0.24\textwidth}
			\centering
			\includegraphics[width=\linewidth]{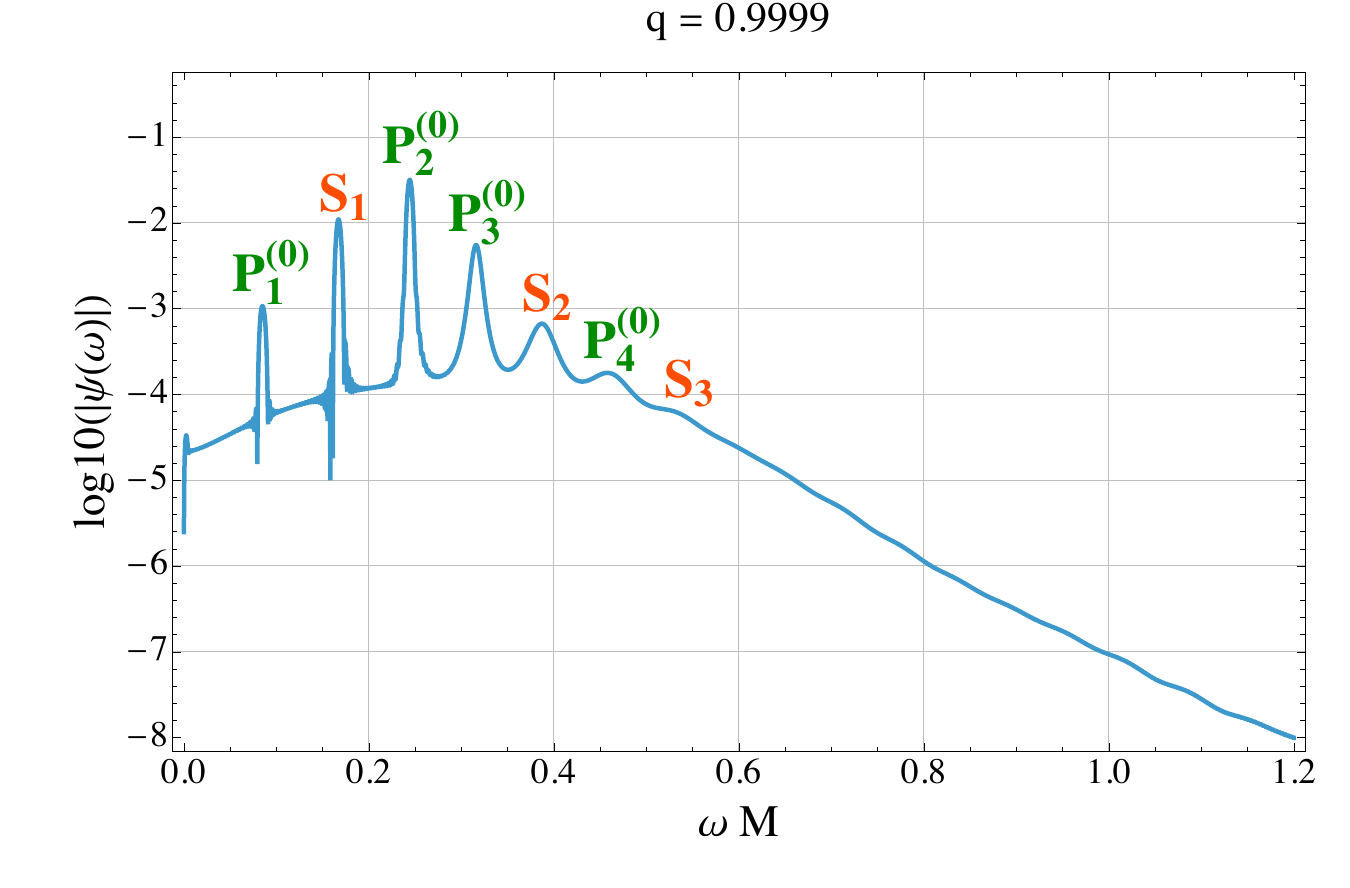}
		\end{minipage}
		\hfill
		\begin{minipage}{0.24\textwidth}
			\centering
			\includegraphics[width=\linewidth]{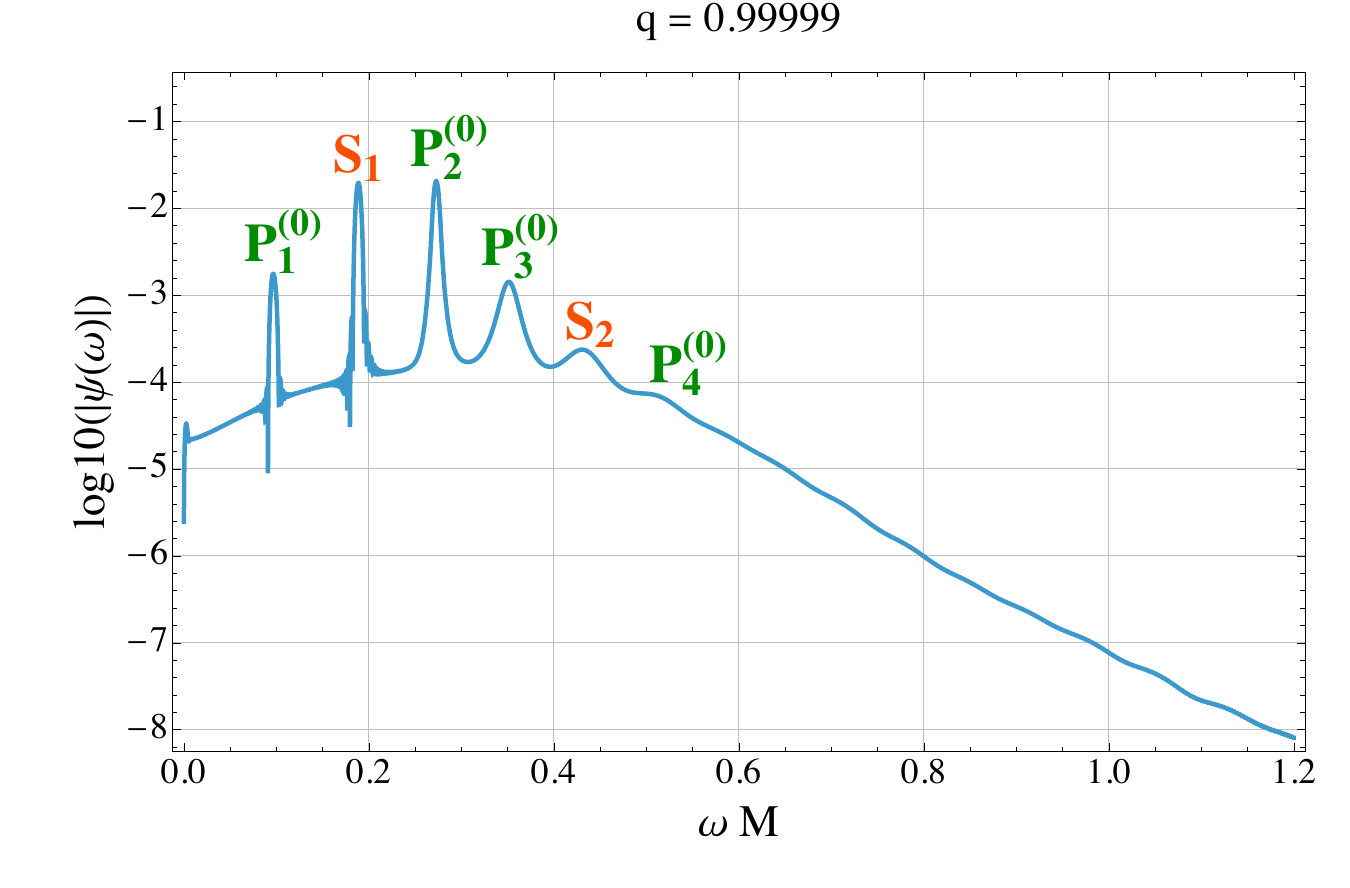}
		\end{minipage}
		\hfill
		\begin{minipage}{0.24\textwidth}
			\centering
			\includegraphics[width=\linewidth]{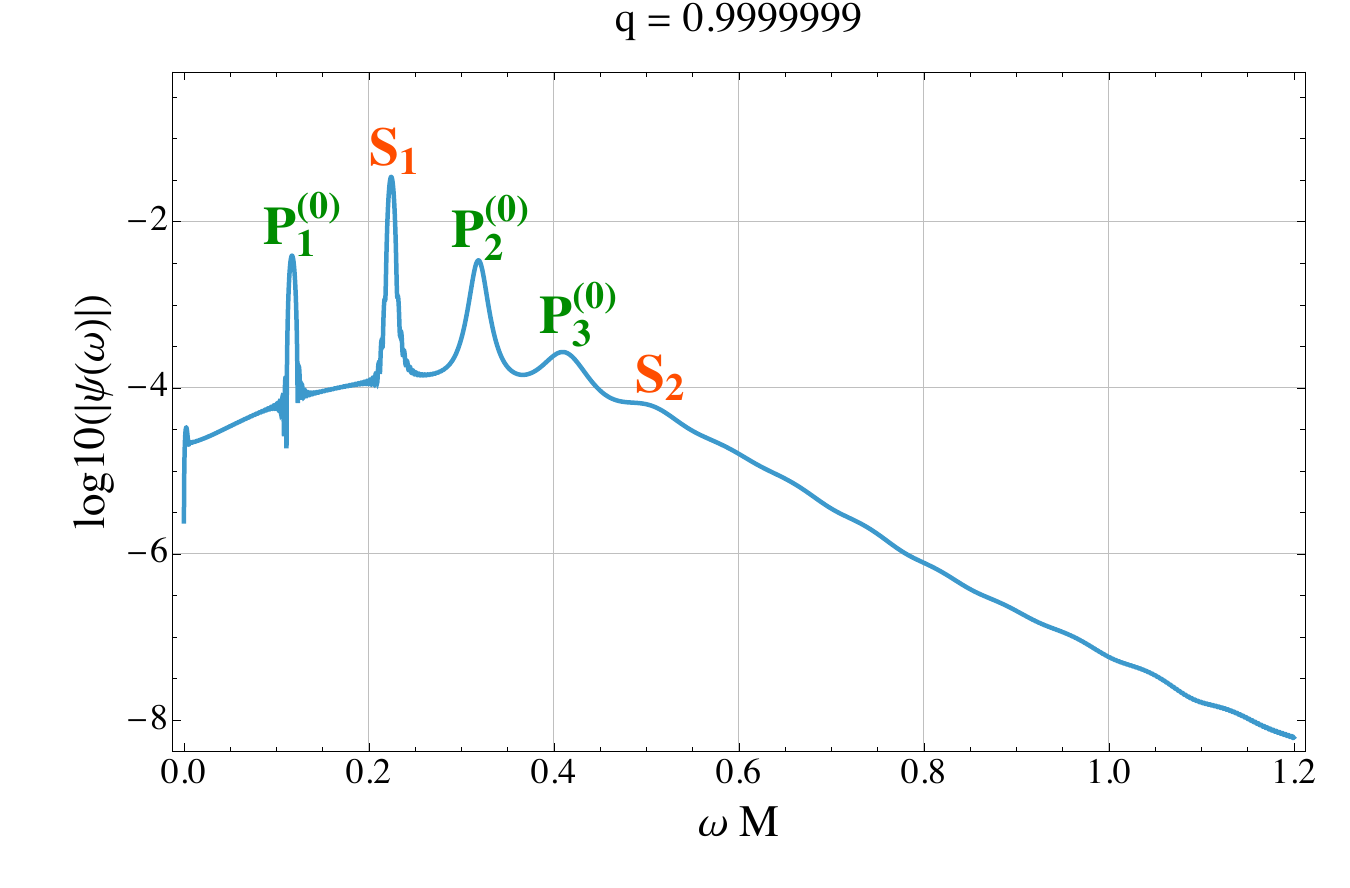}
		\end{minipage}
		\hfill
		\begin{minipage}{0.24\textwidth}
			\centering
			\includegraphics[width=\linewidth]{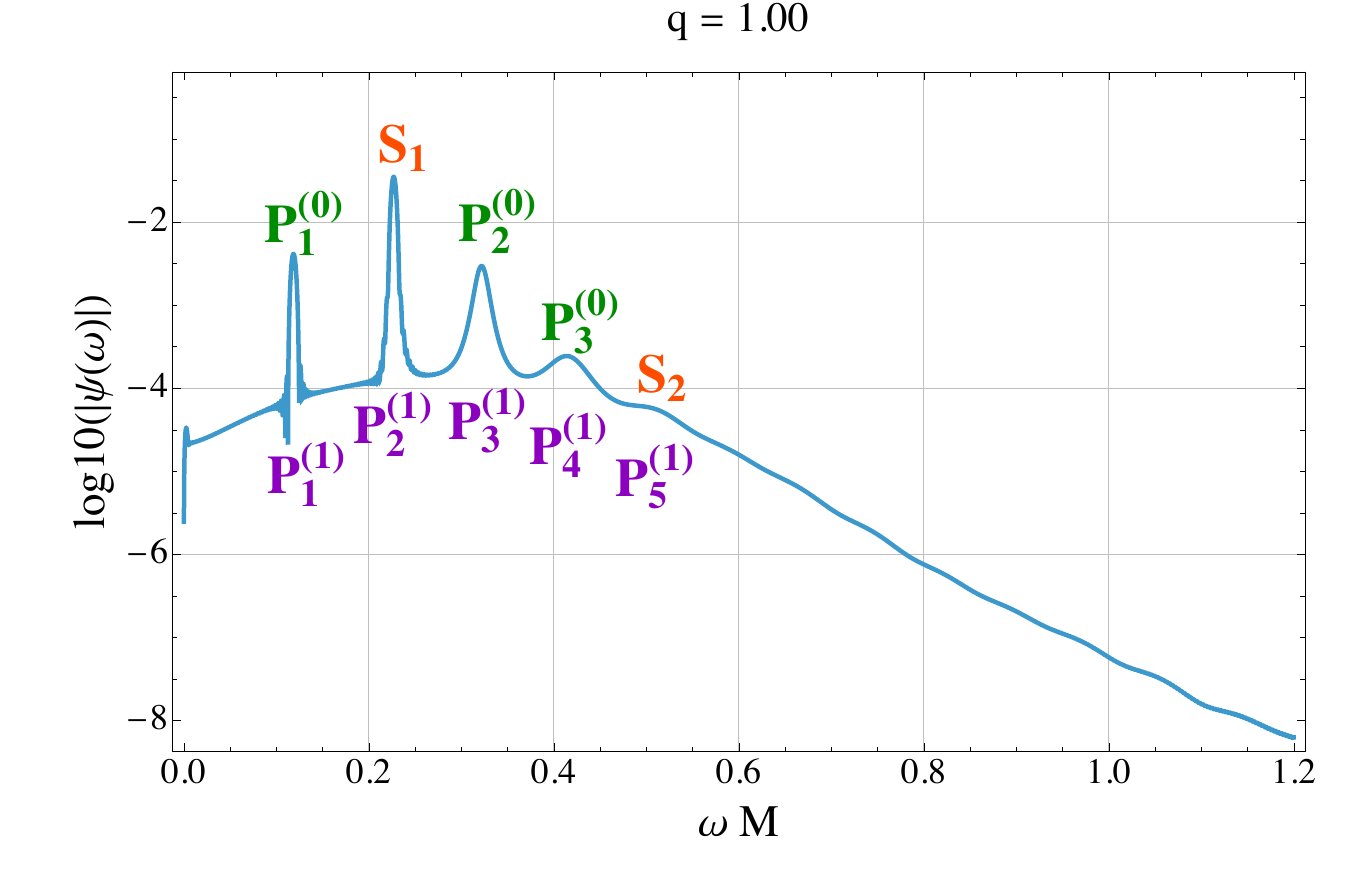}
		\end{minipage}
		
		\caption{
			$q$-dependent waveform spectra exhibiting SQ. The green labels mark the peaks of the $q=0$ spectrum, the red labels identify the entry of the $S_j$ series from the high-frequency side, and the purple labels indicate the $q=1$ spectrum. The figure jointly displays high-frequency series entry, spectral-peak migration, and the first-peak return.
		}
		\label{fig:sp_b}
	\end{figure*}

	Figure~\ref{fig:sp_b} visualizes these three spectral behaviors across the stationary family parameterized by $q$. The logarithmic amplitude scale makes the weaker $S_j$ peaks visible and shows that they form smoothly connected spectral features rather than isolated numerical peaks.
	The frequencies of the main peaks remain stable when the upper limit of the Fourier window is varied among $t_{\max}=1000M$, $1500M$, and $2000M$. Only peaks that can be tracked robustly in frequency and amplitude under refined sampling in $q$ are retained.

\begin{figure}[t]
		\centering
		\begin{minipage}{0.48\textwidth}
			\centering
			\includegraphics[width=\linewidth]{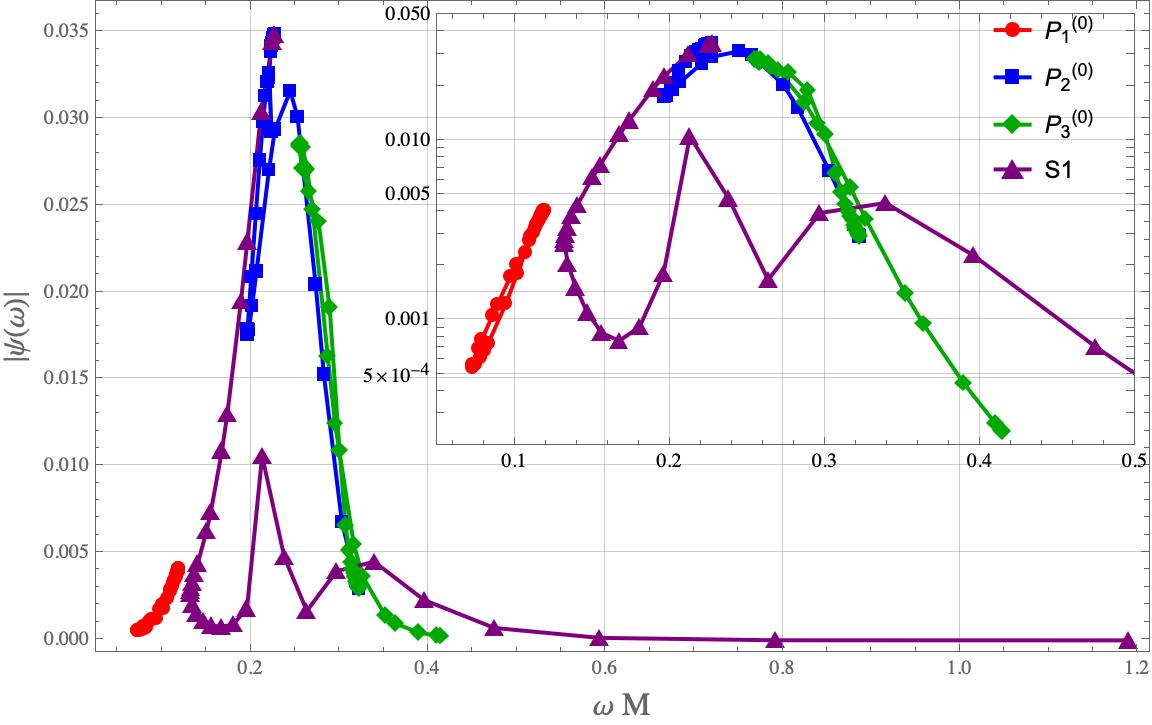}
		\end{minipage}
		
		\caption{
			Peak-tracking diagram showing the three components of SQ: high-frequency series entry represented by $S_1$, spectral-peak migration of the second and later peaks, and first-peak return $P_1^{(0)}$.
		}
		\label{fig:pem}
	\end{figure}

	We denote the $k$th ordered peak of the single-shell spectrum at $q=0$ and $q=1$ by $P_k^{(0)}$ and $P_k^{(1)}$, respectively, and denote the $j$th peak entering the analyzed range from the high-frequency side by $S_j$. The resolved endpoint mappings under SQ are
	\be
	P_1^{(0)}\rightarrow P_1^{(1)}, 
	P_2^{(0)}\rightarrow P_3^{(1)}, 
	P_3^{(0)}\rightarrow P_4^{(1)}, 
	P_4^{(0)}\rightarrow P_6^{(1)}.
	\label{P}
	\ee
	\be
	S_1\rightarrow P_2^{(1)},\qquad
	S_2\rightarrow P_5^{(1)},\qquad
	S_3\rightarrow P_7^{(1)}.
	\label{S}
	\ee
	Thus, the $q=0$ and $1$ spectra are identical as sets of resonance frequencies, but the continuously tracked peaks generally occupy different final peak indices from their $q=0$ peak indices.
	We expect subsequent peaks to follow a similar pattern, but the high-frequency peaks are too weak to track reliably. We therefore restrict our tracking to the series of lower-frequency, higher-amplitude peaks.
	Figure~\ref{fig:pem} displays the trajectories of the main spectral peaks as $q$ varies, showing how the three constituent behaviors of SQ jointly produce the endpoint remapping in Eqs. (\ref{P}) and (\ref{S}). This remapping is a consequence of SQ rather than an additional spectral phenomenon.
	
	To interpret the spectral peak trajectories geometrically, we use the characteristic propagation lengths introduced in Sec.~\ref{sec:potential}. For an isolated or weakly coupled propagation region of length
	$\mathcal{L}$, the resonance frequencies approximately satisfy
	\begin{equation}
		\omega_n\sim\frac{n\pi}{\mathcal{L}}.
	\end{equation}
	We therefore associate the four propagation lengths with the
	inverse-length frequency scales
	\begin{equation}
		f_1=\frac{\pi}{2L_1}, 
		f_2=\frac{\pi}{2L_2}, 
		f_3=\frac{\pi}{L_1+L_2}, 
		f_4=\frac{\pi}{L_2-L_1}.
	\end{equation}
	Tables~\ref{cavity} and~\ref{feq} list these lengths and frequency scales for representative values of $q$. The quantities $f_i$ are geometric reference scales rather than exact eigenfrequencies of four independent effective cavities, because the effective barriers have finite height and the corresponding propagation regions are mutually coupled.
	
	\begin{table}[t]
		\centering
		\caption{
			Characteristic propagation lengths for representative values of $q$,
			in tortoise-coordinate units.
		}
		\label{tab:cavity-lengths}
		\begin{ruledtabular}
			\begin{tabular}{ccccc}
				$q=m_1/M$ & $2L_1$ & $2L_2$ & $L_1+L_2$ & $L_2-L_1$ \\
				\hline
				$0$        & --     & $60.550$  & --      & --      \\
				$0.1$      & $6.055$  & $67.232$  & $36.644$ & $30.589$ \\
				$0.3$      & $18.165$ & $78.977$  & $48.571$ & $30.406$ \\
				$0.5$      & $30.275$ & $89.567$  & $59.921$ & $29.646$ \\
				$0.7$      & $42.252$ & $98.802$  & $70.527$ & $28.275$ \\
				$0.9$      & $48.321$ & $104.871$ & $76.596$ & $28.275$ \\
				$0.99$     & $43.258$ & $99.808$  & $71.533$ & $28.275$ \\
				$0.9999$   & $25.212$ & $81.761$  & $53.486$ & $28.275$ \\
				$0.999999$ & $8.394$  & $64.944$  & $36.669$ & $28.275$ \\
			\end{tabular}
		\end{ruledtabular}
		\label{cavity}
	\end{table}
	
	\begin{table}[t]
		\centering
		\caption{			
			Characteristic inverse-length frequency scales associated with the
			propagation lengths in Table~\ref{cavity}.
		}
		\label{tab:cavity-frequencies}
		
		\begin{ruledtabular}			
			\begin{tabular}{ccccc}				
				$q=m_1/M$ & $f_1$ & $f_2$ & $f_3$ & $f_4$ \\			
				\hline				
				$0$        & --      & $0.052$ & --      & --      \\				
				$0.1$      & $0.519$ & $0.047$ & $0.086$ & $0.103$ \\				
				$0.3$      & $0.173$ & $0.040$ & $0.065$ & $0.103$ \\				
				$0.5$      & $0.104$ & $0.035$ & $0.052$ & $0.106$ \\				
				$0.7$      & $0.074$ & $0.032$ & $0.045$ & $0.111$ \\				
				$0.9$      & $0.065$ & $0.030$ & $0.041$ & $0.111$ \\				
				$0.99$     & $0.073$ & $0.031$ & $0.044$ & $0.111$ \\				
				$0.9999$   & $0.125$ & $0.038$ & $0.059$ & $0.111$ \\				
				$0.999999$ & $0.374$ & $0.048$ & $0.086$ & $0.111$ \\				
			\end{tabular}			
		\end{ruledtabular}		
		\label{feq}
	\end{table}

	The resolved trajectories can be separated into the three constituent behaviors of SQ, discussed below in the same order as introduced above.
	These behaviors can be associated primarily with the $q$-dependent variations of three characteristic propagation lengths: high-frequency series entry with $2L_1$, spectral-peak migration with $L_2-L_1$, and the first-peak return with $2L_2$. 
	
	First, the $S_j$ peaks undergo high-frequency series entry. They are primarily associated with the inner propagation scale $2L_1$. For small $q$, $2L_1$ is short and the associated resonances lie above the displayed range. As $2L_1$ grows with $q$, these resonances shift toward lower frequencies and enter the analyzed frequency range from its high-frequency side. Before the inner barrier becomes strongly truncated, the resolvable peaks approximately satisfy
	\begin{equation}
		\begin{split}
			\omega_{S_1}/f_1 \simeq 2.3,\qquad
			\omega_{S_2}/f_1 \simeq 4.4,\qquad
			\omega_{S_3}/f_1 \simeq 6.3,\\
			\omega_{S_4}/f_1 \simeq 8.3,\qquad
			\omega_{S_5}/f_1 \simeq 10.2
		\end{split}
	\end{equation}
	Their nearly uniform spacing and approximate scaling with $f_1$ identify them primarily as high-order resonances associated with the inner propagation region. Their measured frequencies and ratios are listed in Table~\ref{special}.
	
	\begin{table*}[t]
		\centering
		\caption{
			Frequencies of the peaks entering the analyzed range from the high-frequency side.
		}
		\label{tab:special-branches}
			\begin{tabular}{ccccccccccc}
				\hline
				$q$ &
				$\omega_{S_1}$ & $\omega_{S_1}/f_1$ &
				$\omega_{S_2}$ & $\omega_{S_2}/f_1$ &
				$\omega_{S_3}$ & $\omega_{S_3}/f_1$ &
				$\omega_{S_4}$ & $\omega_{S_4}/f_1$ &
				$\omega_{S_5}$ & $\omega_{S_5}/f_1$ \\
				\hline
				0.10 & 1.19 & 2.29 & --   & --   & --   & --   & --   & --   & --   & --   \\
				0.20 & 0.59 & 2.27 & 1.14 & 4.39 & --   & --   & --   & --   & --   & --   \\
				0.30 & 0.40 & 2.31 & 0.76 & 4.39 & 1.08 & 6.24 & --   & --   & --   & --   \\
				0.40 & 0.30 & 2.31 & 0.57 & 4.39 & 0.81 & 6.24 & 1.04 & 8.02 & --   & --   \\
				0.50 & 0.24 & 2.31 & 0.46 & 4.43 & 0.65 & 6.26 & 0.85 & 8.19 & 1.05 & 10.12 \\
				0.60 & 0.20 & 2.31 & 0.38 & 4.39 & 0.55 & 6.36 & 0.72 & 8.33 & 0.88 & 10.18 \\
				0.70 & 0.17 & 2.29 & 0.33 & 4.44 & 0.47 & 6.32 & 0.63 & 8.47 & 0.78 & 10.49 \\
				0.80 & 0.15 & 2.20 & 0.32 & 4.68 & 0.44 & 6.44 & 0.56 & 8.20 & 0.70 & 10.25 \\
				0.90 & 0.13 & 2.00 & 0.31 & 4.77 & 0.42 & 6.46 & 0.54 & 8.31 & 0.66 & 10.15 \\
				0.99 & 0.14 & 1.93 & 0.32 & 4.41 & 0.44 & 6.06 & 0.57 & 7.85 & --   & --   \\
				0.9999 & 0.17 & 1.36 & 0.39 & 3.13 & 0.54 & 4.33 & --   & --   & --   & --   \\
				0.999999 & 0.21 & 0.56 & 0.49 & 1.31 & --   & --   & --   & --   & --   & --   \\
				\hline
			\end{tabular}
		\label{special}
	\end{table*}
	
	Second, the second and later single-shell peaks undergo spectral-peak migration toward higher frequencies. This behavior is broadly consistent with the decrease of $L_2-L_1$ and the corresponding increase of $f_4$. Their final ordered frequencies are also affected by the peaks undergoing high-frequency series entry, so the detailed trajectories are not determined by $L_2-L_1$ alone. These entering peaks occupy several low-order final peak indices of the single-shell spectrum and thereby push the corresponding $q=0$ peaks to higher final indices.
	
	Third, the first spectral peak exhibits a distinct return behavior. Its trajectory is primarily associated with the global propagation scale $2L_2$, which first increases due to the near-horizon logarithmic stretching of the Schwarzschild tortoise coordinate and then returns toward its single-shell value as $q \to 1$. Accordingly, the first peak initially shifts toward lower frequencies and subsequently returns to its $q=0$ position.
	
	The mixed scale $L_1+L_2$ may contribute to interference and local peak-frequency shifts, but the present spectra do not isolate a peak family controlled uniquely by this scale. The four propagation lengths should therefore not be interpreted as four independent spectral sequences.
	
	For $q\gtrsim2/3$, the nominal inner photon-sphere-like potential peak is truncated by the outer shell and the exterior Schwarzschild region.
	The position defining $L_1$ is then pinned to the truncation boundary, explaining the apparent saturation of $L_2-L_1$ and the nearly constant $f_4$ in Tables~I and~II. In this regime, the spectra are no longer governed by a single propagation length, but show stronger potential peak-height variations and more complex interactions among the three components of SQ.
	
	As $q \to 1$, the inner potential peak disappears as an independent barrier, but the resolved peaks remain identifiable in the final single-shell spectrum: the $S_j$ peaks connect to lower-frequency peaks of the final single-shell spectrum, while several $q=0$ peaks arrive at higher frequencies. This limiting behavior completes the endpoint mappings in Eqs. (\ref{P}) and (\ref{S}) and constitutes the endpoint manifestation of SQ.
	
	\section{\label{sec:robust}ROBUSTNESS ACROSS ANGULAR MOMENTA AND PERTURBATION CHANNELS}
	
\begin{figure*}[t]
		\centering
		
		\begin{minipage}{0.24\textwidth}
			\centering
			\includegraphics[width=\linewidth]{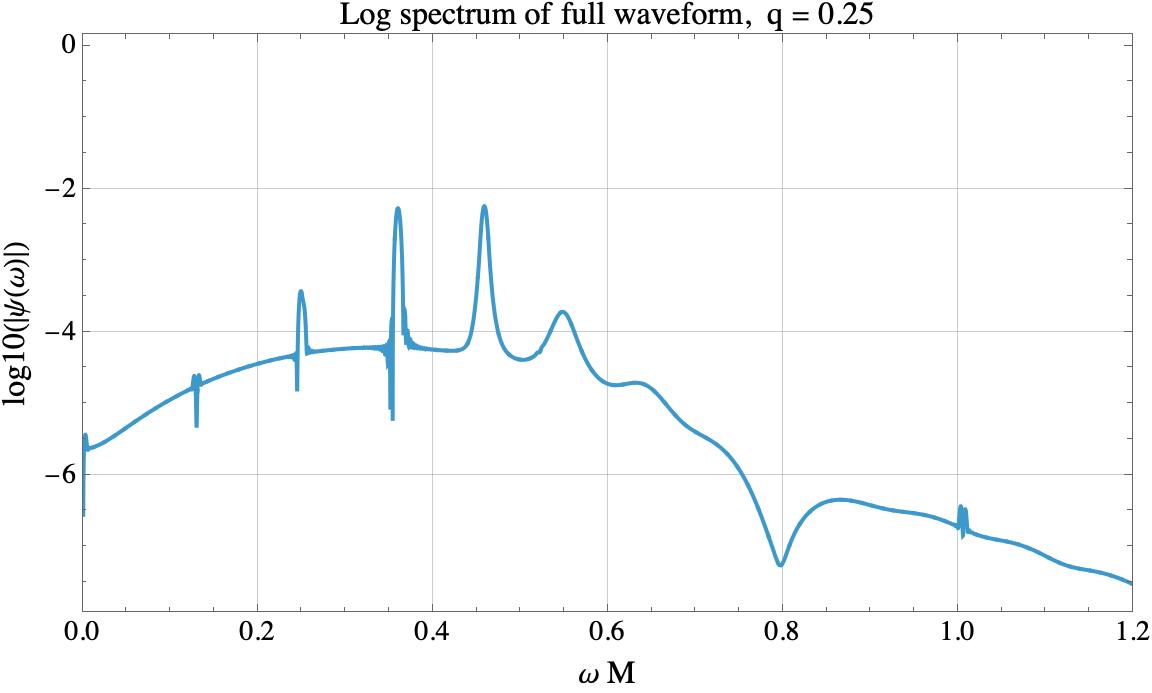}
		\end{minipage}
		\hfill
		\begin{minipage}{0.24\textwidth}
			\centering
			\includegraphics[width=\linewidth]{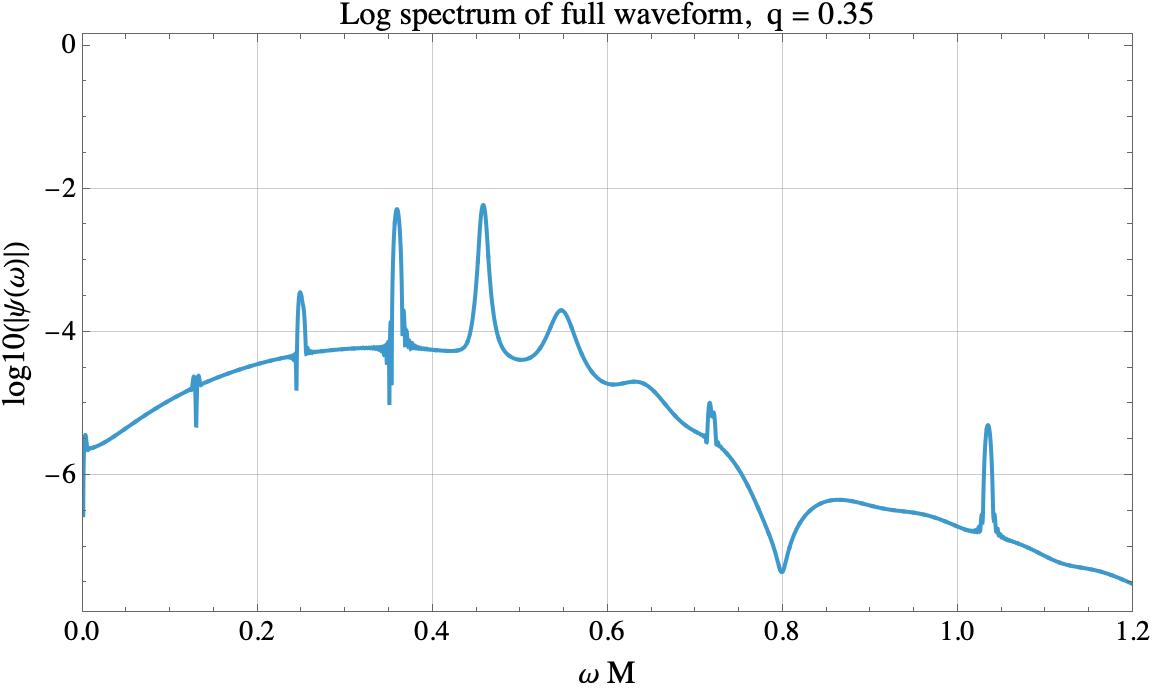}
		\end{minipage}
		\hfill
		\begin{minipage}{0.24\textwidth}
			\centering
			\includegraphics[width=\linewidth]{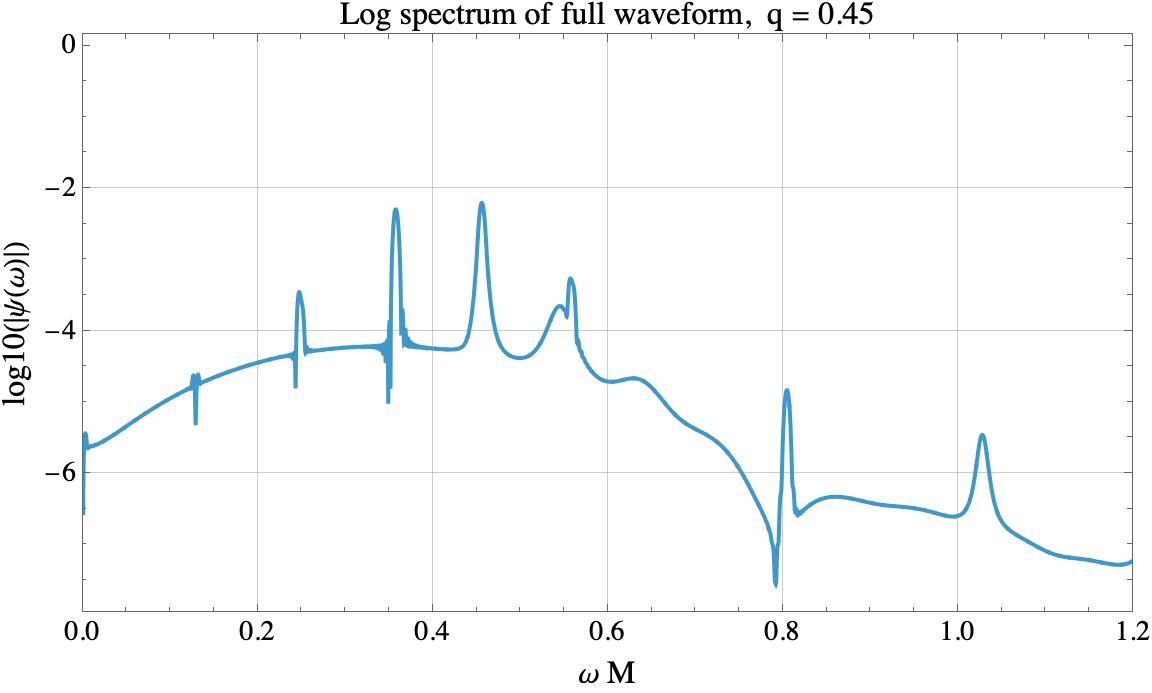}
		\end{minipage}
		\hfill
		\begin{minipage}{0.24\textwidth}
			\centering
			\includegraphics[width=\linewidth]{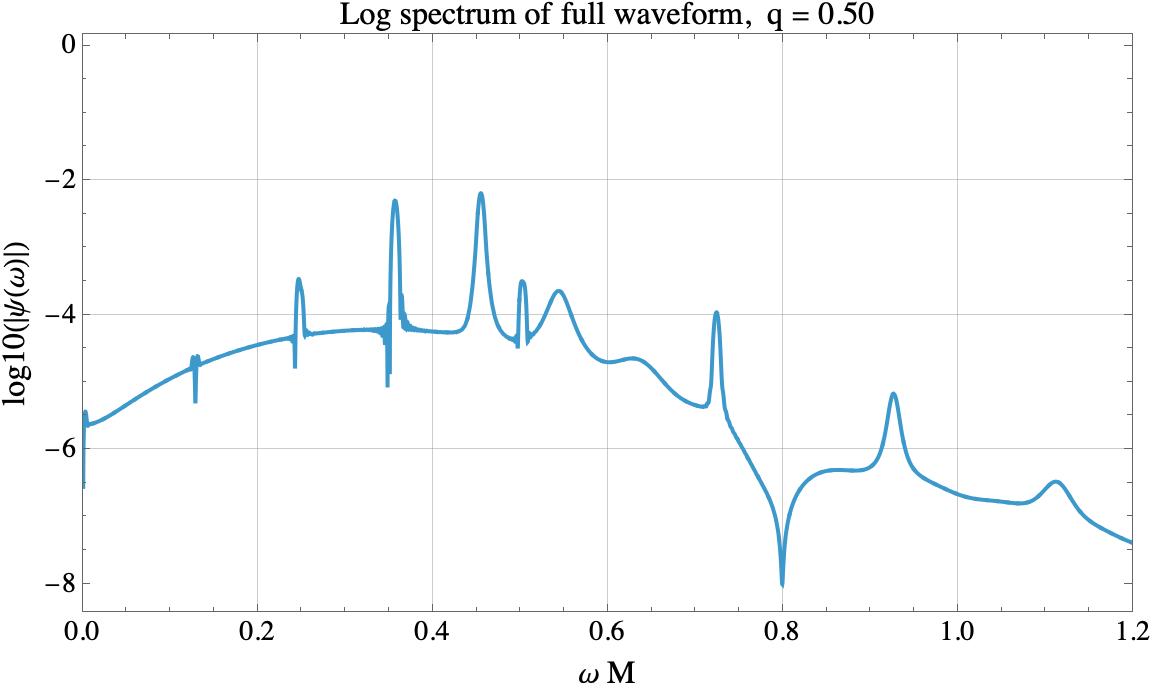}
		\end{minipage}
		
		\vspace{0.35cm}
		\begin{minipage}{0.24\textwidth}
			\centering
			\includegraphics[width=\linewidth]{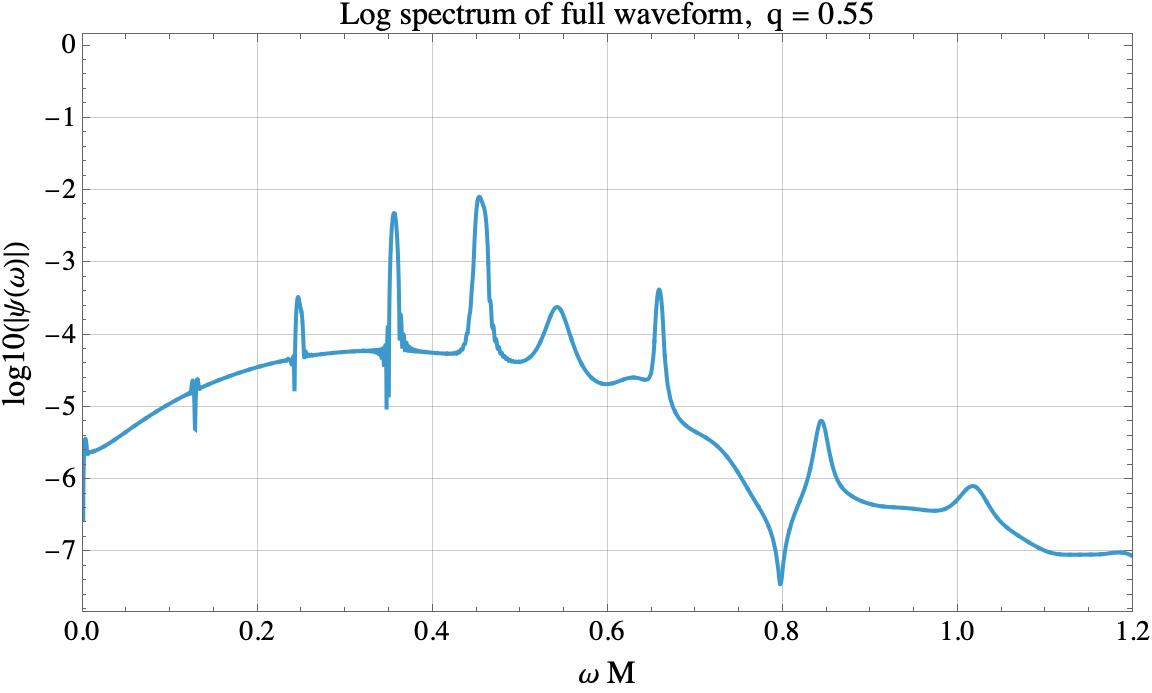}
		\end{minipage}
		\hfill
		\begin{minipage}{0.24\textwidth}
			\centering
			\includegraphics[width=\linewidth]{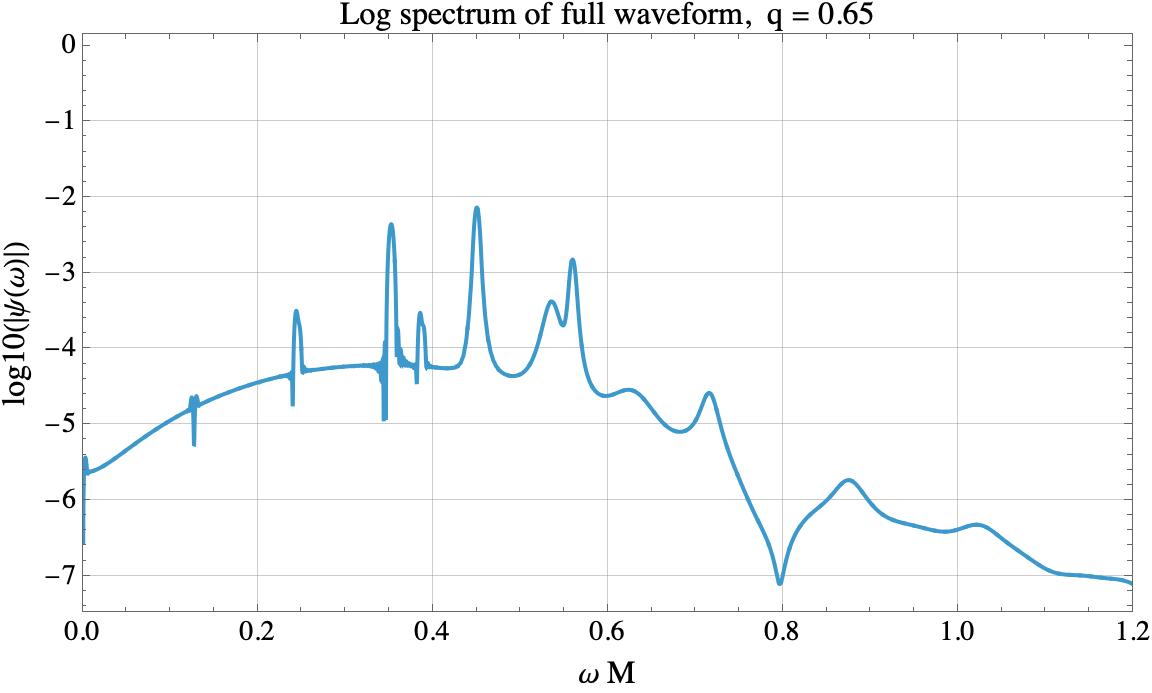}
		\end{minipage}
		\hfill
		\begin{minipage}{0.24\textwidth}
			\centering
			\includegraphics[width=\linewidth]{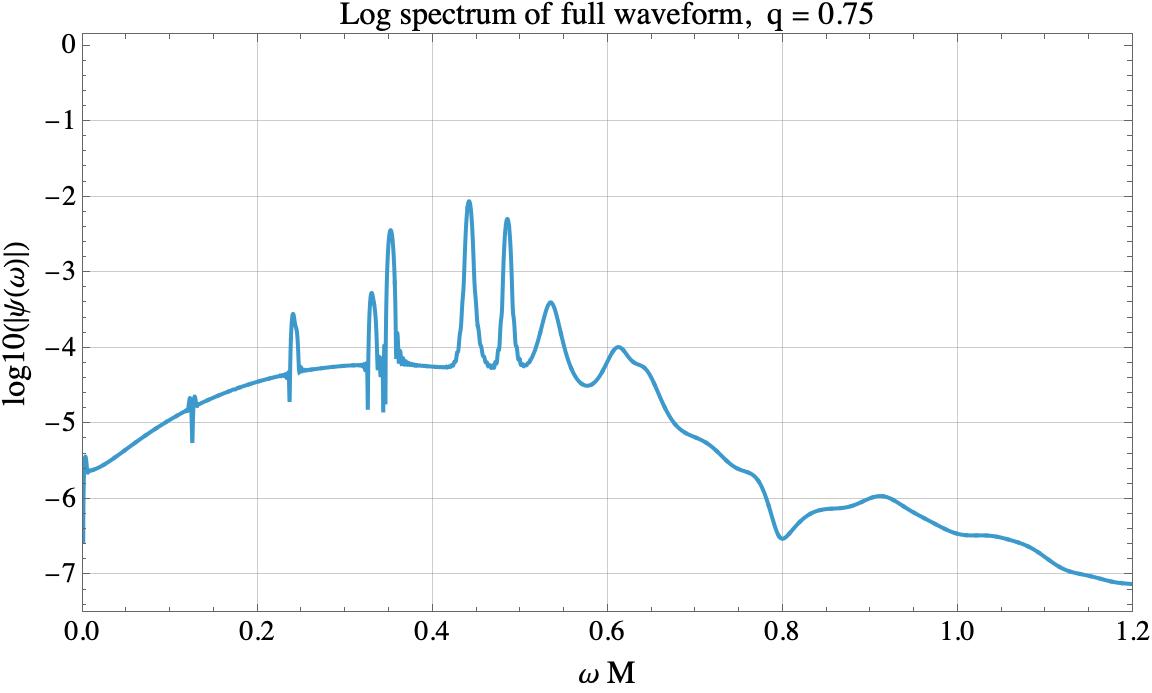}
		\end{minipage}
		\hfill
		\begin{minipage}{0.24\textwidth}
			\centering
			\includegraphics[width=\linewidth]{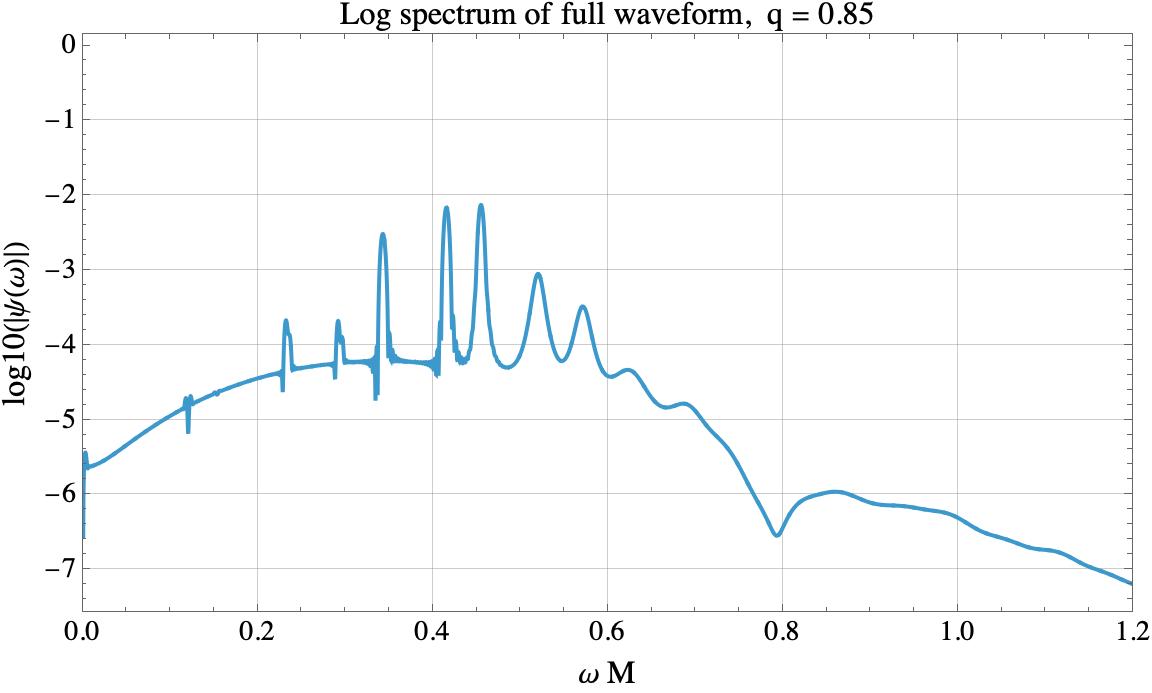}
		\end{minipage}
		
		\vspace{0.35cm}
		\begin{minipage}{0.24\textwidth}
			\centering
			\includegraphics[width=\linewidth]{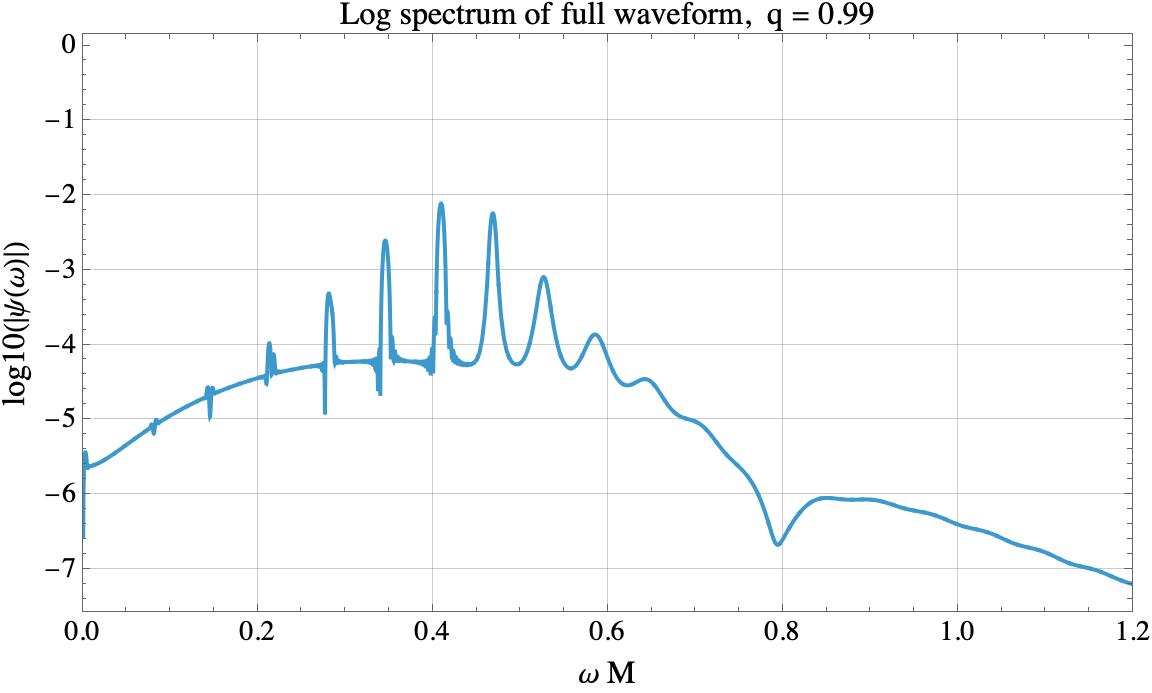}
		\end{minipage}
		\hfill
		\begin{minipage}{0.24\textwidth}
			\centering
			\includegraphics[width=\linewidth]{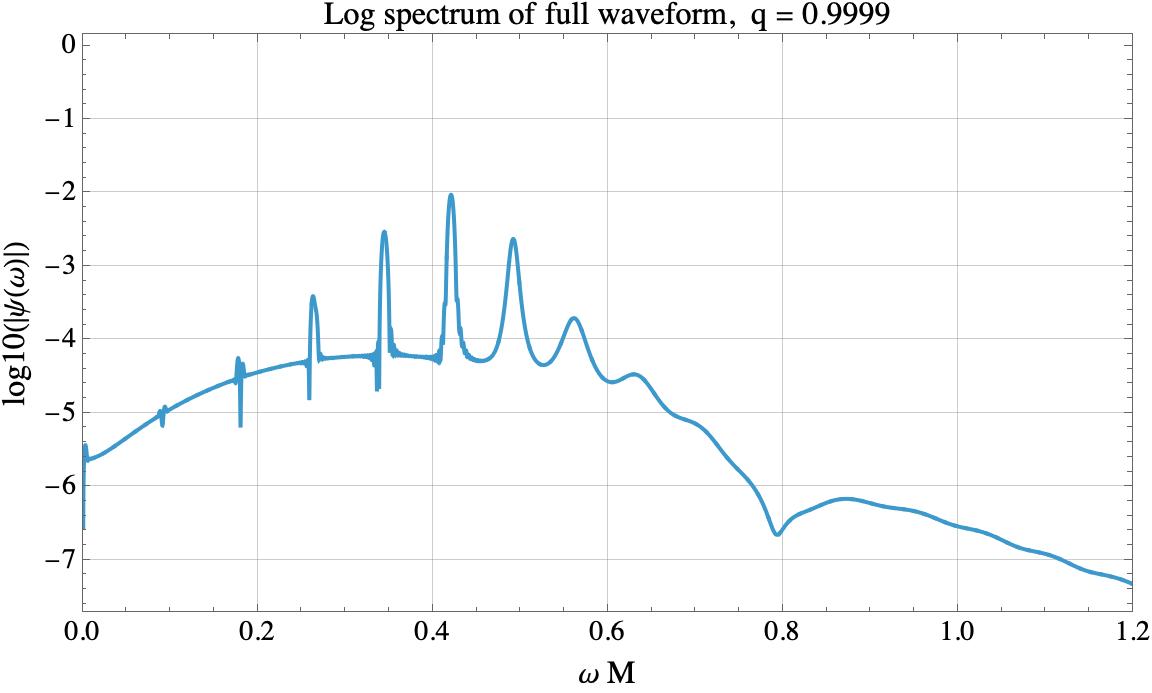}
		\end{minipage}
		\hfill
		\begin{minipage}{0.24\textwidth}
			\centering
			\includegraphics[width=\linewidth]{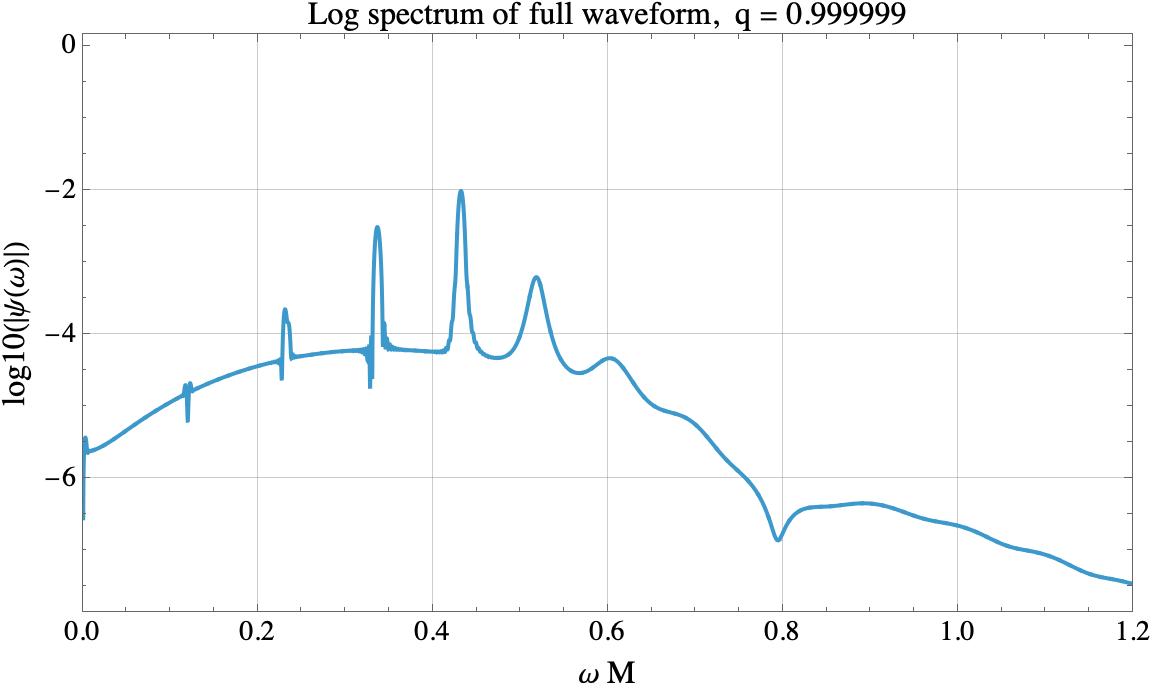}
		\end{minipage}
		\hfill
		\begin{minipage}{0.24\textwidth}
			\centering
			\includegraphics[width=\linewidth]{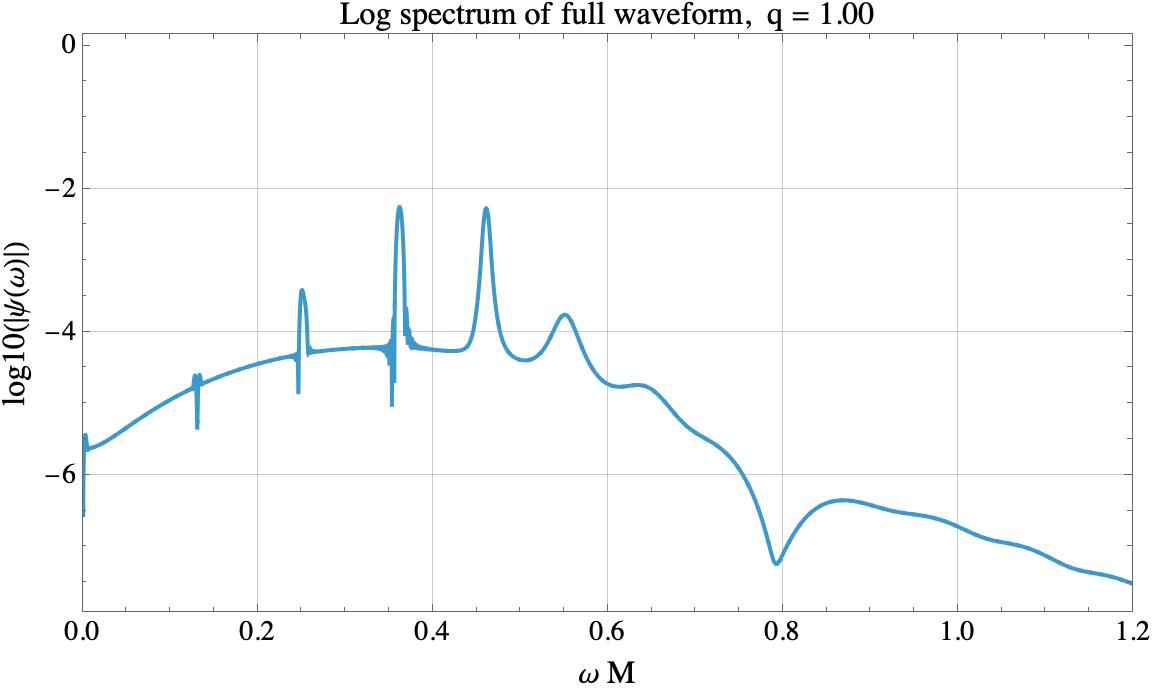}
		\end{minipage}
		
		\caption{
			$q$-dependent waveform spectra for scalar $\ell=2$ perturbations, exhibiting all three SQ components. The resolved endpoint mappings are given in Eqs. (\ref{scaP}) and (\ref{scaS}).
		}
		\label{fig:l=2}
	\end{figure*}

\begin{figure*}[t]
		\centering
		
		\begin{minipage}{0.24\textwidth}
			\centering
			\includegraphics[width=\linewidth]{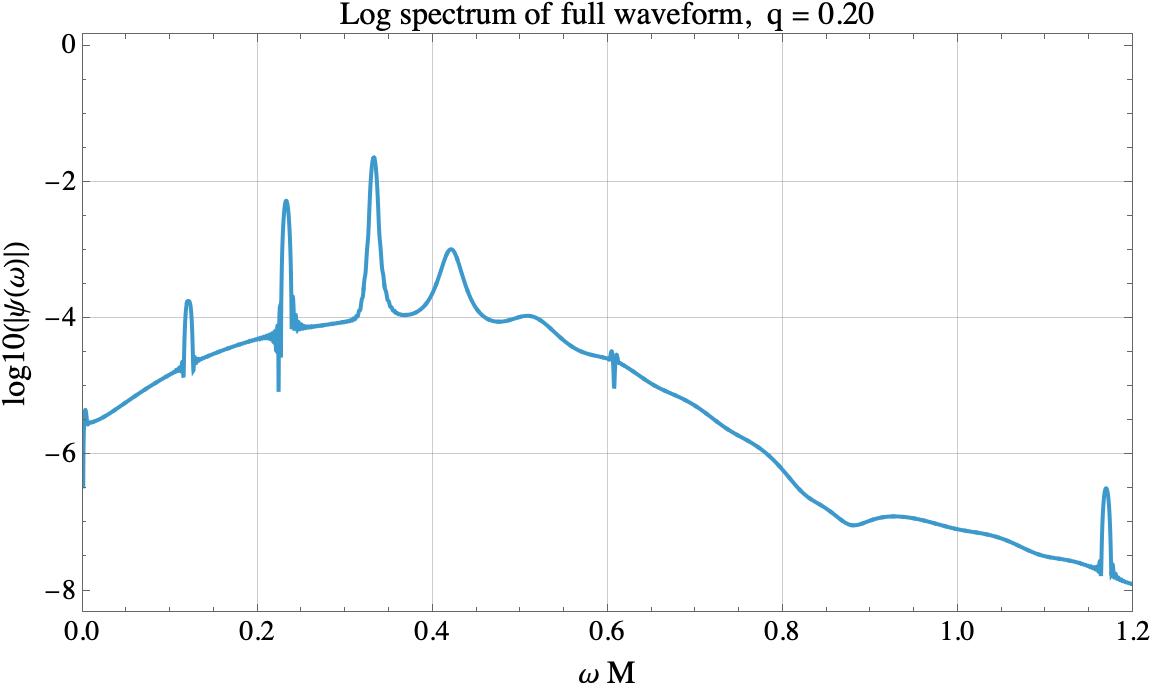}
		\end{minipage}
		\hfill
		\begin{minipage}{0.24\textwidth}
			\centering
			\includegraphics[width=\linewidth]{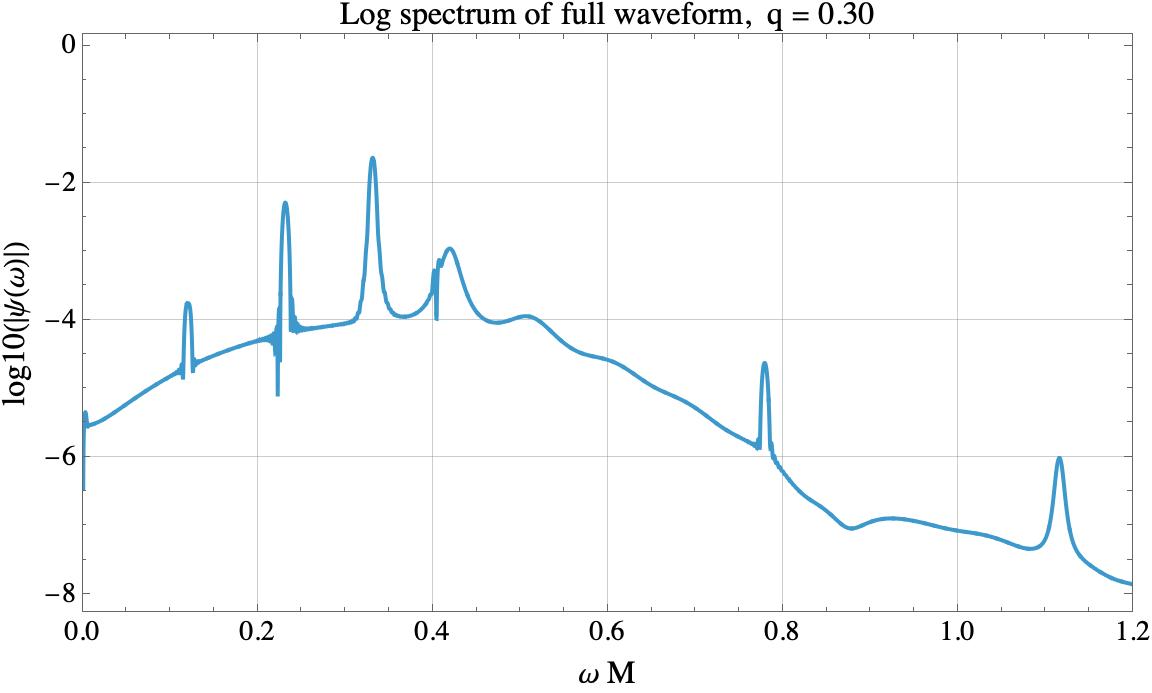}
		\end{minipage}
		\hfill
		\begin{minipage}{0.24\textwidth}
			\centering
			\includegraphics[width=\linewidth]{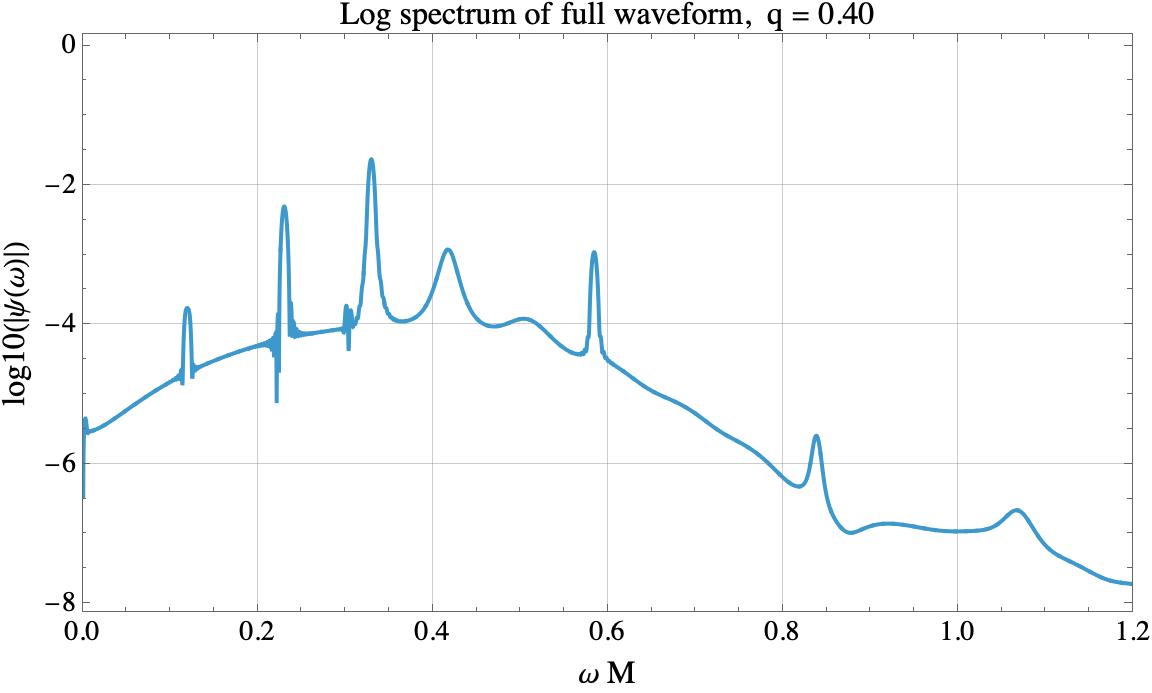}
		\end{minipage}
		\hfill
		\begin{minipage}{0.24\textwidth}
			\centering
			\includegraphics[width=\linewidth]{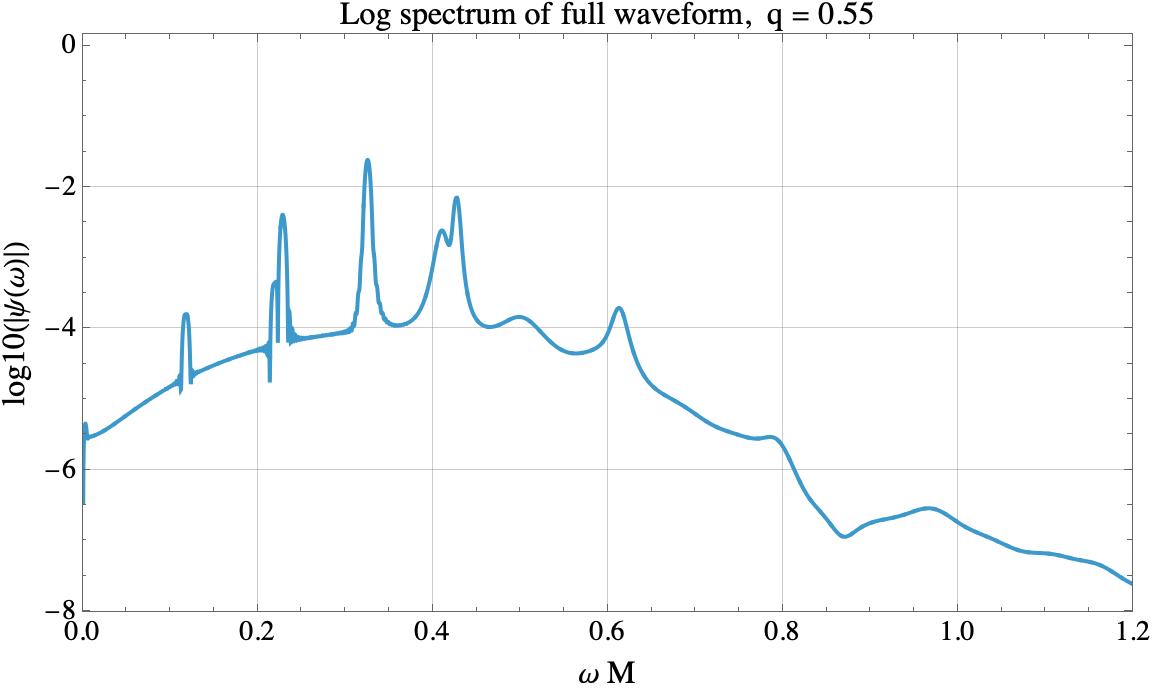}
		\end{minipage}
		
		\vspace{0.35cm}
		\begin{minipage}{0.24\textwidth}
			\centering
			\includegraphics[width=\linewidth]{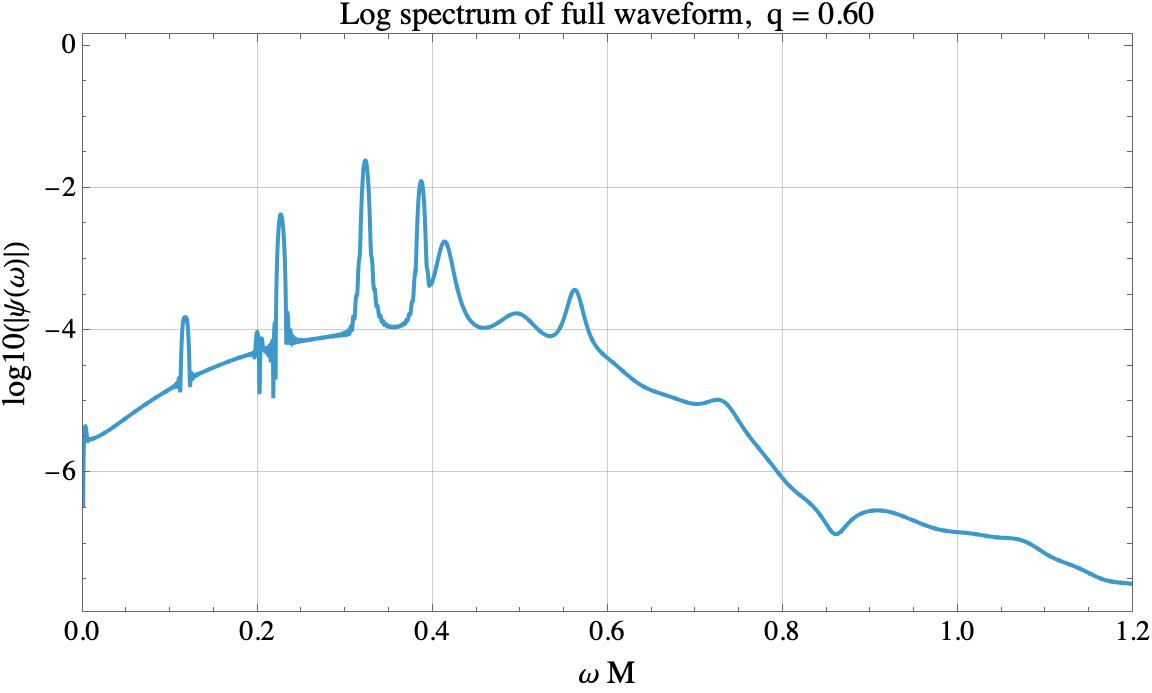}
		\end{minipage}
		\hfill
		\begin{minipage}{0.24\textwidth}
			\centering
			\includegraphics[width=\linewidth]{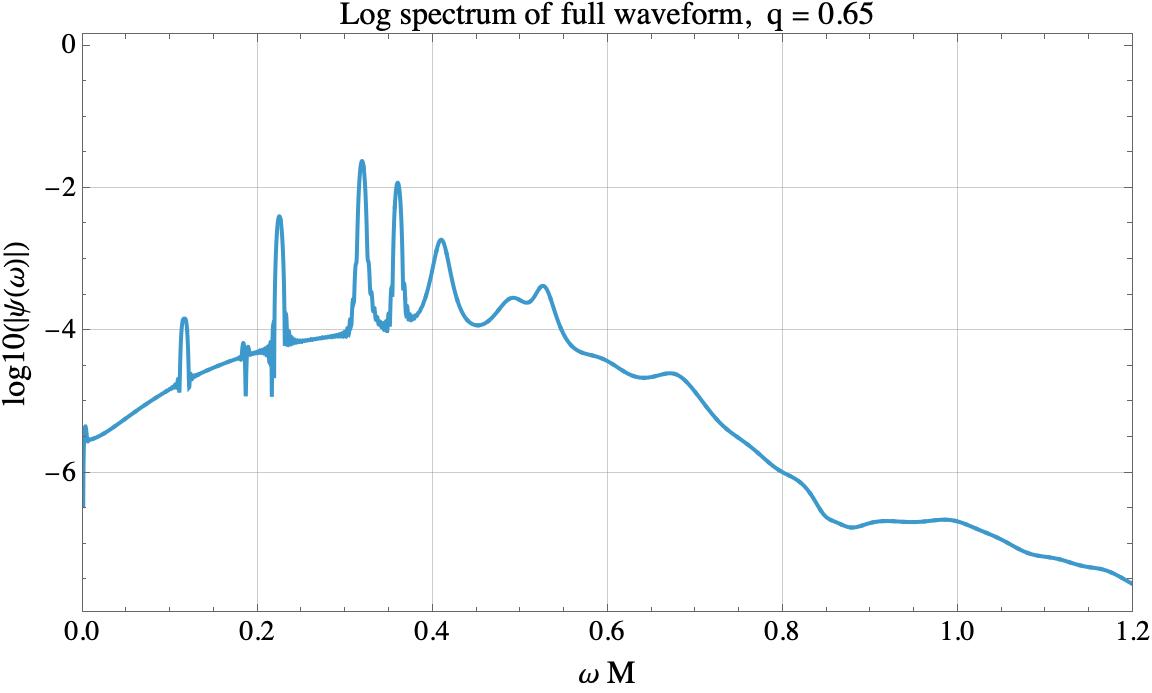}
		\end{minipage}
		\hfill
		\begin{minipage}{0.24\textwidth}
			\centering
			\includegraphics[width=\linewidth]{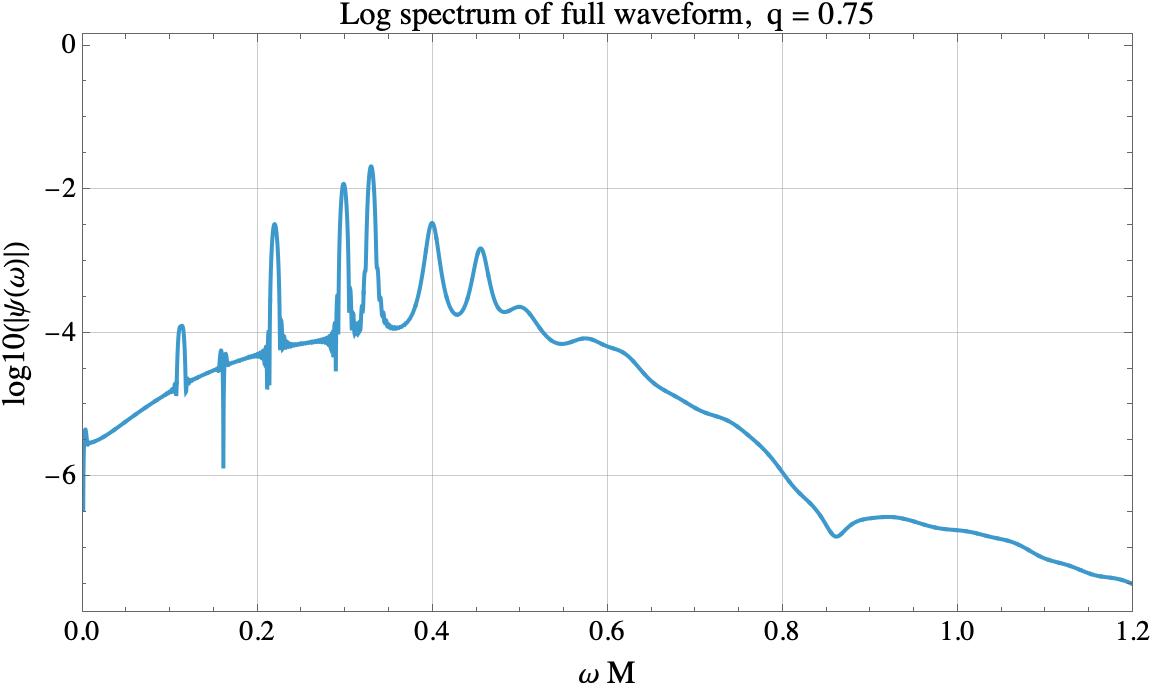}
		\end{minipage}
		\hfill
		\begin{minipage}{0.24\textwidth}
			\centering
			\includegraphics[width=\linewidth]{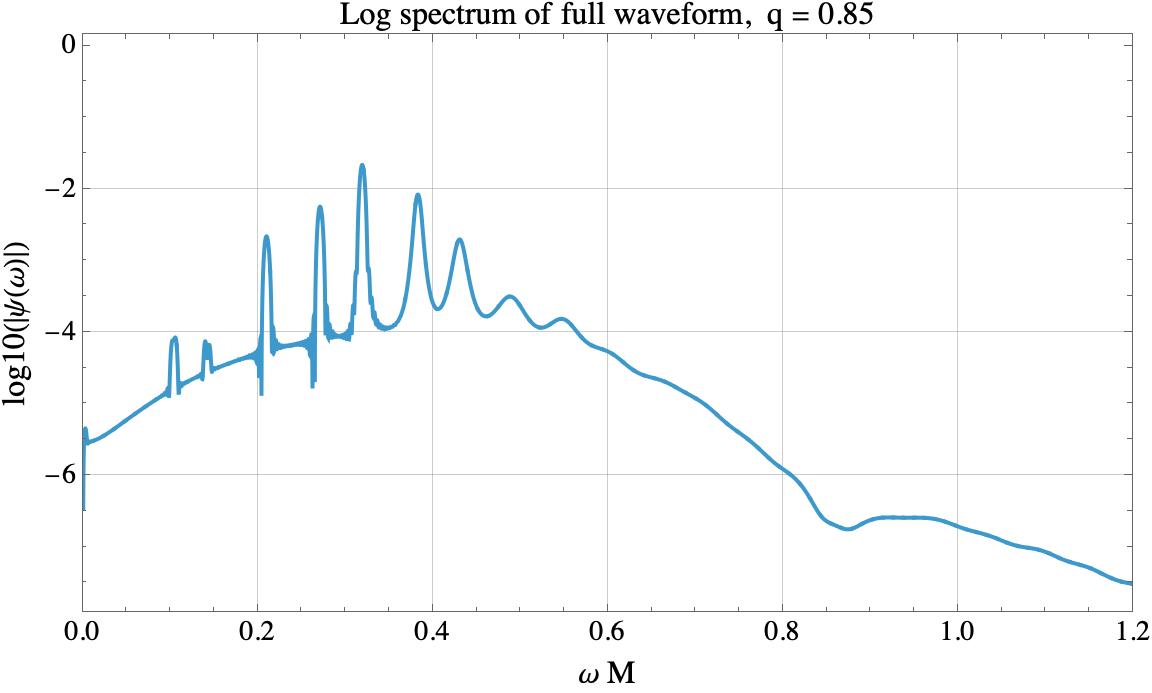}
		\end{minipage}
		
		\vspace{0.35cm}
		\begin{minipage}{0.24\textwidth}
			\centering
			\includegraphics[width=\linewidth]{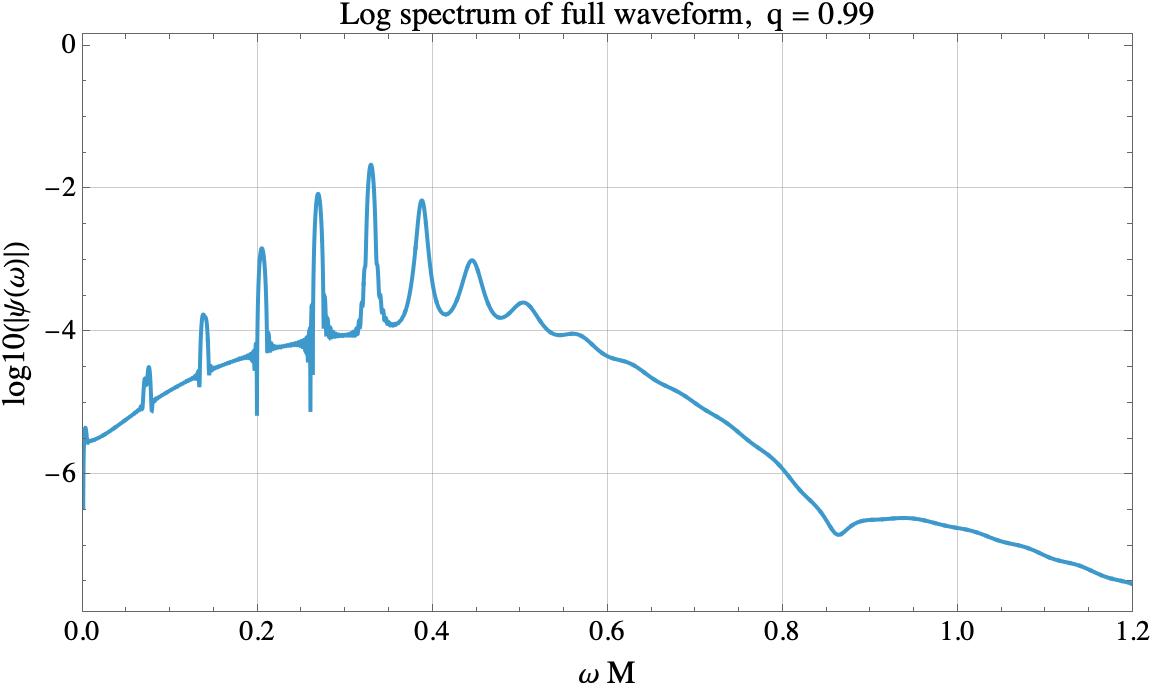}
		\end{minipage}
		\hfill
		\begin{minipage}{0.24\textwidth}
			\centering
			\includegraphics[width=\linewidth]{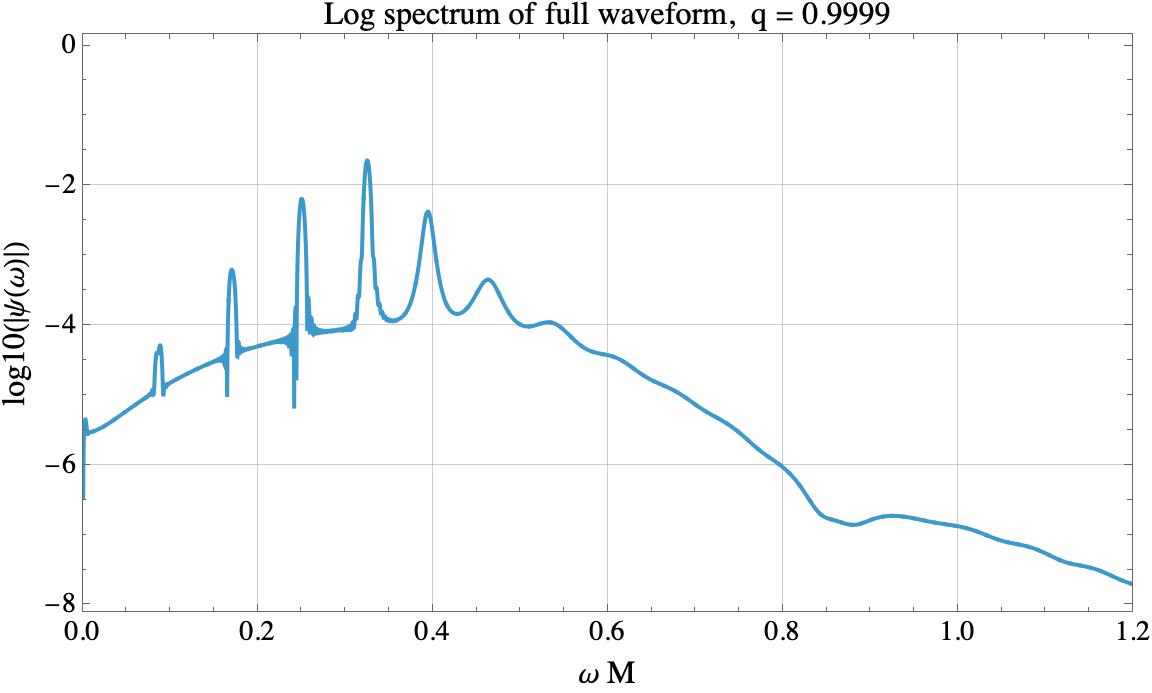}
		\end{minipage}
		\hfill
		\begin{minipage}{0.24\textwidth}
			\centering
			\includegraphics[width=\linewidth]{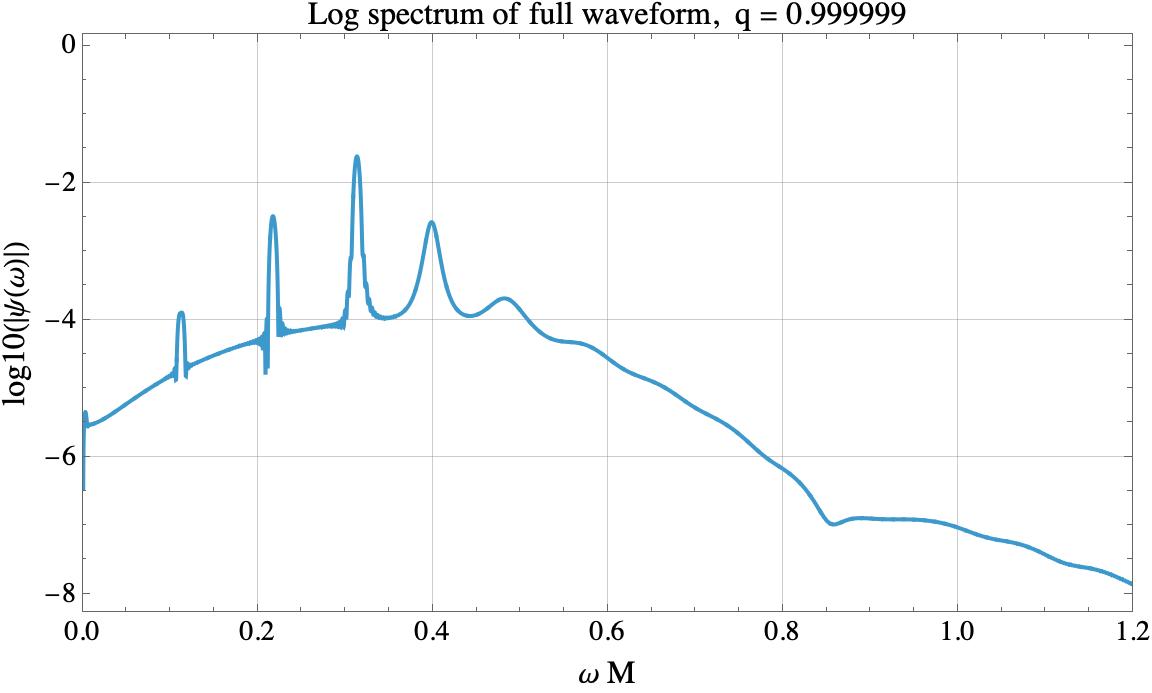}
		\end{minipage}
		\hfill
		\begin{minipage}{0.24\textwidth}
			\centering
			\includegraphics[width=\linewidth]{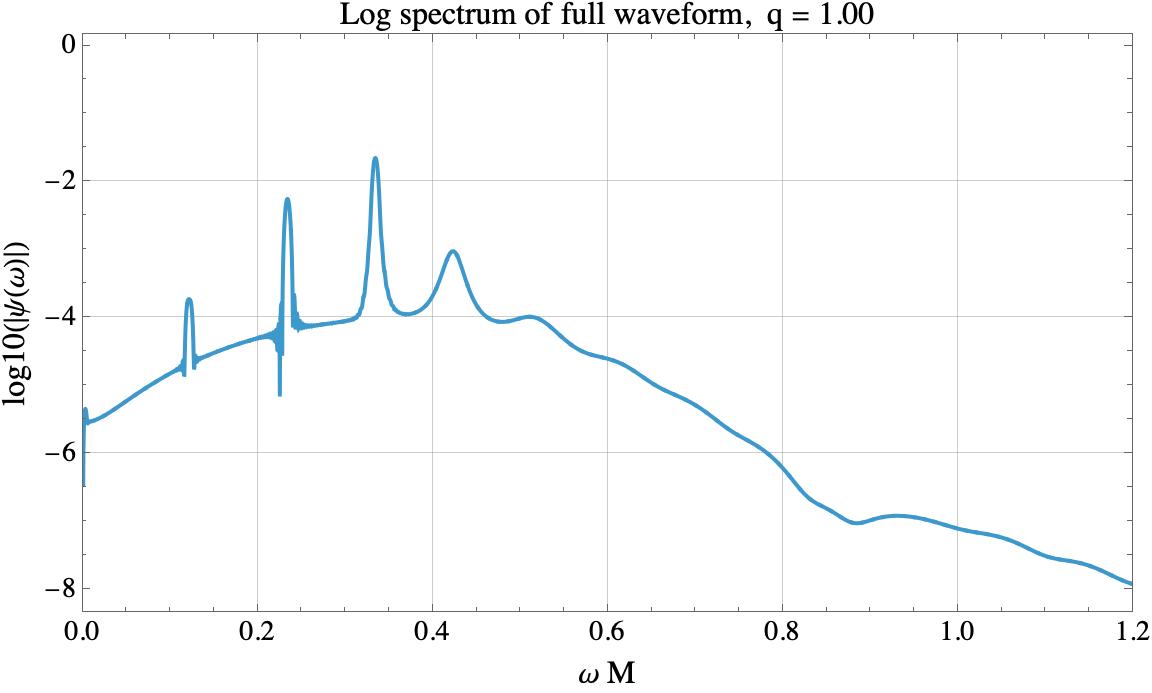}
		\end{minipage}
		
		\caption{
			$q$-dependent waveform spectra for the RW $\ell=2$ perturbations, exhibiting all three SQ components. The resolved endpoint mappings are given in Eqs. (\ref{tenP}) and (\ref{tenS}).
		}
		\label{fig:tenser}
	\end{figure*}

	To test whether SQ is specific to the fiducial scalar $\ell=1$ channel, we analyze scalar $\ell=2$ and RW $\ell=2$ perturbations on the same family of stationary backgrounds. The corresponding waveform spectra are shown in Fig.~\ref{fig:l=2} and Fig.~\ref{fig:tenser}, respectively.
	Compared with the scalar $\ell=1$ case, the scalar $\ell=2$ effective potential is higher and differs in shape. Correspondingly, its main resonance peaks occur at higher frequencies. The high-frequency peaks are also relatively more prominent, whereas the low-frequency peaks are weaker, allowing a larger number of spectral peaks to be resolved.
	
	Although its peak frequencies and amplitudes differ from those of the scalar $\ell=1$ spectrum, the scalar $\ell=2$ spectrum exhibits the same three components of SQ. Additional $S_j$ peaks undergo high-frequency series entry, the second and later peaks undergo spectral-peak migration, and the first spectral peak returns to its $q=0$ position. Within the resolvable frequency range, the corresponding endpoint mappings are
	\be
	\begin{split}
	P_1^{(0)}\to P_1^{(1)},\qquad
	P_2^{(0)}\to P_3^{(1)},\qquad
	P_3^{(0)}\to P_5^{(1)},\\
	P_4^{(0)}\to P_6^{(1)},\qquad
	P_5^{(0)}\to P_8^{(1)},\qquad
	P_6^{(0)}\to P_{10}^{(1)}.
	\label{scaP}
	\end{split}
	\ee
	At the same time, the peaks entering from the high-frequency region satisfy
	\be
	\begin{split}
	S_1\to P_2^{(1)},\qquad
	S_2\to P_4^{(1)},\qquad
	S_3\to P_7^{(1)},\\
	S_4\to P_9^{(1)},\qquad
	S_5\to P_{11}^{(1)}.
	\label{scaS}
	\end{split}
	\ee
	It should be noted that some of the weaker spectral peaks are affected, in intermediate parameter regions, by neighboring strong peaks and the spectral envelope, so that their peak strength may be significantly reduced. We therefore restrict the endpoint assignments to the dominant peaks that can be identified robustly across the sampled stationary configurations.
	
	The RW $\ell=2$ channel provides a stronger robustness test because its effective potential differs from the scalar potential. Nevertheless, the peak frequencies of the RW $\ell=2$ spectrum are nearly identical to those of the scalar $\ell=1$ spectrum, suggesting that the resonance-frequency structure is governed primarily by the characteristic propagation lengths and coupling among multiple effective cavities rather than by the detailed form of the effective potential. The RW $\ell=2$ spectrum also exhibits all three components of SQ. Within the resolvable range, the endpoint mappings are
	\be
	\begin{split}
	P_1^{(0)}\to P_1^{(1)},\qquad
	P_2^{(0)}\to P_3^{(1)},\qquad
	P_3^{(0)}\to P_4^{(1)},\\
	P_4^{(0)}\to P_6^{(1)},\qquad
	P_5^{(0)}\to P_8^{(1)}.
	\label{tenP}
	\end{split}
	\ee
	At the same time, the peaks undergoing high-frequency series entry satisfy
	\be
	S_1\to P_2^{(1)}, 
	S_2\to P_5^{(1)}, 
	S_3\to P_7^{(1)}, 
	S_4\to P_9^{(1)}.
	\label{tenS}
	\ee
	
	The main difference between the RW and scalar spectra lies in peak amplitude rather than peak frequency. In the single-shell limit, the main peak amplitudes of the RW $\ell=2$ and scalar $\ell=1$ spectra are comparable. In the double-shell configurations, however, the low-frequency RW peaks are relatively weaker, whereas the high-frequency peaks are slightly enhanced. This enhancement is more pronounced for the $S_j$ series undergoing high-frequency series entry. Thus, the two perturbation channels weight the same geometry-induced resonances differently while preserving the frequency structure of the three SQ components.
	
	We have also verified that the same SQ effect persists for $\epsilon=10^{-7}$, indicating that SQ is robust under an order-of-magnitude variation of the near-horizon parameter.
	Taken together, these results show that the SQ effect is a robust and nontrivial phenomenon.

\section{\label{sec:conclusion}Conclusion}
	
We have investigated the propagation of scalar and tensor perturbations and the resulting gravitational-wave echoes in a double-shell compact-object model, extending the single-shell setup to the minimal configuration capable of representing internal layering. The mass-distribution parameter $q=m_1/M$ continuously redistributes the total mass between the two shells, with $q=0$ and $q=1$ reducing to the same single-shell configuration, thus providing a common reference spectrum. By varying $q$ while keeping the total mass fixed, we have explored how changes in the internal mass distribution and inner-shell radius modify the echo waveform and spectrum across a family of stationary backgrounds.

For intermediate values of $q$, the layered structure introduces an additional potential barrier in the probe field propagation region, leading to systematic changes in both the time-domain echoes and their spectral structure. The central feature is the phenomenon we called spectral-peak queueing (SQ): although the $q=0$ and $q=1$ configurations have the same resonance spectrum, individual spectral peaks can undergo nontrivial rearrangement as the $q$-dependent internal configuration is varied. This shows that the spectral structure of gravitational-wave echoes can retain information about how mass is distributed within a layered compact object.

The SQ pattern persists in both scalar and Regge–Wheeler perturbations and for different angular indices examined in this work, with the main channel dependence appearing in the peak amplitudes rather than their frequency structure. This robustness suggests that SQ is primarily a consequence of the layered geometry and the associated coupling among multiple effective cavities, rather than a feature specific to a particular perturbation channel.

The parameter $q$ in our discussion labels different stationary compact-object backgrounds with different internal mass distributions and shell locations, rather than describing the time evolution of an individual object. Consequently, the SQ phenomenon should not be interpreted as a continuous frequency drift of a single source. Instead, SQ characterizes how the echo spectrum changes across a family of compact objects with different internal mass distributions. One possible observational application of the $q$-dependent SQ is to compare observed GW echo spectra with the corresponding theoretical templates. Within this model, such a comparison may constrain the coupled inner mass and shell location encoded by $q$.

The present work focuses on a double-shell model with a stationary mass distribution. A natural direction for future work is to extend the analysis to multilayered compact objects and dynamically contracting layered structures. These extensions would allow us to explore how the dynamical evolution of the layered structure modifies the effective potential, the coupled effective cavity structure, and the resulting echo waveforms and spectra. They would also provide a framework for assessing whether the SQ patterns identified in the present stationary model persist, evolve, or are replaced by new spectral signatures during gravitational collapse.
	
\FloatBarrier
\bibliography{refs}

\end{document}